\documentclass[aps,pre,reprint,showpacs,superscriptaddress,10pt]{revtex4-1}
\usepackage{amsmath,amssymb,amsfonts,graphicx,bm,dcolumn,mathrsfs}
\newcommand{\bea}{\begin{eqnarray}}
\newcommand{\eea}{\end{eqnarray}}
\newcommand{\be}{\begin{equation}}
\newcommand{\ee}{\end{equation}}
\newcommand{\dd}{\mbox{d}}
\usepackage{epsfig}

\begin{document}

\title{Complete analysis of ensemble inequivalence in the Blume-Emery-Griffiths model}

\author{V. V. Hovhannisyan}
\affiliation{A. I. Alikhanyan National Science Laboratory, 0036
Yerevan, Armenia}
\author{N. S. Ananikian}
\affiliation{A. I. Alikhanyan National Science Laboratory, 0036
Yerevan, Armenia}
\author{A. Campa}
\email{Corresponding author: campa@iss.infn.it}
\affiliation{National Center for Radiation Protection and
Computational Physics, Istituto Superiore di Sanit\`{a}, and INFN
Roma1, Viale Regina Elena 299, 00161 Roma, Italy}
\author{S. Ruffo}
\affiliation{SISSA, INFN and ISC-CNR, Via Bonomea 265, I-34136 Trieste, Italy}

\begin{abstract}
We study inequivalence of canonical and microcanonical ensembles in the mean-field
Blume-Emery-Griffiths model. This generalizes previous results obtained
for the Blume-Capel model. The phase diagram strongly depends on the value of the 
biquadratic exchange interaction $K$, the additional feature present in the Blume-Emery-Griffiths
model. At small values of $K$, as for the Blume-Capel model, lines of first and second order phase 
transitions between a ferromagnetic and a paramagnetic phase are present, separated by
a {\it tricritical point} whose location is different in the two ensembles. 
At higher values of $K$ the phase diagram changes substantially, with the appearance
of a {\it triple point} in the canonical ensemble which does not find any correspondence
in the microcanonical ensemble. Moreover, one of the first order lines that starts from the
triple point ends in a {\it critical point}, whose position in the phase diagram is
different in the two ensembles. This line separates two paramagnetic phases characterized
by a different value of the quadrupole moment. These features were not previously studied for other
models and substantially enrich the landscape of ensemble inequivalence, identifying new
aspects that had been discussed in a classification of phase transitions
based on singularity theory. Finally, we discuss {\it ergodicity breaking}, which is highlighted by 
the presence of gaps in the accessible values of magnetization at low energies: it also displays new 
interesting patterns that are not present in the Blume-Capel model.
\end{abstract}
\pacs{05.20.Gg; 64.60.Bd; 64.60.De}
\maketitle

\section{Introduction}
In the last couple of decades it has become progressively clear that
systems with long-range interactions show peculiar
properties~\cite{Campa14}. For such systems, the interaction between
constituents decays, at large distance $r$, as $r^{-\alpha}$, with
$\alpha$ smaller than space dimension $d$.

The most prominent property of these systems at thermodynamic equilibrium is {\it ensemble
inequivalence}~\cite{Barre01}: phase diagrams can be different in the canonical and microcanonical ensembles,
This property directly stems from the {\it non additivity} of the energy. There are cases, however, in presence
of only second order transitions in both ensembles, where inequivalence does not occur~\cite{Barre01,touch04}.

Self-gravitating systems~\cite{Binney} can be considered as the
paradigmatic example of long-range interactions. One of the most
important consequences of ensemble inequivalence, i.e. negative
specific heat in the microcanonical ensemble, has been found in
self-gravitating systems already about half a century
ago~\cite{Antonov62,Lydnen-Bell68,Thirring70}. When an isolated system has a negative specific heat,
an increase in its temperature is associated to a decrease of its energy. In a self-gravitating system
this occur when, e.g, a contraction of a globular cluster is accompanied by an increase of the average kinetic energy
of its members.
However, it is now evident that there are other numerous examples of long-range
systems in nature: Coulomb systems like plasmas~\cite{nicholson},
trapped cold atoms~\cite{Morigi}, two-dimensional geophysical and
hydrodynamic flows~\cite{bouchet}, dipolar
media~\cite{Bramwell}, nuclear matter~\cite{chomaz,foot6}.

The investigation of the properties of long-range interacting systems has often been pursued with the use of simplified models,
more amenable to a complete study, that can present the most important features of more realistic systems. Spin systems with mean-field
interactions (i.e., with each spin interacting with equal strength with
all the others) constitute a good example. Indeed, ensemble inequivalence was first discussed for the spin-1 mean-field Blume-Capel (BC)
model~\cite{Blume66} in Ref.~\cite{Barre01}.
It has been shown that the two-dimensional phase diagram $(\Delta,T)$, where $T$ is temperature and
$\Delta$ the single-spin energy parameter, exhibits a tricritical point where a line of second order
transitions meets a line of first order transitions. However, both the location of the tricritical point
and the one of first order transitions are different in the canonical and microcanonical ensembles.
Regions of negative specific heat are present in the microcanonical ensemble in correspondence to a first order
transition in the canonical ensemble. It can be seen that these features are
quite common in mean-field systems~\cite{Campa09}.

A physically relevant generalization of the Blume-Capel model is the
Blume-Emery-Griffiths (BEG) model~\cite{Blume71}. This latter was
introduced as a simplified model to study phase separation and
superfluid transitions in $He^3-He^4$ mixtures. It consists of a
system of spin-1 variables $S_i$ on a three-dimensional lattice with
nearest neighbour interactions. The value $S_i=0$ corresponds to the
presence on site $i$ of a $He^3$ atom, while the values $S_i=1$ and
$S_i=-1$ are associated with the presence of a $He^4$ atom (the
additional degree of freedom is justified by physical observations~\cite{Blume71}).
The analytical study was performed in the mean-field approximation
in the canonical ensemble. The same results can be obtained considering
mean-field interactions, i.e. putting the model on a fully connected
infinite-range lattice.

The solution in the microcanonical ensemble of the mean-field BEG model has never been performed and the corresponding ensemble
inequivalence never discussed in the literature. Moreover, in spite of the numerous studies on ensemble inequivalence, the investigation
has been mostly restricted to phase diagrams with tricritical points. However, it is well known
that more complex phase diagrams showing inequivalence can exist~\cite{bouchet05}.

Recently, the solution of the Thirring model~\cite{Thirring70} has been presented in full
detail~\cite{campa2016} and it has been shown that the phase diagram presents lines of first order transitions
terminating at critical points. However, the location of the critical point is not the same in
the canonical and microcanonical phase diagrams.

The motivation of this work is to extend the study of ensemble inequivalence to more complex situations like the
BEG model. The presence of a further parameter with respect to the BC model, i.e. the coefficient of the biquadratic interaction,
determines the occurrence of a much more complex phase diagram made of critical, tricritical and triple points,
lines of second and first order transitions. In the canonical ensemble these features have been
already studied in Refs.~\cite{Blume71,Hoston}. By studying the system also in the
microcanonical ensemble, our purpose here is to show that ensemble inequivalence manifests
itself in various different interesting ways.

The paper is organized as follows: in Sec.~\ref{model} we describe
in detail the solution of the BEG model in the canonical ensemble by
means of the Gaussian identity. In Sec.~\ref{Microcanonical} the
solution of the model in the microcanonical ensemble is presented.
In Sec.~\ref{results} the most interesting features of
the model are discussed and the analysis of ensemble inequivalence for the
canonical and microcanonical ensembles is completed. In
Sec.~\ref{ergbreak} we study ergodicity breaking, another
important feature often present in long-range systems. Finally, some
conclusions are briefly mentioned in Sec.~\ref{conclusions}.

\section{The model and its solution in the canonical ensemble}
\label{model}

Let us consider the infinite range Blume-Emery-Griffiths~\cite{Blume71} model in
the absence of external magnetic field. The Hamiltonian of the
model is
\begin{eqnarray}
H = \Delta \sum_{i=1}^{N} S_i^2 -\frac{J}{2N}\left(\sum_{i=1}^{N}
S_i\right)^2 -\frac{K}{2N}\left(\sum_{i=1}^{N} S_i^2\right)^2,
\label{hamil}
\end{eqnarray}
where $\Delta$ is the single-spin energy parameter (single-ion anisotropy)
controlling the energy difference between the ferromagnetic
($S_i=\pm 1$) and the paramagnetic ($S_i=0$) states, while $J$ and
$K$ are the bilinear and the biquadratic exchange interaction
parameters, respectively. The partition function reads
\begin{equation}
Z (\beta, N)= \sum_{\{S_{i}\}} e^{ - \beta \Delta
\sum_{i=1}^{N} S_i^2 + \frac{\beta J}{2N}\left(\sum_{i=1}^{N}
S_i\right)^2 +\frac{\beta K}{2N}\left(\sum_{i=1}^{N} S_i^2\right)^2} \, ,
\label{PF}
\end{equation}
where $\beta=(k_{B}T)^{-1}$, $k_{B}$ is Boltzmann's constant and $T$
is the absolute temperature. In the following we use units in which $k_B=1$.
Furthermore, without loss of generality, we can take $J=1$ (this formally amounts
to the substitutions $\beta \to \beta J$, $\Delta \to \Delta/J$ and $K \to K/J$).
Using the Gaussian identity
\begin{eqnarray}
\mathrm{exp} (ba^2)=\sqrt{\frac{b}{\pi}} \int^{+\infty}_{-\infty}
\dd x \, \mathrm{exp} (-bx^2 +2abx) , \label{HST}
\end{eqnarray}
one can then rewrite the partition function of the system as (Hubbard-Stratonovich transformation)
\begin{eqnarray}
&Z (\beta, N)&= \frac{N\beta \sqrt{K}}{2\pi} \sum_{\{S_{i}\}} e^{ - \beta \Delta
\sum_{i=1}^{N} S_i^2 }
 \\
& \times & \!\!\!\!\!\! \int^{+\infty}_{-\infty} \dd x \, \int^{+\infty}_{-\infty} \dd y \,
e^{ -\frac{\beta N }{2}x^2 -\frac{\beta N K}{2}y^2 +\beta N  m x
+\beta N K q y } \, , \nonumber
\label{PFh}
\end{eqnarray}
where $m=\sum_{i} S_i/ N$ and $q=\sum_{i} S_i^2 / N$ are the
magnetization and the quadrupole moment per particle, respectively. Performing the
sum over ${\{S_{i}\}}$ we get
\begin{equation}
Z (\beta, N)= \frac{N\beta \sqrt{K}}{2\pi} \int
\int^{+\infty}_{-\infty} \dd x \dd y \, e^{ -N \beta \tilde{f} (\beta, x, y) } \, ,
\label{PF1}
\end{equation}
where
\begin{eqnarray}
\beta \tilde{f} (\beta, x, y) &=& \frac{\beta}{2} \left( x^2 + K y^2 \right)
\nonumber \\ &-&
\ln \left[ 1+ 2 e^{-\beta \Delta + \beta K y} \cosh \beta  x \right] \, .
\label{free_energy}
\end{eqnarray}
The integral in Eq.~\ref{PF1} can be performed using Laplace (saddle point)
method, which in the limit $N \rightarrow
\infty$ gives the expression of the free energy per particle
\begin{eqnarray}
f(\beta) = \min_{x,y} \tilde{f} (\beta, x, y). \label{free_energy1}
\end{eqnarray}
The partial derivatives of $\tilde{f}$ with respect
to $x$ and $y$ vanish in the saddle point, giving:
\begin{eqnarray}
x = \frac{2 \sinh \beta x}{ \exp \left( \beta \Delta -
\beta K y \right) + 2\cosh \beta x},
\label{mag} \\
y = \frac{2 \cosh \beta  x}{ \exp \left(\beta \Delta -
\beta K y \right) + 2\cosh \beta x}.
 \label{quad}
\end{eqnarray}
Furthermore the Hessian matrix of $\tilde{f}$ must be positive
definite at the values of $x$ and $y$ that solve these equations. If there is more than
one solution, the relevant one for the equilibrium state is that
realizing the absolute minimum of $\tilde{f}$. One can easily see that these values
of $x$ and $y$ correspond to the equilibrium
magnetization $m$ and quadrupole moment $q$, respectively.
Using Eqs.~(\ref{mag}) and (\ref{quad}), it can be shown that, when $x \ne 0$, the
magnetization and the quadrupole moment at equilibrium are related
by
\begin{eqnarray}
y = x \coth \beta  x.
 \label{Q}
\end{eqnarray}
This relation will turn out to be useful in the following.

We begin the study of the thermodynamic phase diagram of the system by considering
two particular cases, i.e. $\Delta=0$ and $T=0$.

Let us first consider the case $\Delta=0$. It is not difficult to
see that in this case the system exhibits a second order phase
transition, by increasing $T$, from a ferromagnetic ($m \ne 0$)
to a paramagnetic ($m=0$) phase. In fact, for $\Delta=0$ the
right hand side of Eq.~(\ref{mag}) has, as a function of $x$, a
positive definite first derivative and a negative definite second
derivative for $x>0$. This implies that there is a second order
phase transition at the value of $\beta$ where the $x$ derivative of
this function at $x=0$ is equal to $1$. In evaluating this
derivative we have to take the value of $y$ equal to the limit of
Eq.~(\ref{Q}) for $x\to 0$, i.e., $y=1/\beta$. Then, the critical
$\beta$ value is obtained as $\beta_c(\Delta=0) = 1+e^{-K}/2$, i.e.,
$T_c(\Delta=0)=(1+e^{-K}/2)^{-1}$. This is, consistently, the value
of $\beta$ for which Eq.~(\ref{quad}) has the solution $y=1/\beta$
for $x=0$.

On the other hand, for $T=0$ the system exhibits a first order phase transition. In fact, for $T=0$
the equilibrium state is the one of minimum energy. It is easy to see that the minimum energy state
is the fully magnetized state with $m=1$ or $m=-1$ (i.e., $S_i= 1$ $\forall i$ or $S_i=-1$ $\forall i$)
and $q=1$ when $\Delta < (K+1)/2 \equiv \Delta_c$, while it is the paramagnetic state with $m=q=0$
(i.e., $S_i=0$ $\forall i$) when $\Delta > (K+1)/2$. Thus, in the $(\Delta,T)$ phase diagram we have, for
any given $K$, a second order transition at the point $(0,T_c)$ and a first order transition at the point
$(\Delta_c,0)$. We should then expect to find, in the phase diagram, both a line of second order
transitions and a line of first order transitions as a continuation of these two
extreme cases. We now show that this is the case.

\begin{figure}[ht]
\begin{center}
\includegraphics[width=6.5cm]{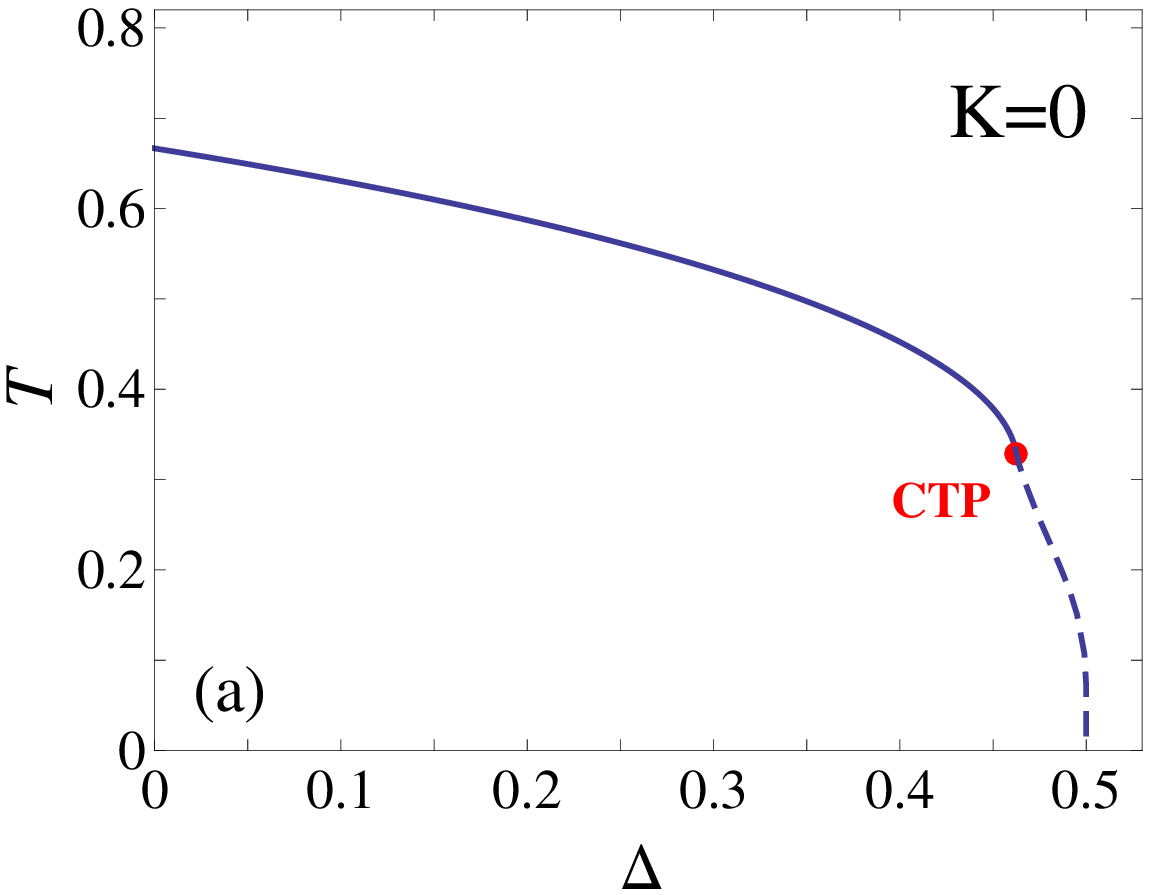}\\
\includegraphics[width=6.5cm]{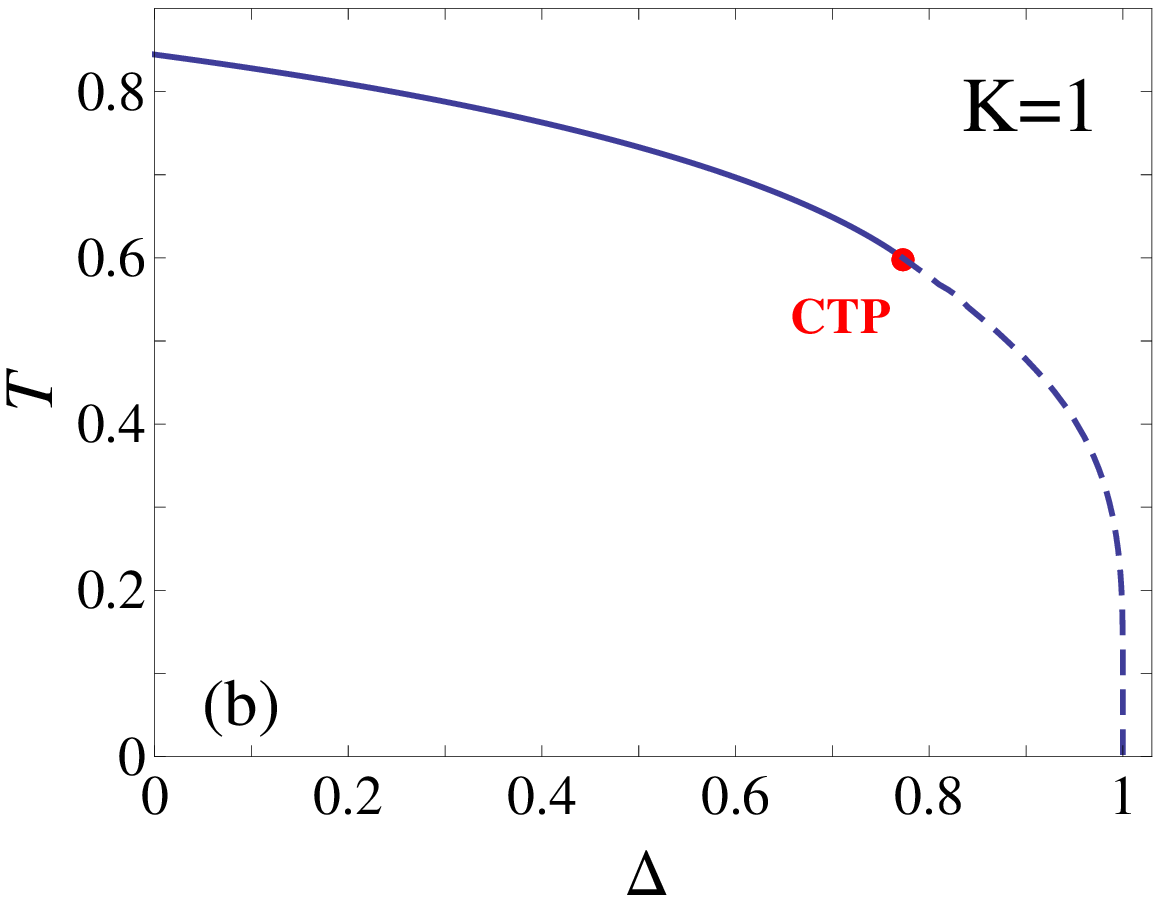}
\caption {$(\Delta,T)$ phase diagram of the BEG model in the
canonical ensemble for two different values of the biquadratic
interaction parameter $K$. Solid and dashed lines are second and
first order phase transition lines, respectively, while the red dot
denotes the canonical tricritical point (CTP). (a) $K=0$, i.e. the BC model,
$\Delta_{CTP} \backsimeq 0.4621$, $T_{CTP}=1/3$; (b) $K=1$,
$\Delta_{CTP} \backsimeq 0.7726$, $T_{CTP}=0.6$.} \label{Phasec}
\end{center}
\end{figure}

The critical line of second order transitions from a ferromagnetic ($m\ne 0$) state to a paramagnetic ($m=0$) state
can be obtained following Landau theory of phase transitions. If $\tilde{f}$ would depend, as far as order parameters are concerned, only
on $x$, we know that the critical points would be given by $\partial \tilde{f}/\partial x =
\partial^2 \tilde{f}/\partial x^2 = \partial^3 \tilde{f}/\partial x^3 = 0$, with
$\partial^4 \tilde{f}/\partial x^4 > 0$. 
Since in our case $\tilde{f}$ depends on $x$ and $y$, the critical points, besides
satisfying Eqs.~(\ref{mag}) and (\ref{quad}), should make the determinant of the Hessian
matrix vanish, i.e. $(\partial^2 \tilde{f}/\partial x^2)(\partial^2 \tilde{f}/\partial
y^2) -(\partial^2 \tilde{f}/\partial x \partial y)^2=0$. Besides that, we require
the vanishing of the third order variation and the positiveness of the fourth order variation of 
$\tilde{f}$ along a particular path on the $(x,y)$ plane passing through the given equilibrium 
point.
Details of this procedure are given in Appendix~\ref{appcantric};
here we just give the result. The third order variation, for the
solutions of Eqs.~(\ref{mag}) and (\ref{quad}) with $x=0$,
identically vanishes since $\tilde{f}$ is an even function of $x$.
We are left with the vanishing of the Hessian and the
positiveness of the fourth order variation; these conditions are
expressed by
\begin{eqnarray}
A_c = \exp (\beta \Delta -K)+2 - 2\beta &=& 0
\label{accan} \\
B_c = -\beta \left( 1+2K \right) + \left( 3 + 2K \right) &>&0
\label{bccan}
\end{eqnarray}
From Eq. (\ref{accan}) we get the equation of the critical line in the
$(\Delta,T)$ phase diagram as
\begin{equation}
\beta = \frac{1}{2}e^{\beta \Delta - K}+1 \, .
\label{Second-order}
\end{equation}
For the particular case $\Delta=0$ we obtain the value $\beta_c=1/T_c$
given above. Eq.~(\ref{bccan}) gives
\begin{equation}
\label{bccan0} \beta < \frac{3+2K}{1+2K} \, .
\end{equation}
The equality in the last equation gives the value of $\beta$ at the canonical tricritical point (CTP), i.e., the point where
the line of second order transitions ends and the line of first order transitions begins. It is characterized by
the vanishing of both the second order and the fourth order variations of $\tilde{f}$.
In terms of the temperature $T$
\begin{equation}
T_{CTP}=\frac{1+2K}{3+2K} \, ,
\label{TCTP}
\end{equation}
and the value of $\Delta$ at the tricritical point is then obtained
by Eq.~(\ref{Second-order}) as
\begin{equation}
\Delta_{CTP}=T_{CTP} \ln\frac{4 e^{K}}{1+2K} \, .
\label{DeltaCTP}
\end{equation}
After the tricritical point the first order transition line has to
be computed numerically: the function $\tilde{f}(\beta,x,y)$ has local minima, in the $(x,y)$ plane, both
at $x=0$ and at $x\ne 0$, the former corresponding to a paramagnetic phase and the latter to a ferromagnetic phase. The equilibrium state
is the one corresponding to the global minimum (with the magnetization $m$ equal to the $x$ of the global minimum). The points of the $(\Delta,T)$
phase diagram where the two local minima are equal give the first order transition line. In Fig.~\ref{Phasec} we show the
phase diagram for two cases, $K=0$ and $K=1$.

We will see that for larger values of $K$ the phase diagram is more complex.
The new main feature is the appearance of a {\it triple point} from which three
lines of first order transitions depart. One of them, which separates two
paramagnetic phases, ends in a critical point. For even larger values of $K$ the
tricritical point disappears, since the critical line branches in
two lines of first order transitions before reaching the tricritical
point. Details will be given in Sec.~\ref{results}, while
treating ensemble inequivalence. We now turn to the study of the
microcanonical phase diagram.

\section{Solution in the microcanonical ensemble}
\label{Microcanonical}

Now, let us consider the BEG model in microcanonical ensemble.
Suppose that, in the macroscopic system with $N$ particles, $N_+$,
$N_-$ and $N_0$ are the number of particles with up, down and zero
spins, respectively, with $N=N_+ + N_- + N_0$. The energy $E$ of the system can be expressed
as a function of these occupation numbers as
\begin{eqnarray}
E = \Delta Q -\frac{1}{2N} M^2 -\frac{K}{2N} Q^2, \label{enerDQM}
\end{eqnarray}
where $M=\sum_{i=1}^{N} S_i = N_+ -N_-$ and $Q=\sum_{i=1}^{N}
S_i^2=N_+ +N_-$ are the magnetic and quadrupole moments. 
The number $W$ of microscopic configurations with macroscopic
occupation numbers $N_+$, $N_-$ and $N_0$ is
\begin{eqnarray}
W = \frac{N!}{N_+! N_-! N_0!}. \label{numberW}
\end{eqnarray}
The associated entropy $S=\ln W$ can be computed in the thermodynamic
limit $N \rightarrow \infty$ using Stirling's approximation. We obtain 
\begin{eqnarray}
S &=& - N \left[ (1-q)\ln(1-q)+\frac{1}{2}(q+m)\ln(q+m) \right. \nonumber \\
&+& \left. \frac{1}{2}(q-m)\ln(q-m)-q\ln2 \right] \, ,
\label{entropyqm}
\end{eqnarray}
where $m=M/N$ and $q=Q/N$ are the single-site magnetization and
quadrupole moment. This is not yet the equilibrium
entropy, which is given by optimizing the macroscopic occupation
numbers at fixed energy $E$~\cite{foot7}. To this purpose, let us
introduce the single-site energy $\epsilon = E/N$ and rewrite
Eq.~(\ref{enerDQM}) as
\begin{eqnarray}
q^2 - 2 \frac{\Delta}{K} q + \frac{2\epsilon}{K} + \frac{m^2}{K}=0 \, .
\label{enerqmeps}
\end{eqnarray}
Eq.~(\ref{enerqmeps}) is a quadratic equation with respect to $q$
with solutions
\begin{equation}
q_{\pm} = \frac{\Delta}{K} \pm \sqrt{\left(\frac{\Delta}{K}\right)^2-\frac{2\epsilon }{K} -\frac{m^2}{K}} \, .
\label{qpmsol}
\end{equation}
Obviously the range of variation of the energy $\epsilon$ must be such
that the expression under square root in the last equation is never
negative (the bounds of $\epsilon$ will be discussed in detail in
Sec.~\ref{ergbreak}, dedicated to ergodicity breaking). Moreover, 
at least one between the quantities $q_+$ and $q_-$ must lie in the interval $[0,1]$;
when one of the two is not between $0$ and $1$ only the other can be
accepted as solution of Eq.~(\ref{enerqmeps}). Note that in the limit
$K \rightarrow 0$ only the solution $q=q_{-}$ survives, coinciding
with the expression of the quadrupole moment of the BC
model~\cite{Barre01}. However, for a generic value of $K$ we need to consider
both solutions. Substituting these solutions into
Eq.~(\ref{entropyqm}) we obtain two expressions for the single-site
entropy $S/N$ as a function of the magnetization $m$ and the energy
$\epsilon$: $\tilde{s}_{\pm}(\epsilon,m)$. The equilibrium entropy
is given by $s(\epsilon)=\max \{ s_+(\epsilon),s_-(\epsilon) \}$,
where $s_{\pm}(\epsilon) = \max_{m} \tilde{s}_{\pm}(\epsilon,m)$.
The value of $m$ solving this extremum problem is the equilibrium
spontaneous magnetization. Thus, the maximum of
$\tilde{s}_{\pm}(\epsilon,m)$ is given by
\begin{equation}
\frac{\partial \tilde{s}_{\pm}(\epsilon,m)}{\partial m} =
\ln \left[ \frac{2(1-q_{\pm})}{\sqrt{q_{\pm}^2 -m^2}}\right] \frac{\partial q_{\pm}}{\partial m}
+\frac{1}{2}\ln \left[ \frac{q_{\pm}-m}{q_{\pm}+m} \right] = 0 \, ,
\label{eqmaxs}
\end{equation}
where the derivative of $q_{\pm}$ with respect to $m$ is computed
from Eq.~(\ref{qpmsol}), and with the further condition that
$\frac{\partial^2 \tilde{s}_{\pm}(\epsilon,m)}{\partial m^2}<0$. As
for the canonical case, if there is more than one solution, the
relevant one for the equilibrium state realizes the
absolute maximum.

\begin{figure}[ht]
\begin{center}
\includegraphics[width=7cm]{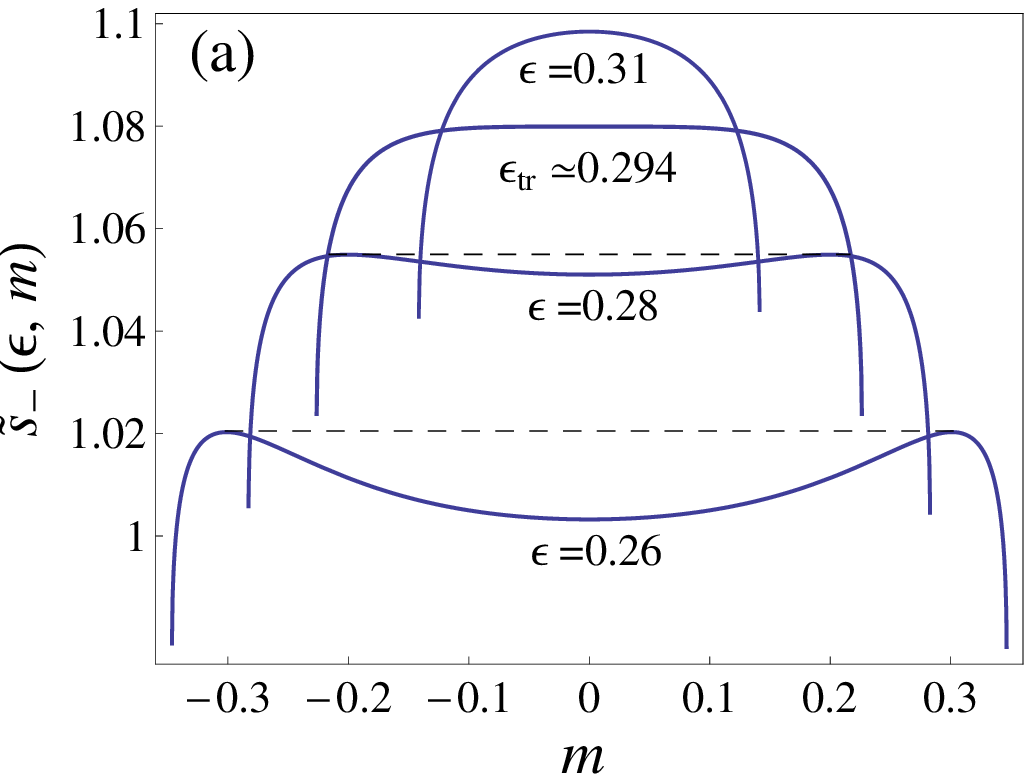}\\
\includegraphics[width=7cm]{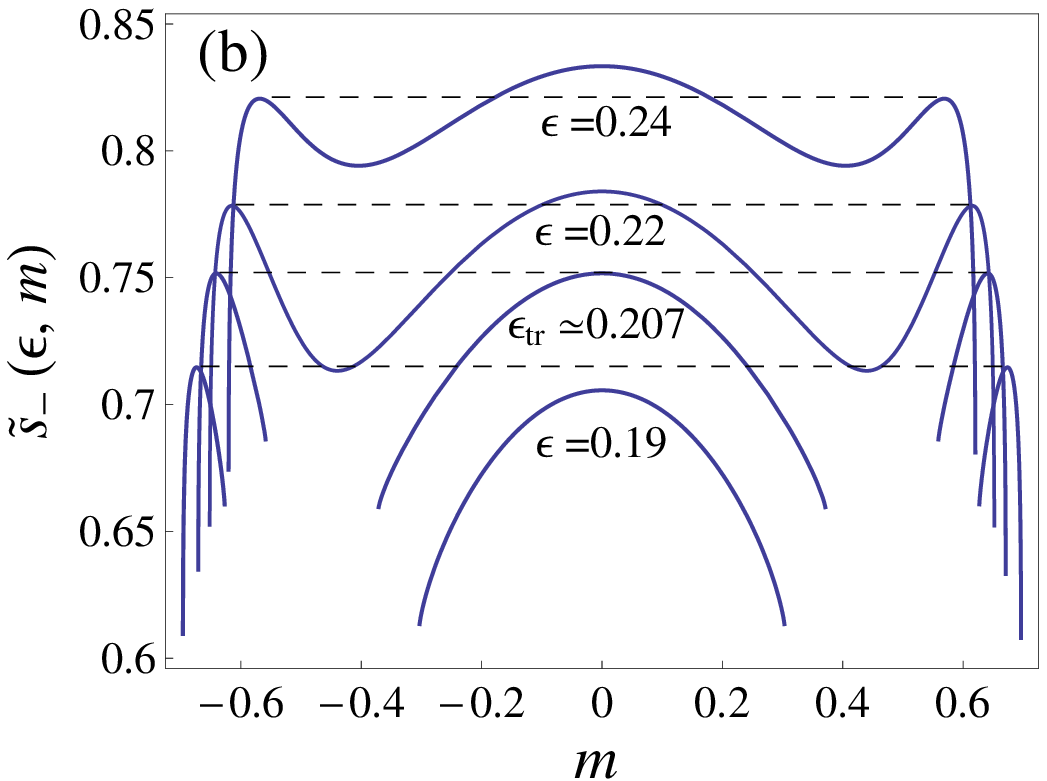}
\caption {Entropy $\tilde{s}_-(\epsilon,m)$ vs. magnetization $m$ for $K=1$
and for two values of the single-spin energy parameter $\Delta$,
plotted for several different values of the energy. (a)
$\Delta=0.8$: second order phase transition. (b) $\Delta=0.93$:
first order phase transition; at the transition energy the maxima at
$m=0$ and at $m \ne 0$ are at the same height.}
\label{Phase2}
\end{center}
\end{figure}

In Fig.~\ref{Phase2} we plot, as an example,
$\tilde{s}_-(\epsilon,m)$ as a function of $m$ for given values of
$\epsilon$, for $K=1$ and for two different values of $\Delta$. In these cases
the function $\tilde{s}_-(\epsilon,m)$ is the one giving the
absolute maximum. The figure shows the behavior of the function at increasing energy
when there is a second order transition ($\Delta=0.8$) or
a first order transition ($\Delta=0.93$).

Similarly to the canonical case, the critical line for the
transition between ferromagnetic and paramagnetic states is found
exploiting the Landau theory of phase transitions. This is done by
expanding $\tilde{s}_{\pm}(\epsilon,m)$ in powers of $m$. Contrary
to the canonical case, where we had to minimize a function,
$\tilde{f}(\beta,x,y)$, with respect to two variables, now we have
to maximize only with respect to the variable $m$, since $q$ is a
function of $\epsilon$ and $m$. Correspondingly, the expansion for
the Landau theory concerns only the variable $m$. Because of the
invariance under change of sign, $m\to -m$, we obtain an expression of the form
\begin{eqnarray}
\tilde{s}_{\pm}(\epsilon,m)=s_0+A_m m^2+B_m m^4 +O(m^6) ,
\label{entropyexpan}
\end{eqnarray}
where the zero magnetization entropy $s_0$ is
\begin{eqnarray}
s_0=-(1-z_{\pm})\ln(1-z_{\pm})-z_{\pm}\ln z_{\pm} + z_{\pm}\ln 2 \, ,
\label{s0expression}
\end{eqnarray}
with $z_{\pm}=\Delta/K \pm \sqrt{(\Delta/K)^2 - 2\epsilon/K}$,
and the coefficients are given by
\begin{eqnarray}
A_m &=& \mp a \ln\frac{2(1-z_{\pm})}{z_{\pm}}-\frac{1}{2z_{\pm}}, \label{entropycoeffa} \\
B_m &=& \mp b \ln\frac{2(1-z_{\pm})}{z_{\pm}} -\frac{a^2}{2z_{\pm}(1-z_{\pm})} \nonumber \\
&&\mp \frac{a}{2z_{\pm}^2} -\frac{1}{12z_{\pm}^3},
\label{entropycoeffb}
\end{eqnarray}
with $a=\left( 4\Delta^2-8K\epsilon \right)^{-\frac{1}{2}}$ and
$b=Ka^3$. We observe that $z_{\pm}$ is equal to $q_{\pm}$ for $m=0$. The critical line is defined by $A_m=0$
with $B_m<0$. To obtain the critical line in the
$(\Delta,T)$ plane we use the expression
\begin{eqnarray}
\frac{1}{T}=\frac{\partial s}{\partial \epsilon} \, . \label{microtemp}
\end{eqnarray}
On the critical line $m=0$, therefore
\begin{eqnarray}
\frac{1}{T}=\frac{\partial s_0}{\partial \epsilon} = \mp 2a \ln\frac{2(1-z_{\pm})}{z_{\pm}} \, .
\label{microtempb}
\end{eqnarray}
For positive temperatures $T$~\cite{foot1}, this equation gives some constraints
on the admissible values of $z_-$ and $z_+$ on the critical line (or, in this respect, for the
values of $q_-$ and $q_+$ in all cases where the equilibrium
magnetization $m$ is equal to zero). In particular, the argument of
the logarithm must be smaller (larger) than $1$ for $z_+$ ($z_-$),
i.e. $z_+ > 2/3$ ($z_- < 2/3$). However, the positivity
of $T$ is assured by the condition that $A_m$ must vanish on the
critical line. In fact, the condition $A_m=0$ gives, using
Eqs.~(\ref{entropycoeffa}) and (\ref{microtempb}), $z_{\pm}=T$.
Substituting $z_{\pm}$ with $T=1/\beta$ in Eq.~(\ref{microtempb})
and in the expression of $a$ we obtain
\begin{equation}
\beta = \frac{1}{2}e^{\beta \Delta - K}+1 \, ,
\label{Second-order_micr}
\end{equation}
i.e., the same expression of the canonical ensemble, that in addition
shows that $\beta$ is indeed positive. As explained above, the further
condition $B_c>0$, valid for the canonical case, is replaced by
$B_m<0$, so that the critical line ends when $B_m=0$, that
characterizes the microcanonical tricritical point. It can be proved
that $B_m<0$ whenever $B_c>0$, meaning that the microcanonical
critical line extends further with respect to the canonical critical
line. These findings are consistent with the general result that for
any system the microcanonical critical line includes and extends
beyond the canonical critical line~\cite{Campa09}. This in turn
is consistent with the fact that at the points of the phase
diagram where there is a canonical second order transition
ensembles are equivalent~\cite{Barre01,touch04}. In
Appendix~\ref{appmicrtric} we give some details of the computation
of the microcanonical tricritical point, characterized by the values
$\Delta_{MTP}$ and $T_{MTP}$, explicitly given in Sec.~\ref{results}
for different values of $K$.

As for the canonical case, after the tricritical point the first order transition line has to be
computed numerically. Furthermore, also in this ensemble, by increasing $K$, a line of first
order transitions, ending in an isolated critical point, between different paramagnetic states, appears,
and the tricritical point disappears, with the critical line branching in two first order
transition lines. Details are given in the next Section. However, these relevant points will be different in the two ensembles,
showing a marked ensemble inequivalence. Just to anticipate one of the results, contrary to the canonical case,
in the microcanonical ensemble the appearance of the critical point and the disappearance of the tricritical point
occur at the same value of $K$.

\section{Ensemble inequivalence}
\label{results}

In this Section we show the complete analysis of ensemble
inequivalence. The Section is divided in six Subsections. Depending
on the range where the value of $K$ is situated, we have four
different situations for the structure of the phase diagrams, and in all
cases there is inequivalence. 
The boundaries of these ranges of $K$ are obtained numerically. The first 
notable value of $K$ is $K_1\backsimeq 2.775$,
which is where, in the canonical ensemble, the triple and critical points appear.
The value $K_2=3$ is where the microcanonical tricritical point disappears.
Finally at the value $K_3\backsimeq 3.801$ the canonical tricritical point 
disappears.
In the first four Subsections we treat,
separately, each one of the cases. The fifth Subsection treats
briefly the case of negative temperatures, that can occur in systems
with upper bounded energy~\cite{TGBtemp}. A short discussion of the results
presented in the first four Subsections, in the framework of the
classification of phase transitions as studied in
Ref.~\cite{bouchet05}, is presented in the last Subsection.

\subsection{Low values of $K$: $0\leq K<K_1\backsimeq 2.775$}
\label{lowK}

\begin{figure}[!h]
\begin{center}
\includegraphics[width=6.5cm]{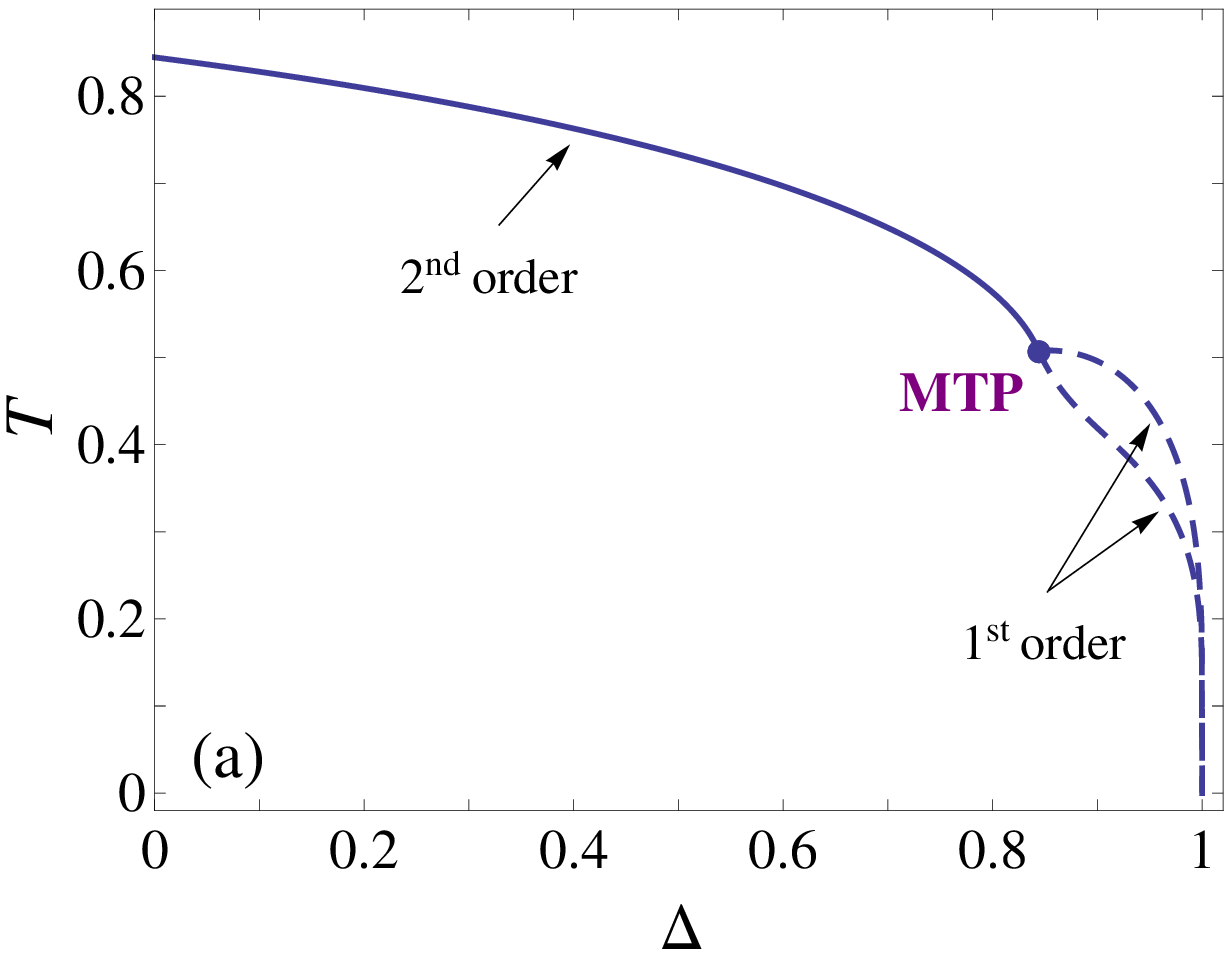}\\  \vskip 0.2cm
\includegraphics[width=6.5cm]{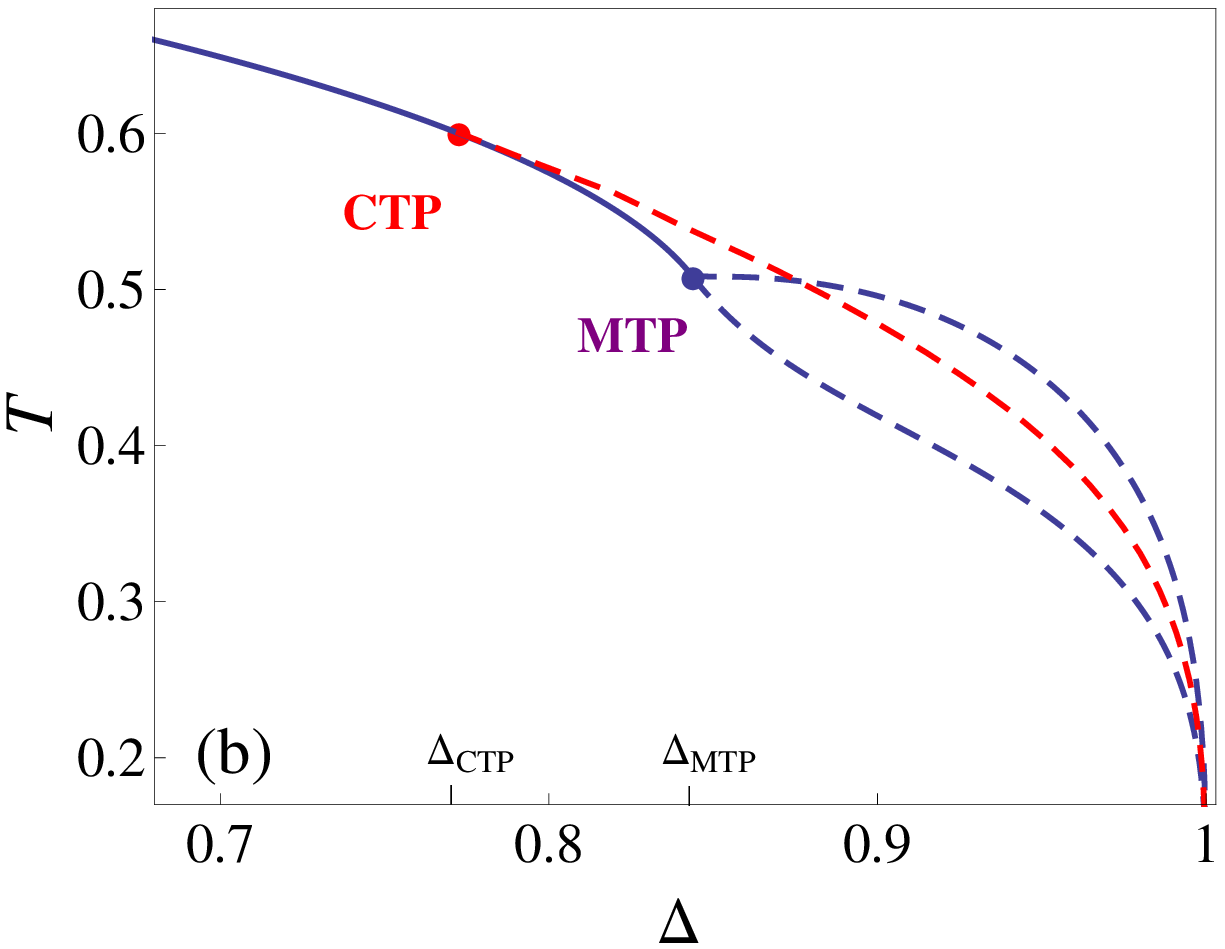}
\caption {(a) $(\Delta,T)$ phase diagram of the BEG model in the microcanonical
ensemble at $K=1$. Solid and dashed lines are the second and
first order phase transition lines, respectively. The second order transition line
ends at the microcanonical tricritical point (MTP), located at $\Delta_{MTP}\backsimeq 0.8439$,
$T_{MTP}\backsimeq 0.5078$. (b) Zoom of the region of the first order transitions, showing also
the canonical tricritical point (CTP), located at $\Delta_{CTP} \backsimeq 0.7726$,
$T_{CTP}=0.6$, and the canonical first order transition line (red dashed line).
The coordinates of the tricritical points in the phase diagram can be calculated from
Eqs.~(\ref{TCTP}), (\ref{DeltaCTP}), (\ref{DeltaMTP}) and (\ref{bm04}). The first order lines
reach the $T=0$ line at $\Delta=(K+1)/2=1$.}
\label{Phase}
\end{center}
\end{figure}

We treat this case by showing the results for $K=1$. The structure
of the phase diagram for other values of $K$ in this range is the
same. The upper panel of Fig.~\ref{Phase} shows the
$(\Delta,T)$ phase diagram in the microcanonical ensemble for $K=1$.
The value of $T$ is obtained from Eq.~(\ref{microtemp}). The
second order transition changes to first order at the microcanonical
tricritical point (MTP). The canonical phase diagram for $K=1$ was
shown in the lower panel of Fig.~\ref{Phasec}. To have a better
comparison the two diagrams are now plotted together in the lower
panel of Fig.~\ref{Phase}, that shows the zoom of the most
interesting region. In this plot the second order transitions are
denoted by full lines, while the first order transitions are
indicated by dashed lines. As anticipated above, the microcanonical
critical line extends further with respect to the canonical one. The
second order phase transition line is common to both ensembles
up the canonical tricritical point (CTP). After this point the
canonical ensemble has first order transitions (red dashed line),
while for the microcanonical ensemble the second order transition
line continues up to the microcanonical tricritical point (MTP). The
microcanonical first order transitions are characterized by a
temperature jump~\cite{bouchet05,Campa09}, and for this reason the
plot has correspondingly two branches.

\begin{figure}[!h]
\begin{center}
\begin{tabular}{cccccc}
\includegraphics[width=4.2cm]{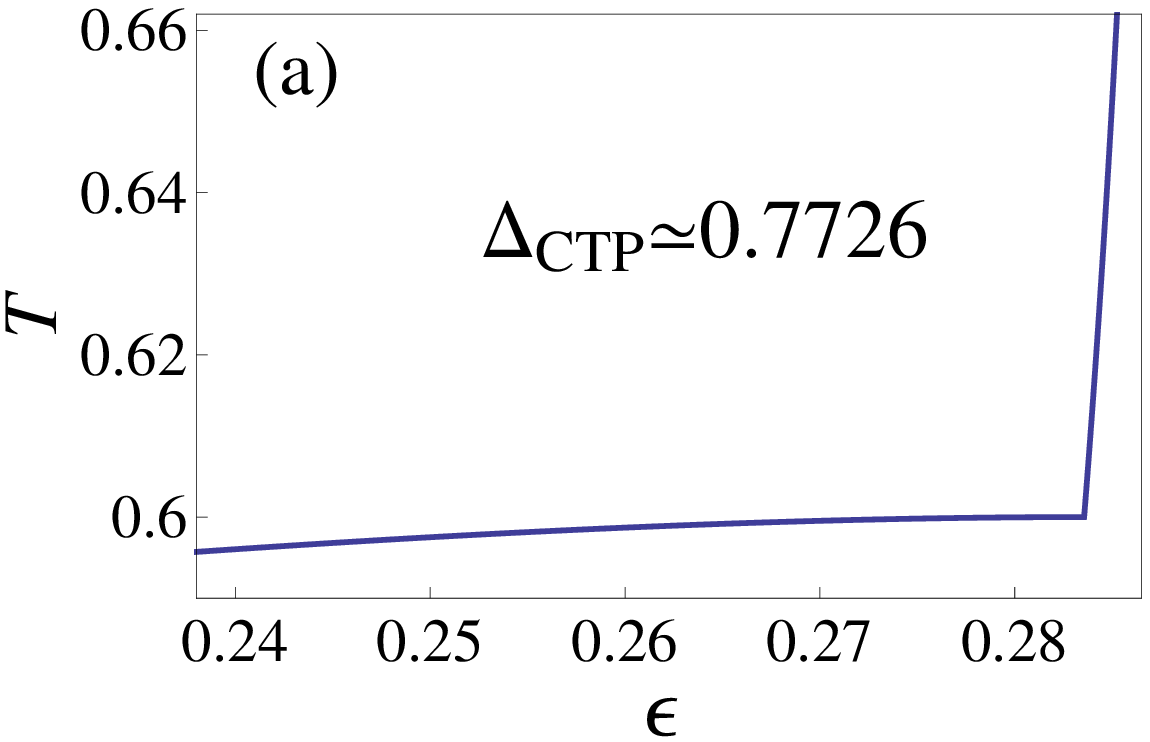} &
\includegraphics[width=4.2cm]{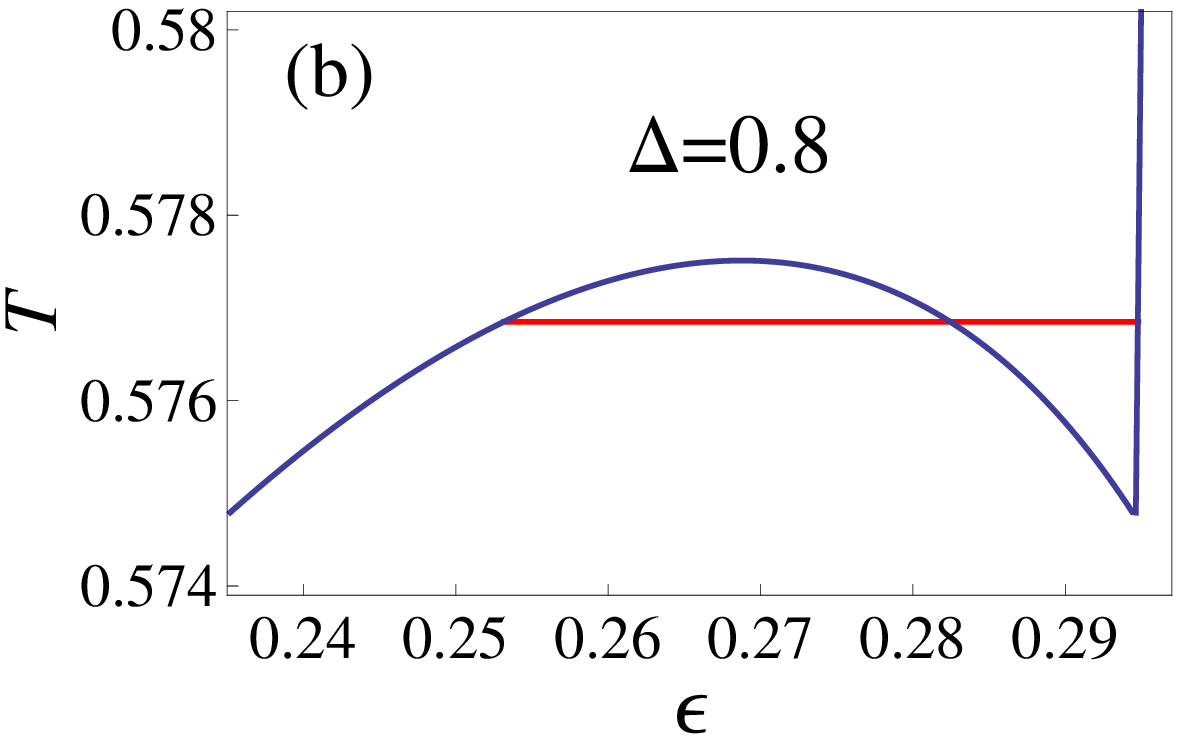}
\nonumber \\
\includegraphics[width=4.2cm]{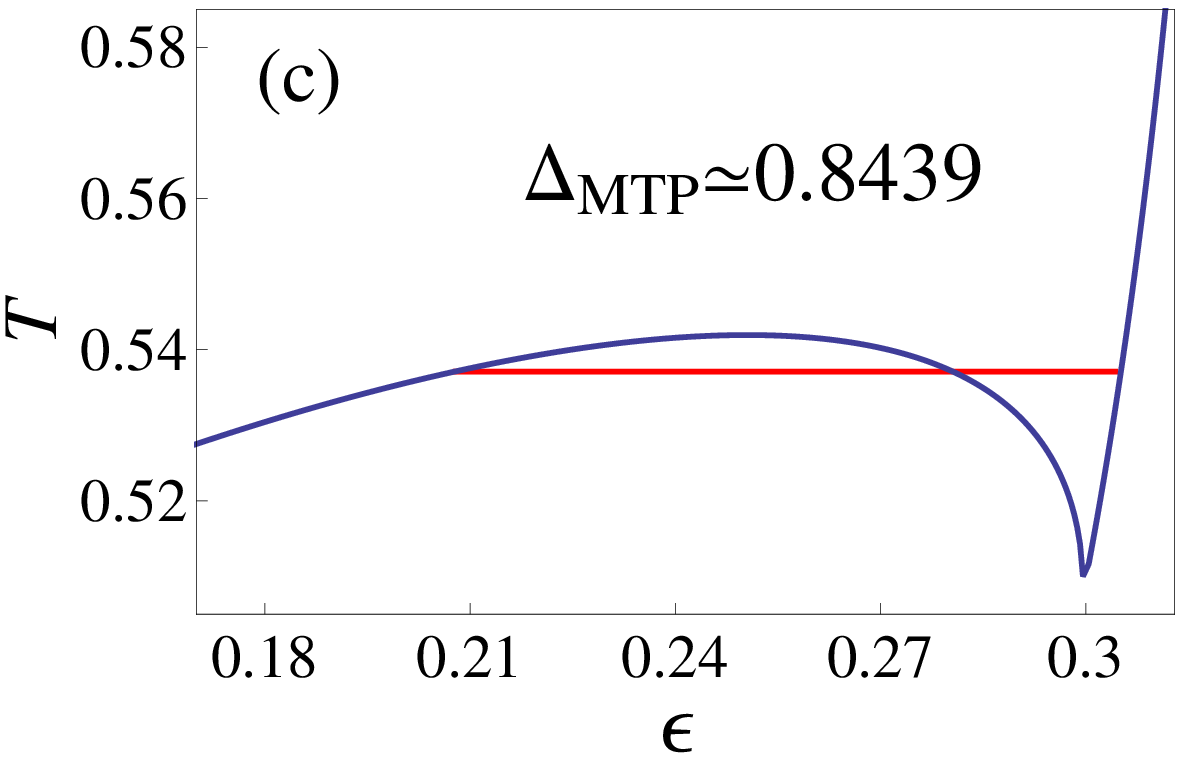} &
\includegraphics[width=4.2cm]{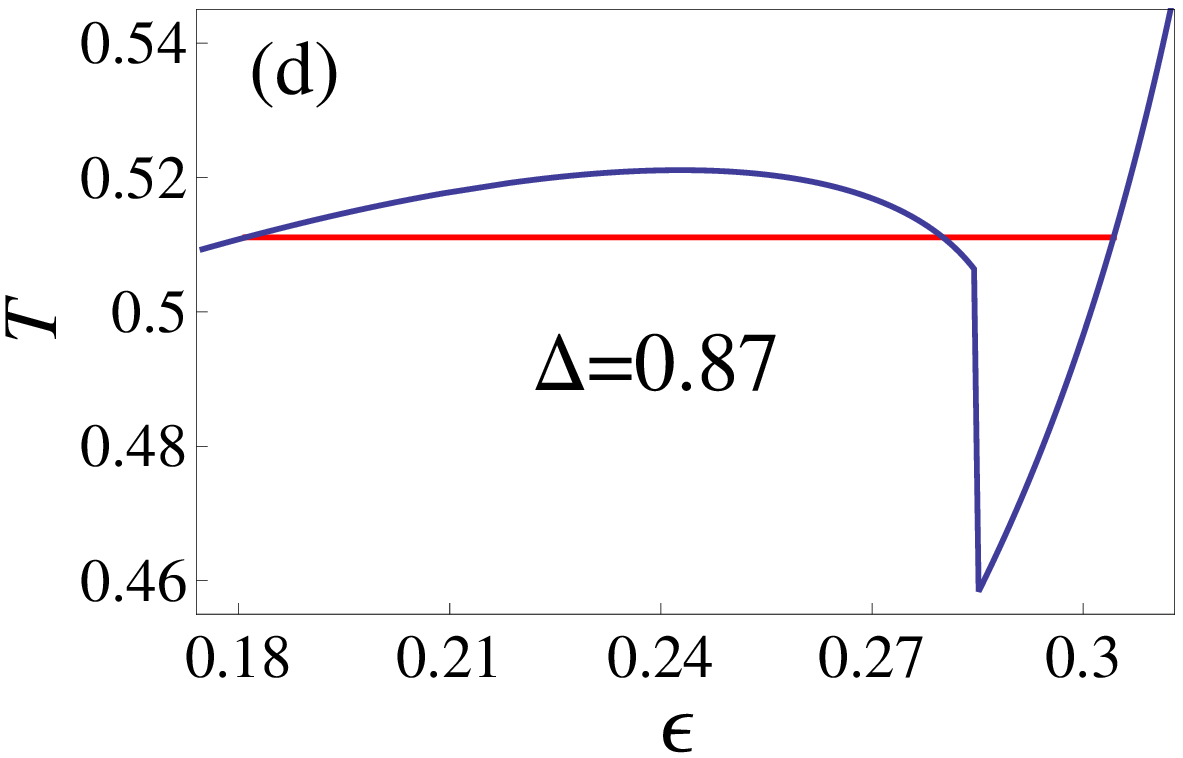}
\nonumber \\
\includegraphics[width=4.2cm]{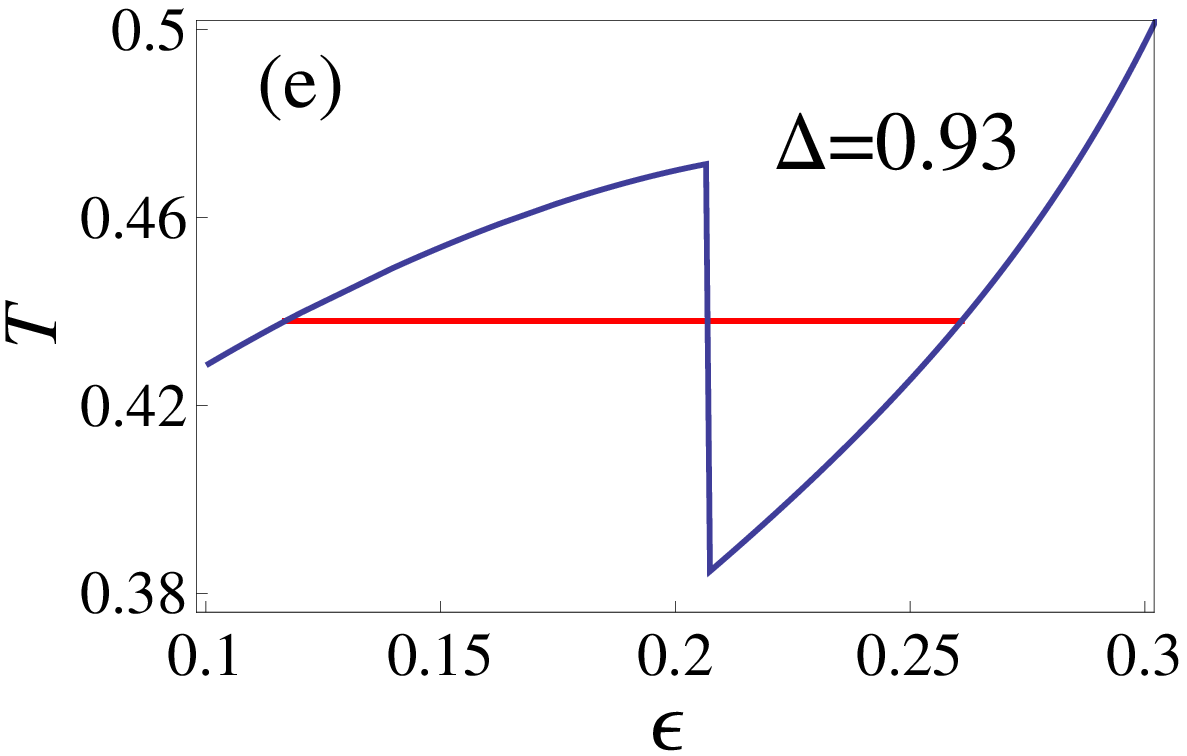} &
\includegraphics[width=4.2cm]{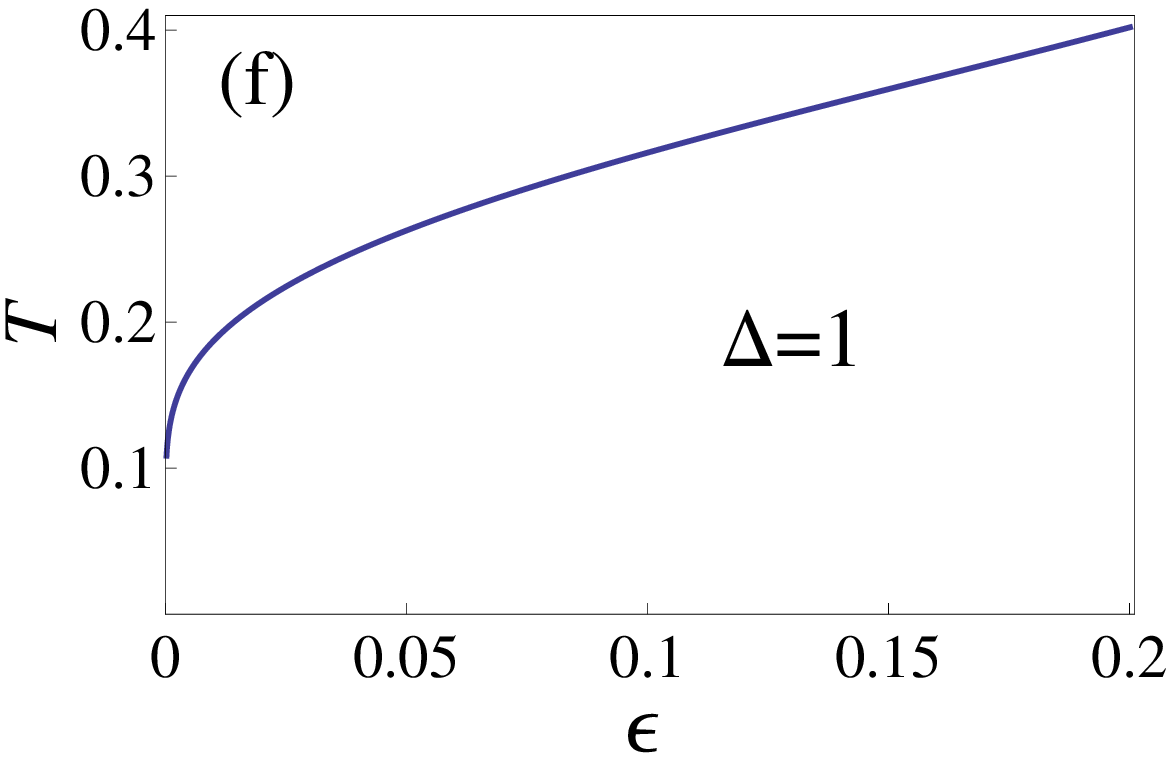}
\end{tabular}
\caption {Energy dependence of the temperature (caloric curve) for $K=1$ and different values of the
single-spin energy parameter $\Delta$. The blue lines are the microcanonical curves; they coincide
with the canonical ones in the energy ranges where ensembles are equivalent. Where instead
the ensembles are inequivalent, the canonical curves are given by the horizontal red lines (Maxwell
construction). The energies within the range covered by these lines are forbidden in the canonical ensemble.}
\label{Energy}
\end{center}
\end{figure}

A further illustration of ensemble inequivalence is given by the
caloric curves, i.e., the plots showing temperature $T$ as a
function of energy $\epsilon$. In Fig.~\ref{Energy} we show the
caloric curves for $K=1$ for several values of $\Delta$, ranging
from the value at the canonical tricritical point to the value where
the first order transition lines meet the $T=0$ axis (see
Fig.~\ref{Phase}). Let us remind that in the canonical ensemble $T$ is the control variable and
$\epsilon$ is obtained from the thermodynamic expression $\epsilon =
\partial \left( \beta f(\beta) \right)/\partial \beta$. In Fig.~\ref{Energy}
the microcanonical curves are plotted in blue; in the energy ranges
where ensembles are equivalent, these curves represent also the
canonical ensemble, while in the inequivalence ranges the canonical
curves are given by the red horizontal lines (Maxwell construction),
showing the canonical first order transitions.

The second order transition is associated to a discontinuity in the
first derivative of the caloric curve. Fig.~\ref{Energy}(a) shows
the caloric curve at the canonical tricritical point. In this case
the low energy branch has a zero slope at the transition point,
denoting a diverging specific heat. Increasing $\Delta$, in the
range between the canonical and the microcanonical tricritical
points, the transition is second order in the microcanonical ensemble
and first order in the canonical ensemble, while one can observe the
appearance of a negative specific heat $(\partial T / \partial
\epsilon < 0)$ in the microcanonical ensemble
(Fig.~\ref{Energy}(b)). At the microcanonical tricritical point
(Fig.~\ref{Energy}(c)) $\partial T / \partial \epsilon$ diverges in
the low energy branch, indicating a vanishing specific heat. For
$\Delta>\Delta_{MTP}$ the transition becomes first order also in the
microcanonical ensemble, characterized by a temperature jump at the
transition energy (Fig.~\ref{Energy}(d)). We observe that the
temperature jump is negative (by increasing energy), a fact that can
be proved on general grounds on the basis of singularity theory for
maximization problems~\cite{bouchet05}; however, this can also be
understood more simply using a heuristic argument, see the last
Subsection. The presence of the negative temperature jump means that
the temperatures between the two blue dashed lines in Fig.~\ref{Phase}(b)
are achieved, in the microcanonical ensemble, at two
different energies, one before and one after the transition.
Increasing further the value of $\Delta$, negative specific heat disappears,
leaving only a temperature jump (Fig.~\ref{Energy}(e)). Finally, at
$\Delta =1$ the system is always paramagnetic, there is no
transition, the two ensembles are again equivalent and the
temperature is monotonically increasing with the energy
(Fig.~\ref{Energy}(f)).

\begin{figure}[t]
\centerline{\psfig{figure=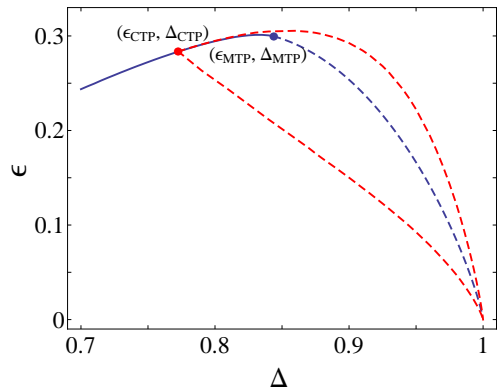,width=6.5 cm}}
\caption{The most interesting region of the $(\Delta,\epsilon)$ phase diagram  for the canonical and
microcanonical ensembles at $K=1$. The solid blue line denotes second order transitions and
ends at the CTP and the MTP for the canonical and microcanonical ensembles, respectively.
Red (blue) dashed lines are the first order transition lines for the canonical
(microcanonical) ensemble. The coordinates of the tricritical points are
$\Delta _{CTP} \backsimeq 0.7726$, $\epsilon_{CTP}\backsimeq 0.2836$ and
$\Delta _{MTP} \backsimeq  0.8439$, $\epsilon_{MTP}\backsimeq 0.2996$. The region inside the two
red dashed lines is forbidden in the canonical ensemble. The first order lines meet at the point
with coordinates $\Delta=1$ and $\epsilon=0$ (the minimum allowed energy for $\Delta=1$).}
\label{Energy-Delta}
\end{figure}

It is instructive to study the behavior of the system also in the
$(\Delta,\epsilon)$ phase diagram. In this case it is the canonical
ensemble that has two branches associated to first order
transitions: each branch gives the energy value of one of the two
extremes of the red horizontal lines representing the Maxwell
construction. Fig.~\ref{Energy-Delta} shows the interesting region
of the phase diagram~\cite{foot2}. The region between the two red
dashed lines is forbidden in the canonical ensemble. We remind that
in a short-range system, on the contrary, there are no ranges of
forbidden energy in the canonical ensemble, since phase coexistence
would allow to achieve any energy by adjusting the relative weight
of each phase.

For convenience, we summarize the coordinates of the tricritical points for $K=1$.

$\bullet$ Canonical Tricritical Point: $T_{CTP} = 0.6$, $\Delta _{CTP} \backsimeq
0.7726$, $\epsilon_{CTP}\backsimeq 0.2836$.

$\bullet$ Microcanonical Tricritical Point: $T_{MTP} \backsimeq 0.5078$, $\Delta _{MTP}
\backsimeq  0.8439$, $\epsilon_{MTP}\backsimeq 0.2996$.

\subsection{Low-Intermediate values of $K$: $2.775 \backsimeq K_1<K<K_2=3$}

\begin{figure}[!h]
\begin{center}
\includegraphics[width=6.5cm]{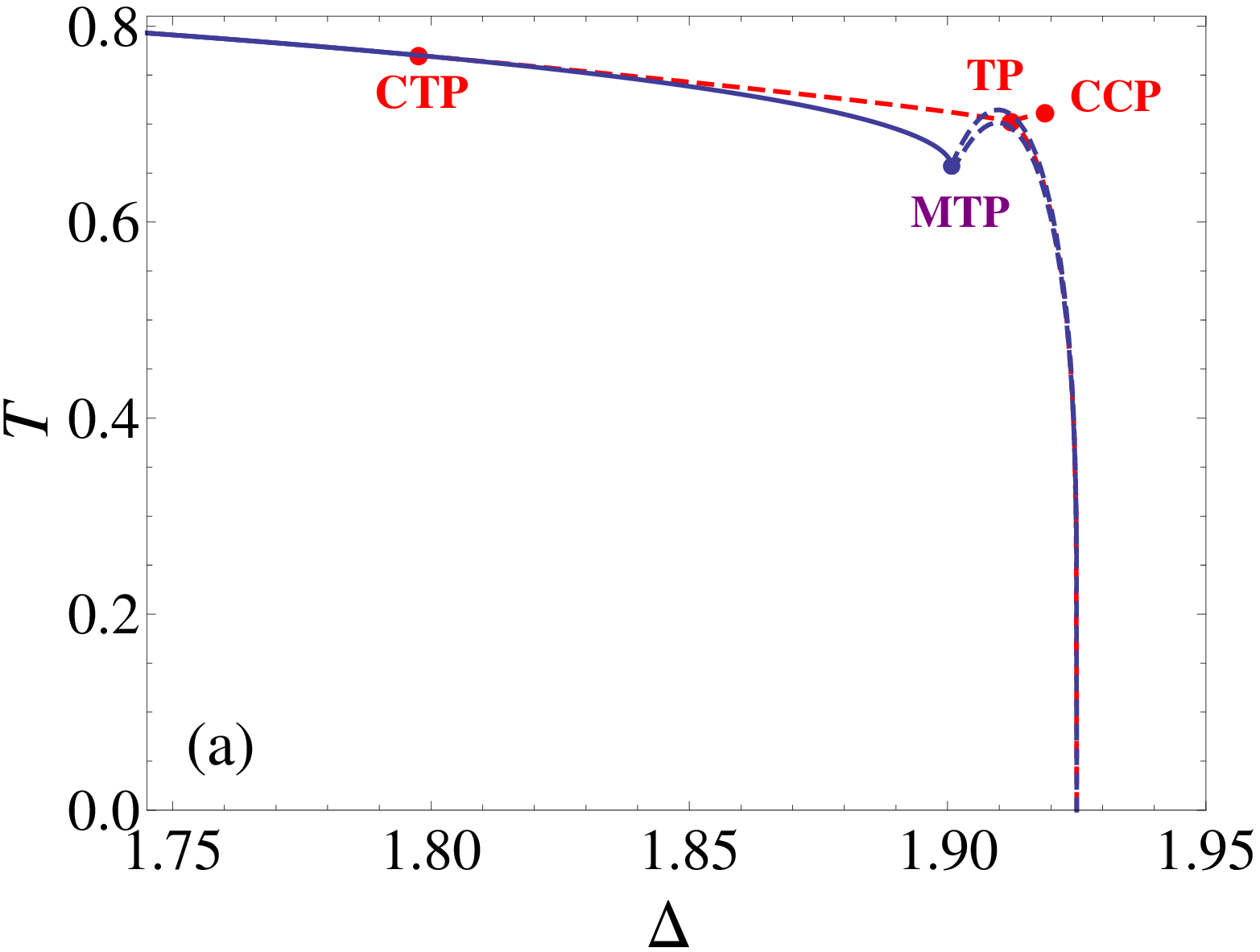}\\ \vskip 0.2cm
\includegraphics[width=6.5cm]{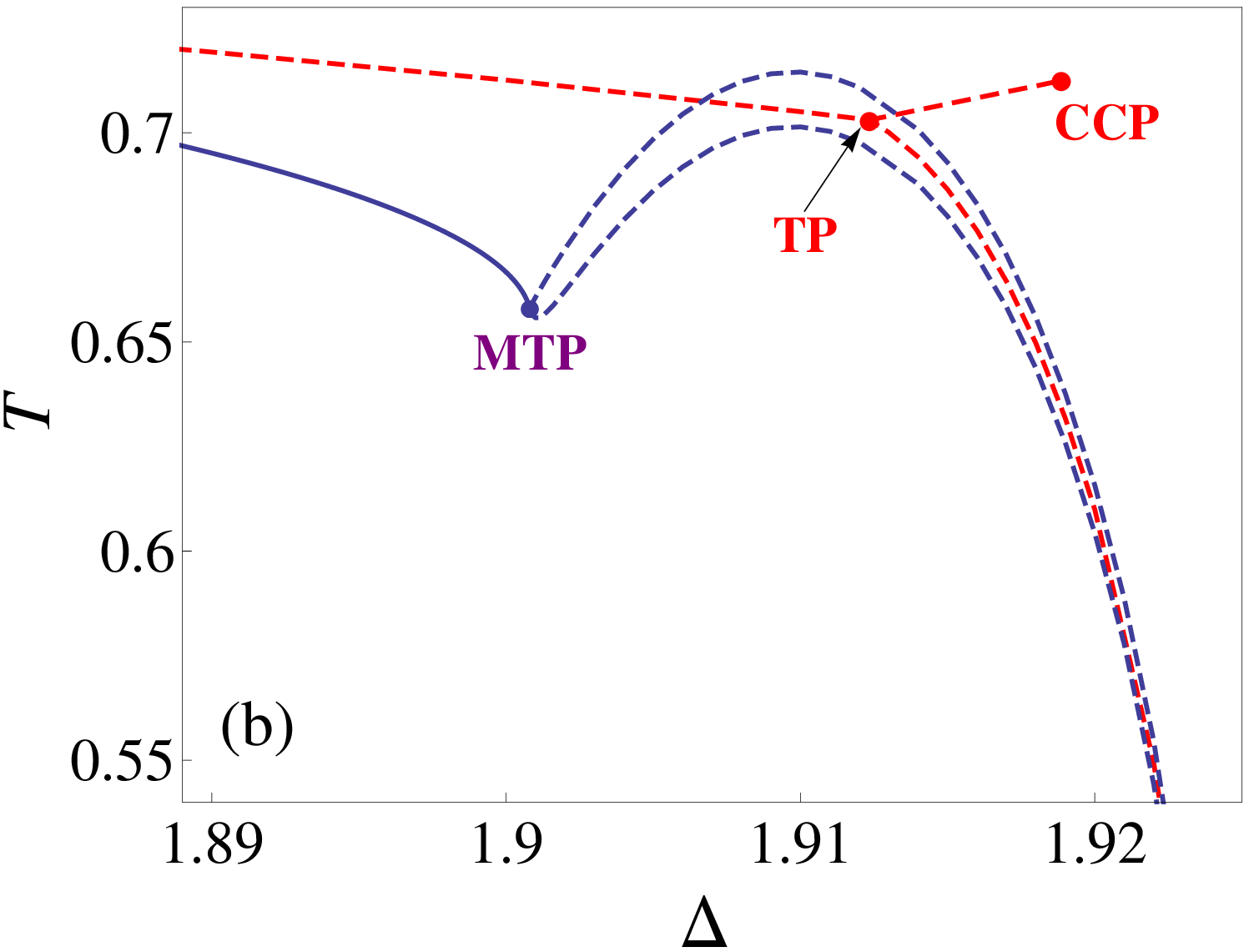}
\caption {$(\Delta,T)$ phase diagram of the canonical and microcanonical ensembles at $K=2.85$:
panel (a) shows the most interesting region, which is further zoomed in panel (b).
The solid blue line denotes the line of second order transitions, ending at the CTP and the MTP for the canonical
and microcanonical ensembles, respectively. Red (blue) dashed lines are the first order transition lines
for the canonical (microcanonical) ensemble. The first order canonical line branches at the canonical
triple point (TP): one branch reaches $T=0$ where it meets the two microcanonical first order lines at
$\Delta=(K+1)/2=1.925$; the other branch ends at the canonical critical point (CCP). The coordinates of the relevant
points, CTP, MTP, TP and CCP, are given in the text.}
\label{TempDeltaK2p85}
\end{center}
\end{figure}

In this range of $K$ the phase diagram of the system becomes more
complex, and ensemble inequivalence is characterized by more
features. In fact, not only are the tricritical points and the
first order transitions different in the two ensembles, but a critical
and a triple point appear in the canonical ensemble, with no
counterparts in the microcanonical ensemble. This is visualized in
Fig.~\ref{TempDeltaK2p85}, that shows the $(\Delta,T)$ phase diagram
for $K=2.85$; this value has been chosen as representative of this
range.
As before, at the canonical tricritical point (CTP) the canonical
transition changes from second order to first order (the red dashed
line), while the microcanonical transition remains second order up
to the microcanonical tricritical point (MTP), from which the two
blue dashed lines denote the two temperature branches associated to
the microcanonical first order transition. However, now the
first order canonical line branches in two lines at the triple point
(TP): one branch extends to $T=0$, while the other ends at a
canonical critical point (CCP). The canonical first order transitions
corresponding to the branch going from the triple point to the
critical point are transitions between paramagnetic states
($m=0$) with different values of the quadrupole moment
$q$. Some details on the computation of the canonical critical point
are given in Appendix~\ref{appcantric}.

On the other hand, the structure of the microcanonical phase diagram
is the same as that in the previous range of $K$. In fact, only a
tricritical point is present, and there is no counterpart, in the
microcanonical ensemble, of the canonical transition between
different paramagnetic states. This is clearly seen in
Fig.~\ref{TempDeltaK2p85}(b).

\begin{figure}[!h]
\begin{center}
\begin{tabular}{cccc}
\includegraphics[width=4.2cm]{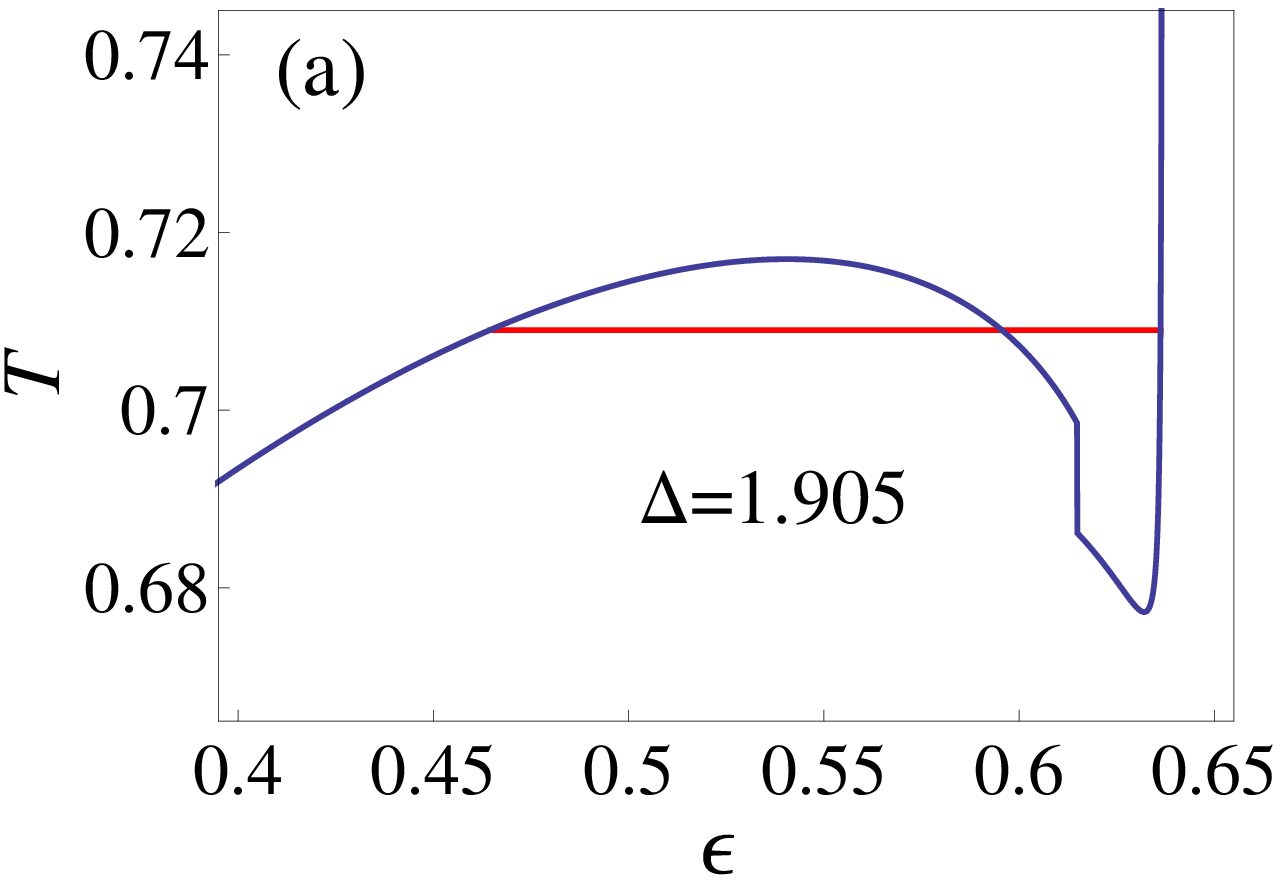} &
\includegraphics[width=4.2cm]{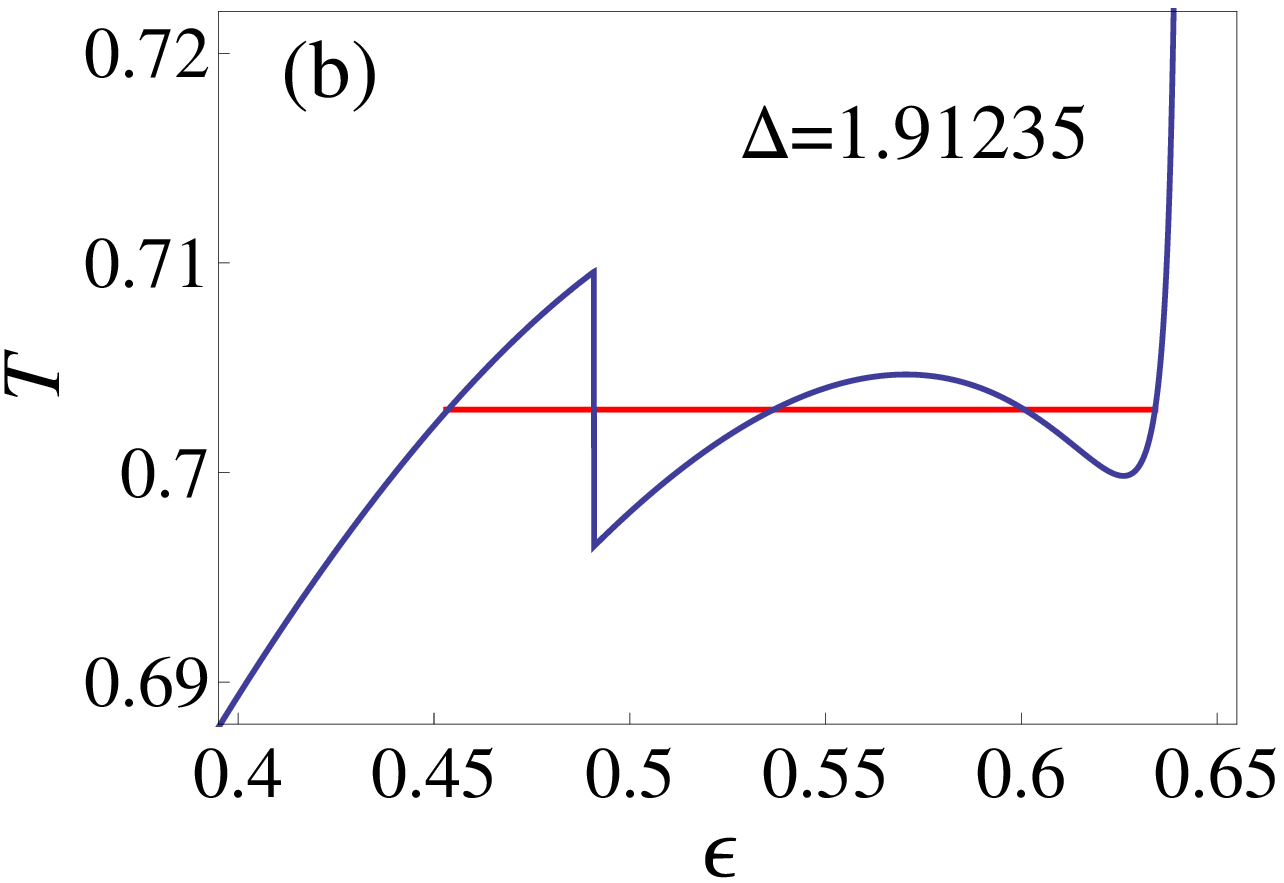}
\nonumber \\
\includegraphics[width=4.2cm]{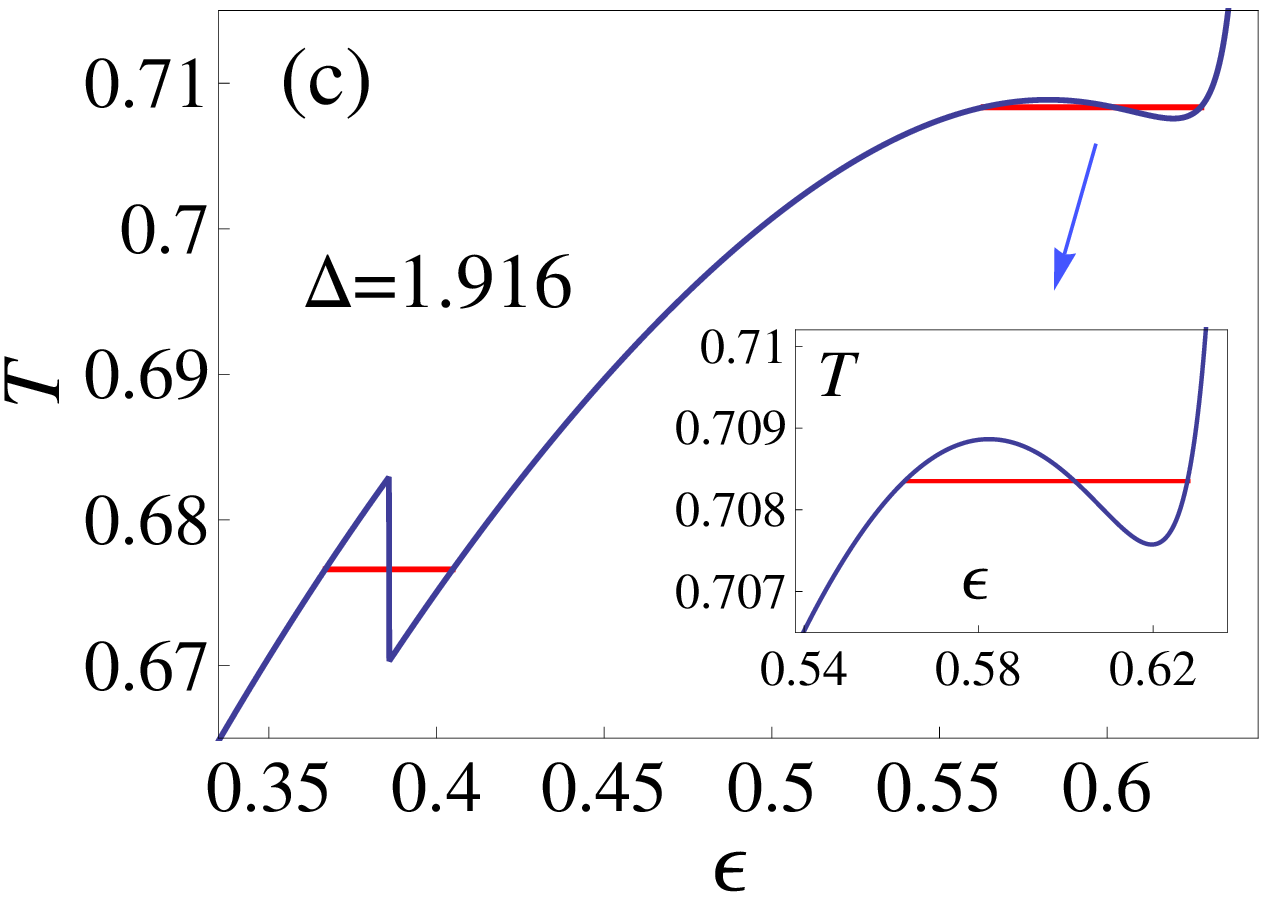} &
\includegraphics[width=4.2cm]{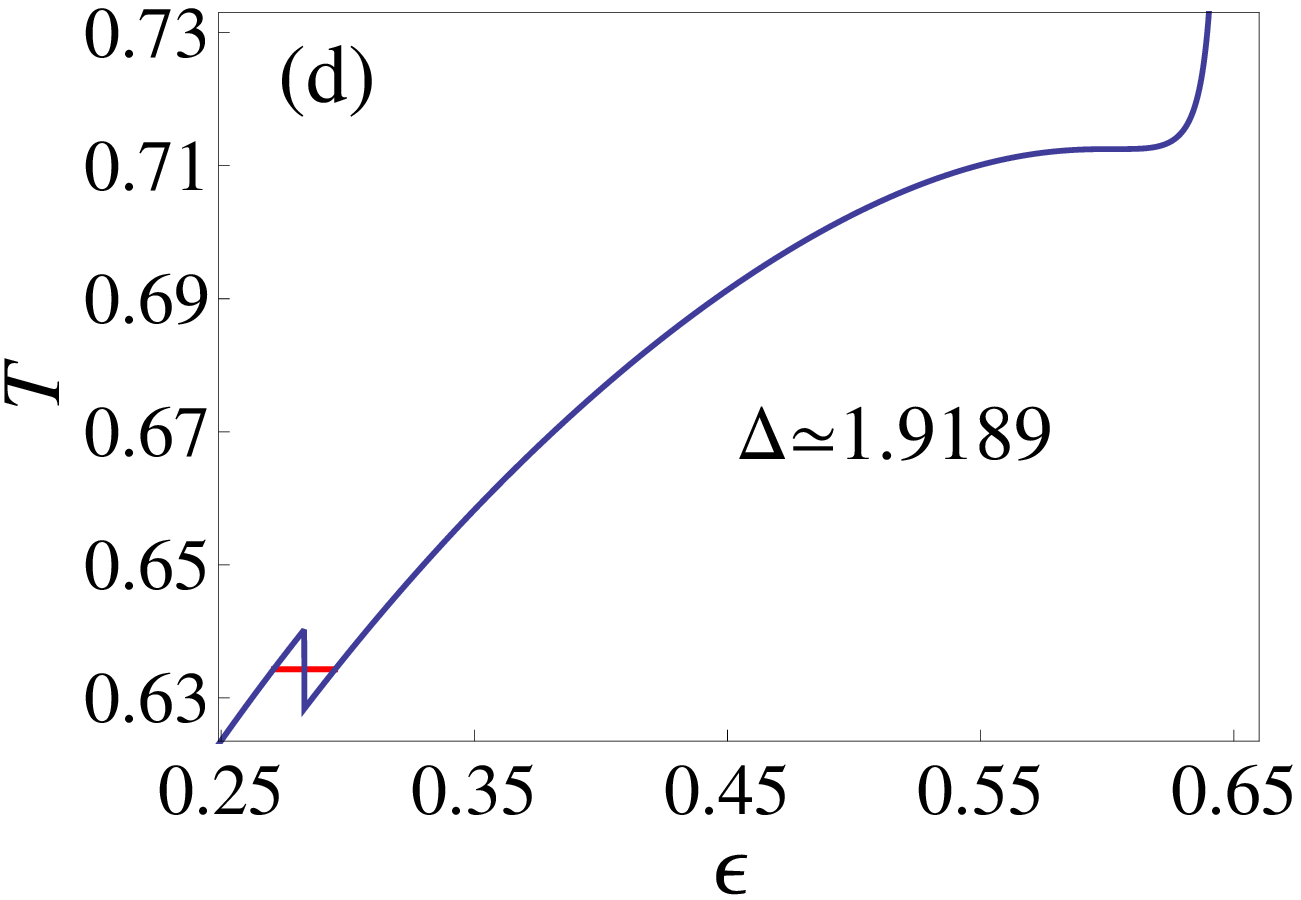}
\end{tabular}
\caption {Caloric curves for $K=2.85$ and different values of the
single-spin energy parameter $\Delta$. The meaning of the lines is the same as in
Fig. \ref{Energy}.}
\label{CaloricK2p85}
\end{center}
\end{figure}

The caloric curves give a better image of what happens. They are shown in
Fig.~\ref{CaloricK2p85} for four different values of $\Delta$. As
for Fig.~\ref{Energy}, the red horizontal lines denote the portion
of the canonical curves in the inequivalence range.
Fig.~\ref{CaloricK2p85}(a) corresponds to a value of $\Delta$ which
is larger than that of the microcanonical tricritical point and
smaller than that of the canonical triple point; in this case there
is a first order transition between a ferromagnetic state and a
paramagnetic state in both ensembles; the transition is accompanied
by a region of negative specific heat in the microcanonical
ensemble.
For a value of $\Delta$ larger than that of the triple point, as in
Fig.~\ref{CaloricK2p85}(c), the canonical caloric curve shows two
first order transitions, the first (by increasing energy) from a
ferromagnetic state to a paramagnetic state and the second between
two paramagnetic states. As shown in the plot, in correspondence to
this second canonical transition, the microcanonical ensemble
presents only a region with negative specific heat, and no
transition. This agrees with the fact that in
Fig.~\ref{TempDeltaK2p85} there is no line of microcanonical
transitions associated to the canonical
line going from the triple point to the critical point.
At exactly the triple point (Fig.~\ref{CaloricK2p85}(b)) the two
first order canonical transitions merge in a single one. Instead, by
increasing $\Delta$ up to the critical point value, as in
Fig.~\ref{CaloricK2p85}(d), the second canonical transition becomes
continuous, and the caloric curve has a vanishing derivative at the
transition point (infinite specific heat).

\begin{figure}[t]
\centerline{\psfig{figure=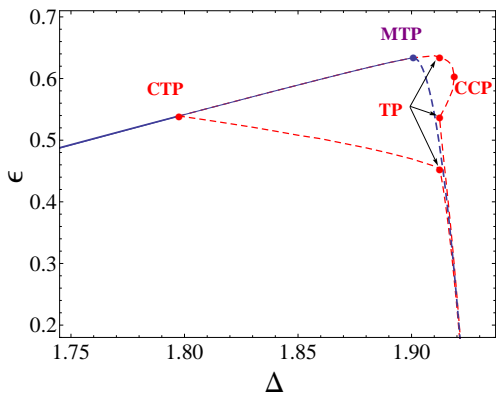,width=6.5 cm}}
\vspace*{8pt}
\caption{
The most interesting region of the $(\Delta,\epsilon)$ phase diagram for the canonical and
microcanonical ensembles at $K=2.85$. The blue solid line denotes the line of second order transitions,
ending at CTP and MTP for the canonical and microcanonical ensembles, respectively.
Red (blue) dashed lines are the first order transition lines for the canonical
(microcanonical) ensemble. Between the CTP and the MTP the second order microcanonical line is extremely
close (from below) to one of the first order canonical lines, so that in the plot they are practically indistinguishable.
The plot shows the three energies of the canonical triple point (TP) and the canonical critical point (CCP).
The first order lines meet at the point with coordinates $\Delta=1.925$ and $\epsilon=0$ (the minimum allowed energy for
$\Delta=1.925$). The region inside the red dashed lines is forbidden in the canonical ensemble.}
\label{EnergyDeltaK2p85}
\end{figure}

The phase diagram in the $(\Delta,\epsilon)$ plane is shown in
Fig.~\ref{EnergyDeltaK2p85}. It happens that the energy of the
second order microcanonical transition, for values of $\Delta$
between those of the canonical and the microcanonical tricritical
points, is very close to the upper energy of the canonical first
order transition, so that the two corresponding lines in the phase
diagram are very close. The region inside the dashed red lines is
forbidden in the canonical ensemble.
The canonical TP in the $(\Delta,T)$ diagram splits in three points of the $(\Delta,\epsilon)$
diagram, which represent the energies of the three phases that meet at this point.

The coordinates of the relevant points of the phase diagrams for $K=2.85$ are:

$\bullet$ Canonical Tricritical Point: $T_{CTP} \backsimeq 0.7701$,
$\Delta _{CTP} \backsimeq 1.7976$, $\epsilon_{CTP}\backsimeq
0.5392$,

$\bullet$ Canonical Triple Point: $T_{TP} = 0.7030$, $\Delta _{TP} =
1.91235$, $\epsilon_{TP}\backsimeq 0.4531; 0.5367; 0.6342$,

$\bullet$ Canonical Critical Point: $T_{CCP} = 0.7125$, $\Delta
_{CCP} \backsimeq 1.9189$, $\epsilon_{CCP}\backsimeq 0.6032$,

$\bullet$ Microcanonical Tricritical Point: $T_{MTP} \backsimeq
0.6581$, $\Delta _{MTP} \backsimeq  1.9008$,
$\epsilon_{MTP}\backsimeq 0.6337$.

\subsection{Intermediate-High values of $K$: $3=K_2<K<K_3\backsimeq 3.801$}

\begin{figure}[!h]
\begin{center}
\includegraphics[width=6.5cm]{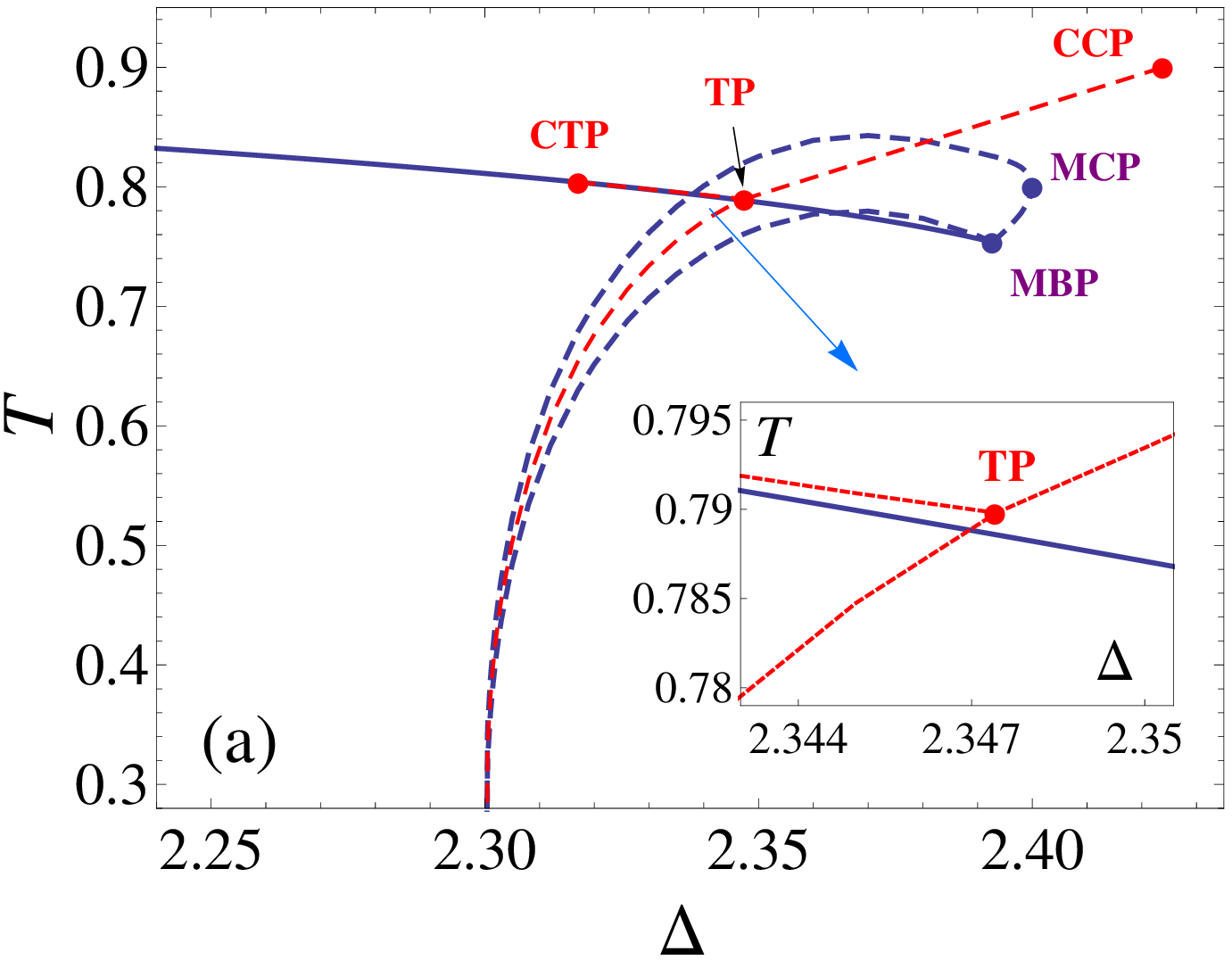}\\ \vskip 0.2cm
\includegraphics[width=6.5cm]{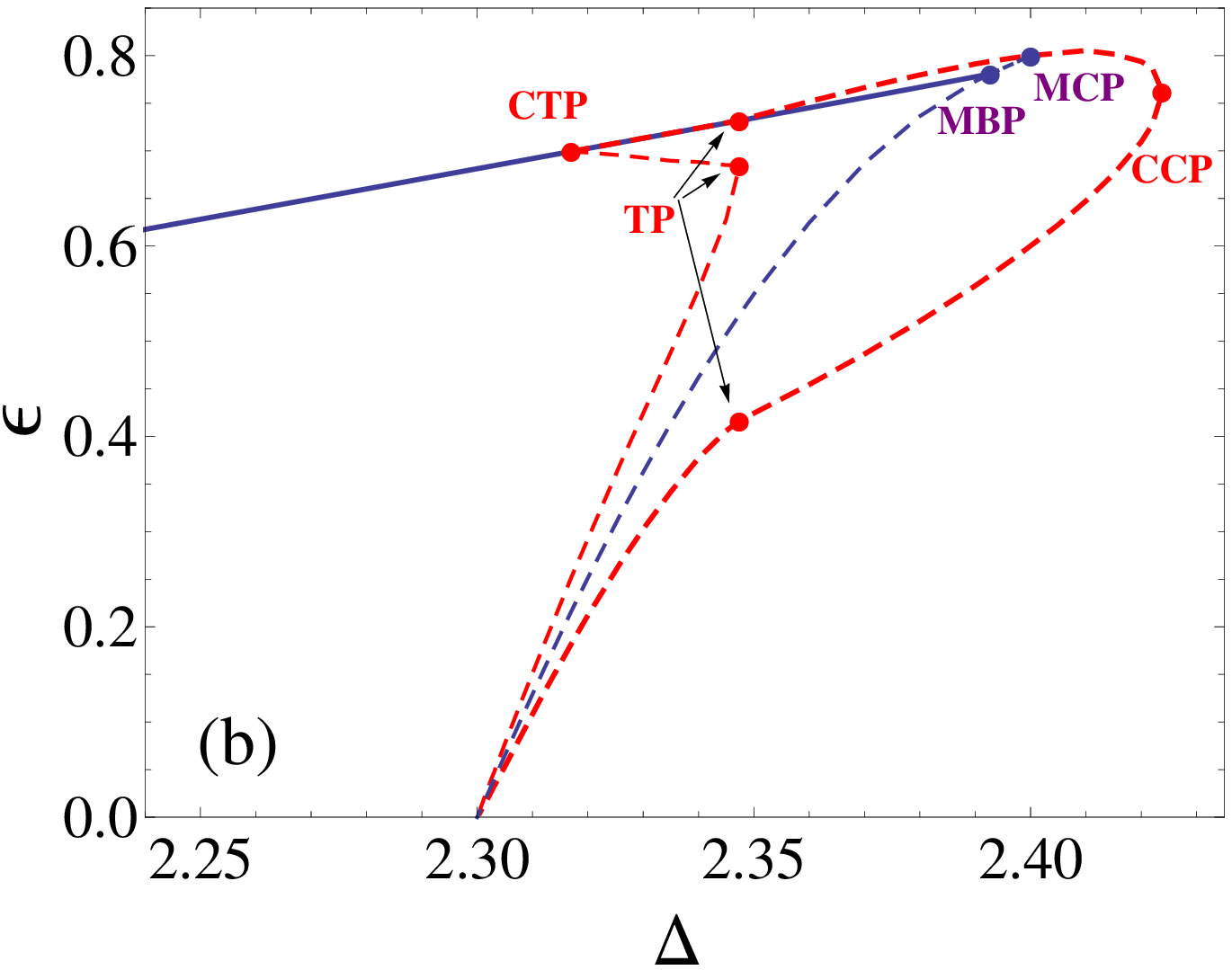}
\caption{The most interesting region of the phase diagrams of the canonical and microcanonical
ensembles at $K=3.6$. The solid blue line denotes the line of second order transitions, ending at the canonical
tricritical point (CTP) and at a microcanonical branching point (MBP) for the two ensembles, respectively.
Red (blue) dashed lines are the first order transition lines for the canonical (microcanonical) ensemble.
(a) $(\Delta,T)$ phase diagram. Between the CTP and the triple point (TP) the first order canonical line is very close
to the second order microcanonical line; the inset shows a zoom of the region near the TP. One canonical first order
branch and the two microcanonical first order branches meet at $T=0$ at $\Delta=(K+1)/2=2.3$.
(b) $(\Delta,\epsilon)$ phase diagram. The three energies of the canonical triple point (TP) are indicated.
The first order lines meet at the point with coordinates $\Delta=2.3$ and $\epsilon=0$ (the minimum allowed energy for
$\Delta=2.3$).
The coordinates of the relevant points, CTP, MBP, MCP, TP and CCP, are given in the text.}
\label{TempDeltaK3p6}
\end{center}
\end{figure}

In this range of $K$ we have a new structure of the phase diagram.
While the canonical structure is similar to that of the previous $K$
range, the microcanonical structure is characterized by the
disappearance of the tricritical point and the appearance of a
{\it branching point}, as explained in detail below, and a critical point.
The representative value of $K$ that we have chosen for this range
is $K=3.6$. In Fig.~\ref{TempDeltaK3p6} we show both the
$(\Delta,T)$ and the $(\Delta,\epsilon)$ phase diagrams. We notice
that, although the structure of the canonical diagram
is the same as before, now the $\Delta$ value where the transition
lines meet the $T=0$ axis is smaller than that of the relevant
points. On the other hand, the structure of the microcanonical
diagram has completely changed. The microcanonical tricritical point
is no more present. The reason is that the line of second order
transition ends before the quantity $B_m$ of
Eq.~(\ref{entropycoeffb}) reaches the value $0$: for $\Delta$
values larger than those where the line ends, the states
corresponding to the critical line do not realize the absolute maximum of
the microcanonical entropy and become metastable. Then, the
critical line branches, at the {\it microcanonical branching point} (MBP),
in two lines of first order transitions. 
We are denoting by MBP a point that in the
literature is usually called microcanonical critical end point; see
also the last Subsection.
In the $(\Delta,T)$ diagram
the lines departing from the MBP are those of the lower temperature
associated to the temperature jump corresponding to the first order
transition. For $\Delta$ values smaller than that of the MBP the
transition is between a ferromagnetic and a paramagnetic state,
while for $\Delta$ values larger than that of the MBP it is between two
paramagnetic states . The line of first order transitions
between paramagnetic states ends at the microcanonical critical
point (MCP). In Appendix~\ref{appmicrtric} we give some details on
the computation of the microcanonical critical point. The inset in
Fig.~\ref{TempDeltaK3p6}(a) shows the region near the canonical
triple point, where the first order canonical line is very close to
the second order microcanonical line. In the $(\Delta,\epsilon)$
diagram (panel (b) of the figure) one canonical first order branch
and the microcanonical second order line are still closer, and a
large magnification would be needed to reveal the gap between them.

\begin{figure}[!h]
\begin{center}
\begin{tabular}{cccc}
\includegraphics[width=4.2cm]{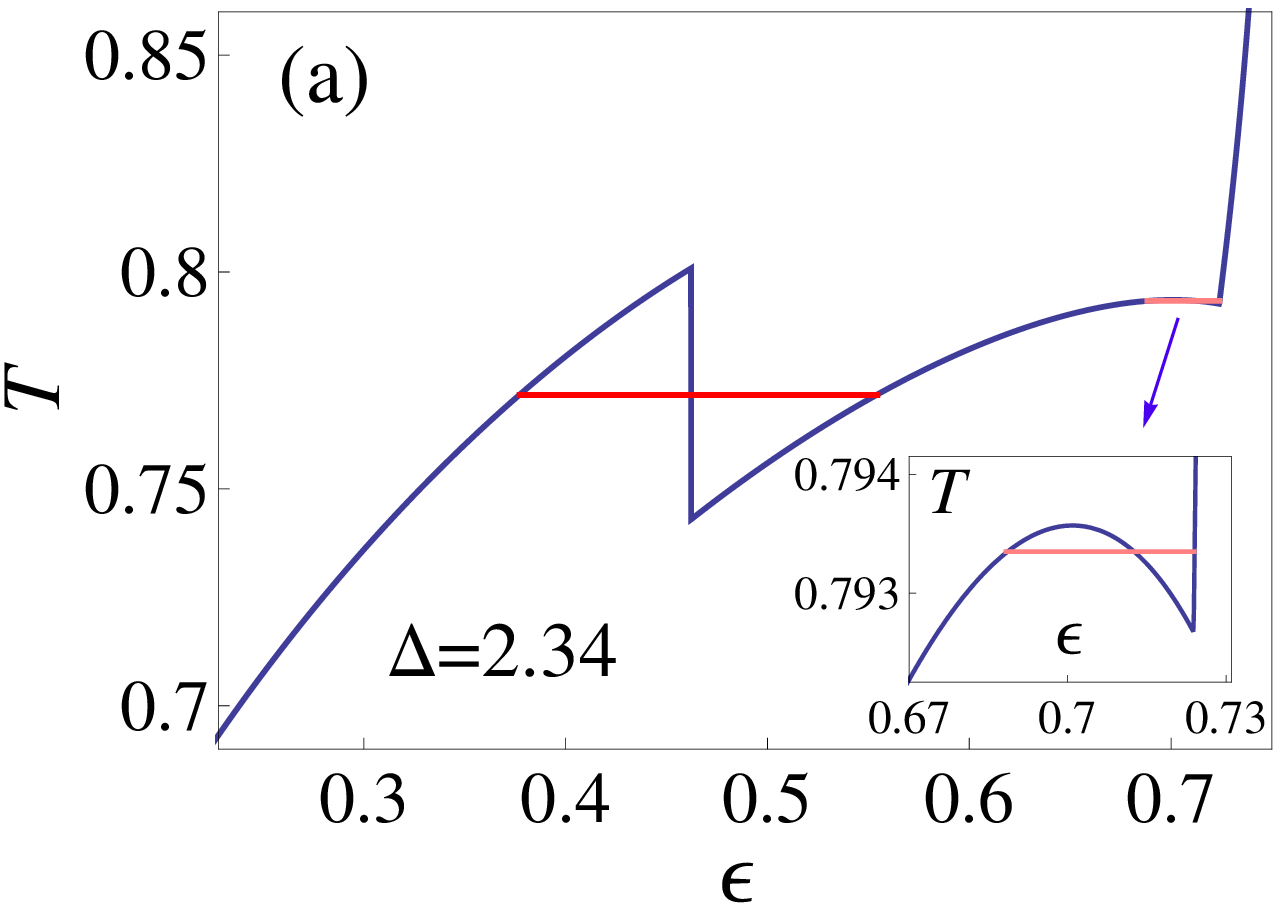} &
\includegraphics[width=4.2cm]{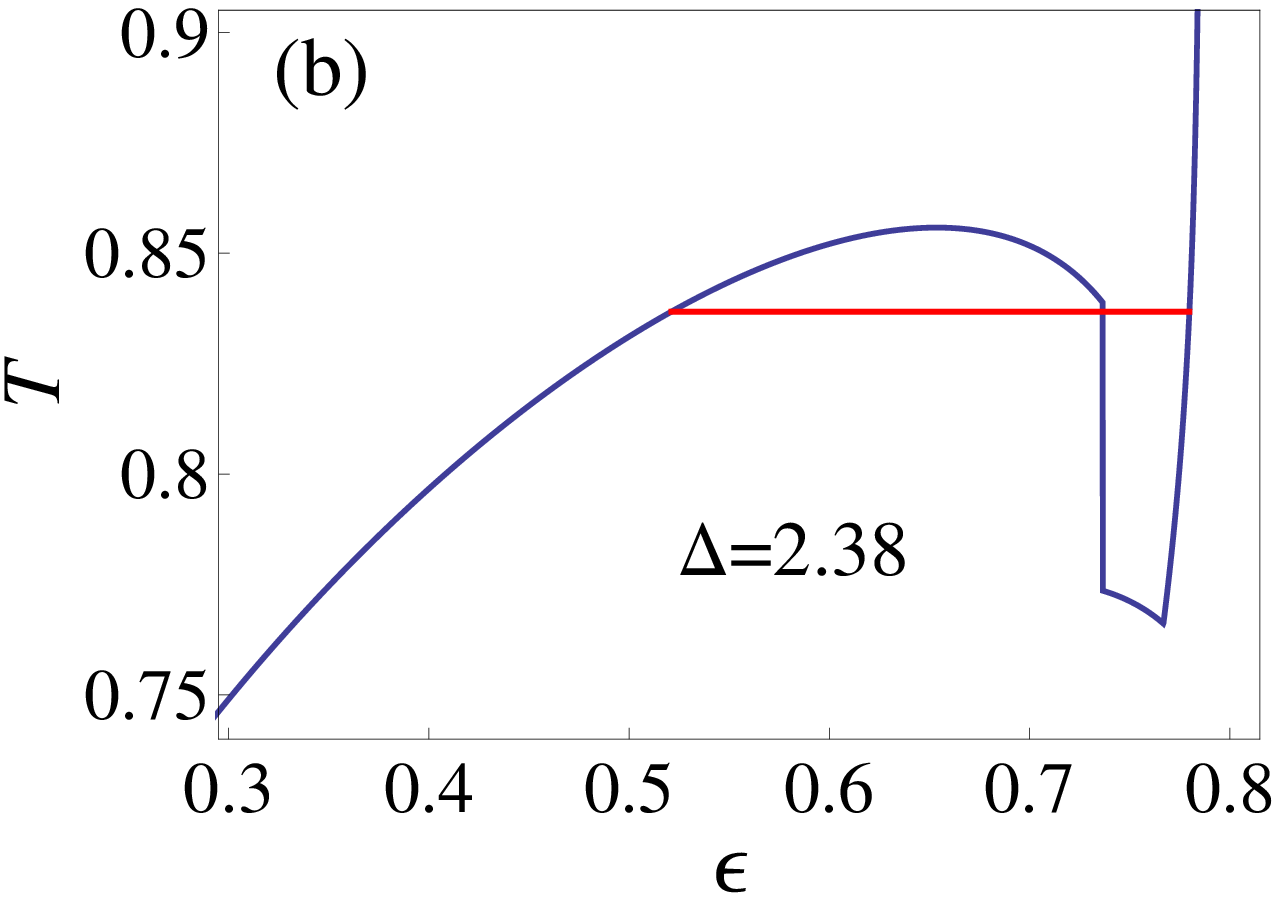}
\nonumber \\
\includegraphics[width=4.2cm]{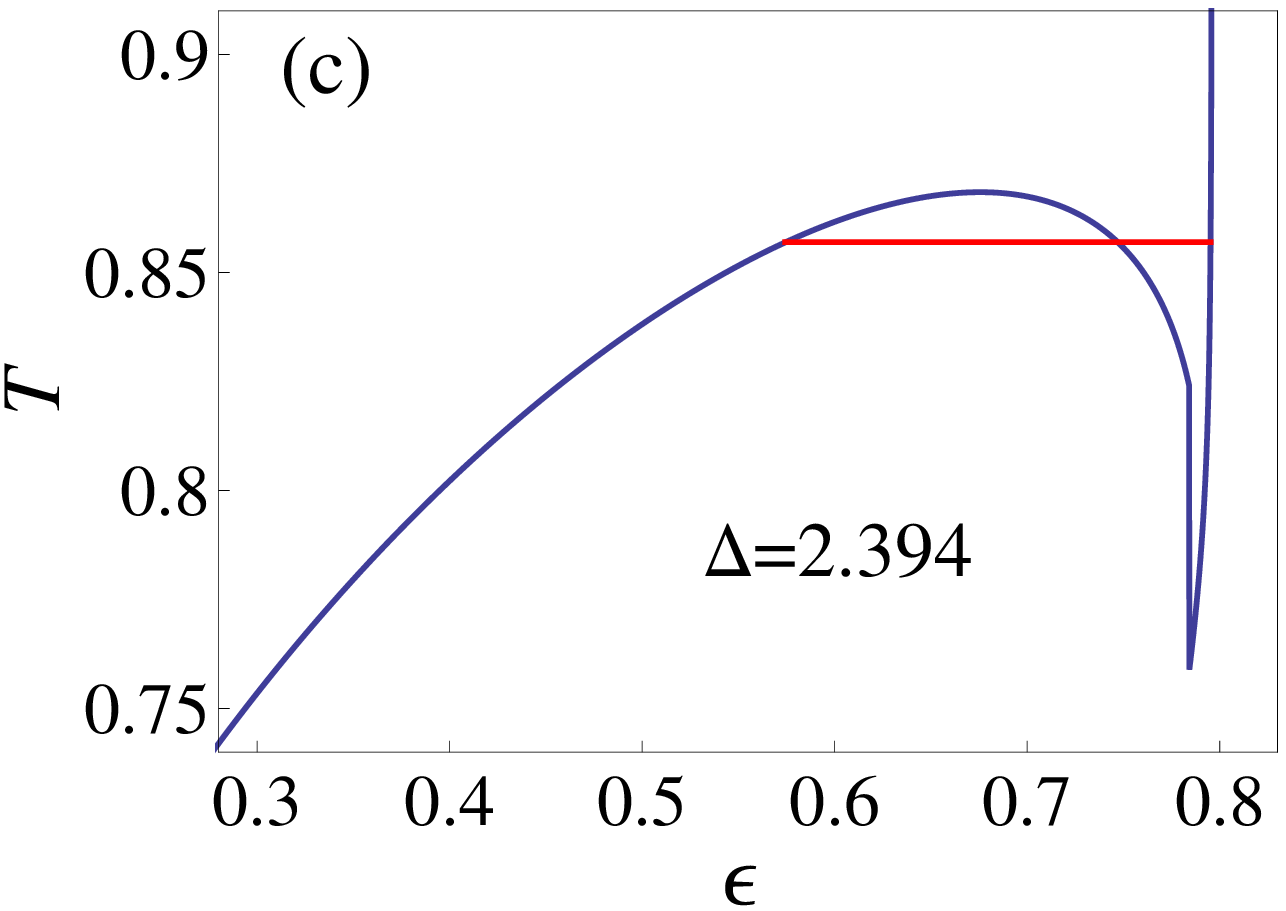} &
\includegraphics[width=4.2cm]{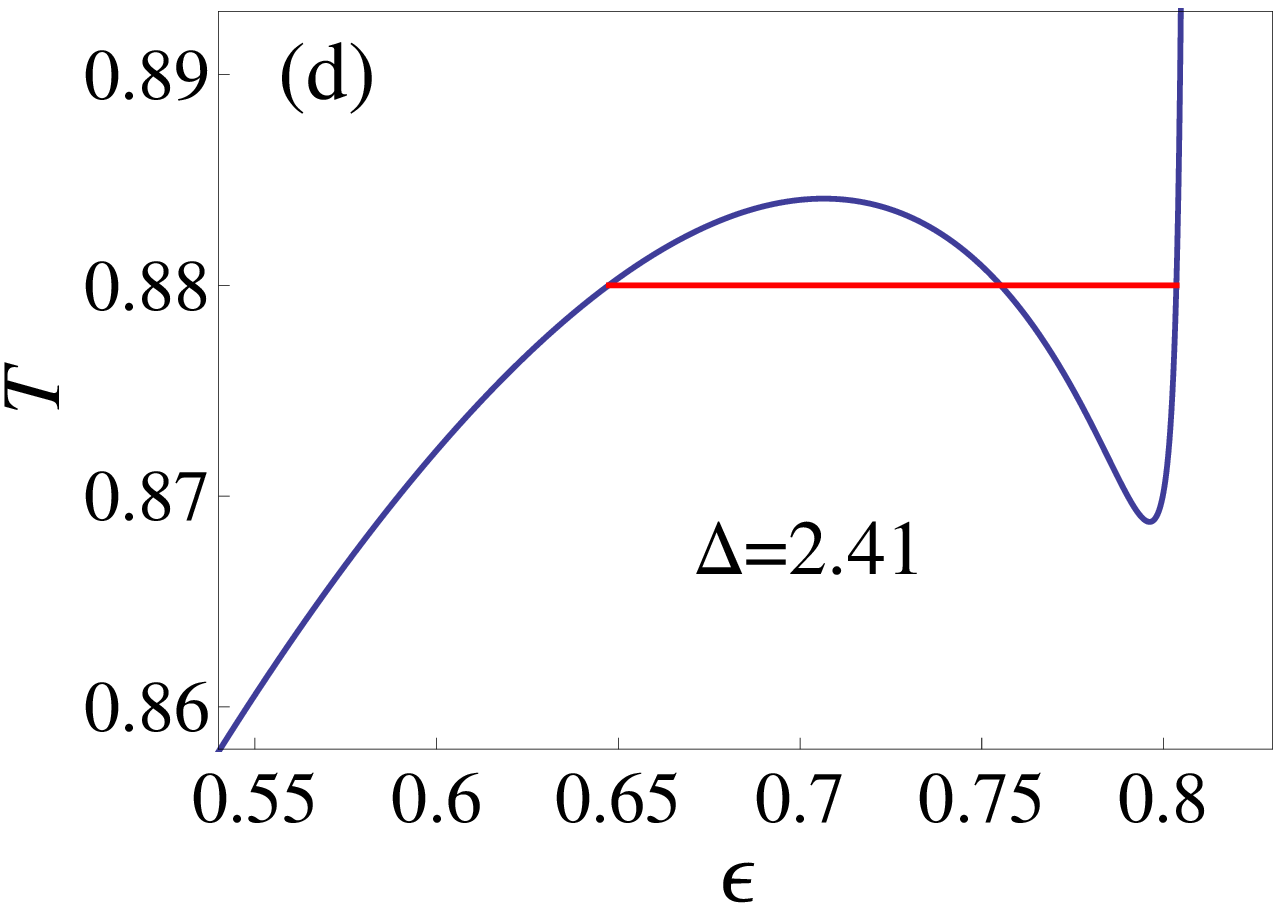}
\end{tabular}
\caption {Caloric curves for $K=3.6$ and different values of the
single-spin energy parameter $\Delta$. The meaning of the lines is the same as in
Fig.~\ref{Energy}.}
\label{CaloricK3p6}
\end{center}
\end{figure}

In Fig.~\ref{CaloricK3p6} we show the caloric curves at four
different values of $\Delta$.
Fig.~\ref{CaloricK3p6}(a) refers to a $\Delta$ value between the
canonical tricritical point and the canonical triple point.
Interestingly, by increasing the temperature or the energy the
system goes from a paramagnetic to a ferromagnetic state and
then back to a paramagnetic state. The first transition is
first order in both ensembles, while the second transition is first
order in the canonical ensemble and second order in the
microcanonical ensemble, accompanied by a region of negative
specific heat (it would be second order in both ensembles for a
$\Delta$ value smaller than that of the CTP). For a $\Delta$ value
between the CTP and the MBP (Fig.~\ref{CaloricK3p6}(b)) there is, in the canonical
ensemble, only a first order transition between two paramagnetic
states; in the microcanonical ensemble the situation is as in the
previous plot, although now the negative specific heat occurs also
before the temperature jump. For a $\Delta$ value between the
MBP and the MCP (Fig.~\ref{CaloricK3p6}(c)) there is, in both ensemble, only a first
order transition between two paramagnetic states. Finally, for
$\Delta$ between the microcanonical and the canonical critical point
(Fig.~\ref{CaloricK3p6}(d)) only the canonical transition is
present, while the microcanonical ensemble shows just a region of
negative specific heat. We will discuss in the Conclusions the analogy 
between this behaviour and the so-called gravitational phase transitions.

The coordinates of the relevant points of the phase diagrams for $K=3.6$ are:

$\bullet$ Canonical Tricritical Point: $T_{CTP} \backsimeq 0.8039$,
$\Delta _{CTP} \backsimeq 2.3170$, $\epsilon_{CTP}\backsimeq 0.6994$,

$\bullet$ Canonical Critical Point: $T_{CCP} = 0.9$, $\Delta _{CCP}
\backsimeq 2.4238$, $\epsilon_{CCP}\backsimeq 0.7619$,

$\bullet$ Canonical Triple Point: $T_{TP} \backsimeq 0.7898$, $\Delta _{TP}
\backsimeq 2.3474$, $\epsilon_{TP}\backsimeq 0.4168; 0.6848;
0.7321$,

$\bullet$ Microcanonical Branching Point: $T_{MBP} \backsimeq 0.7537$,
$\Delta_{MBP} \backsimeq 2.3927$, $\epsilon_{MBP}\backsimeq 0.7809$,

$\bullet$  Microcanonical Critical Point: $T_{MCP} = 0.8$, $\Delta_{MCP} = 2.4$,
$\epsilon_{MCP} = 0.8$.

\subsection{High values of $K$: $K>K_3\backsimeq 3.801$}

\begin{figure}[!h]
\begin{center}
\includegraphics[width=6.5cm]{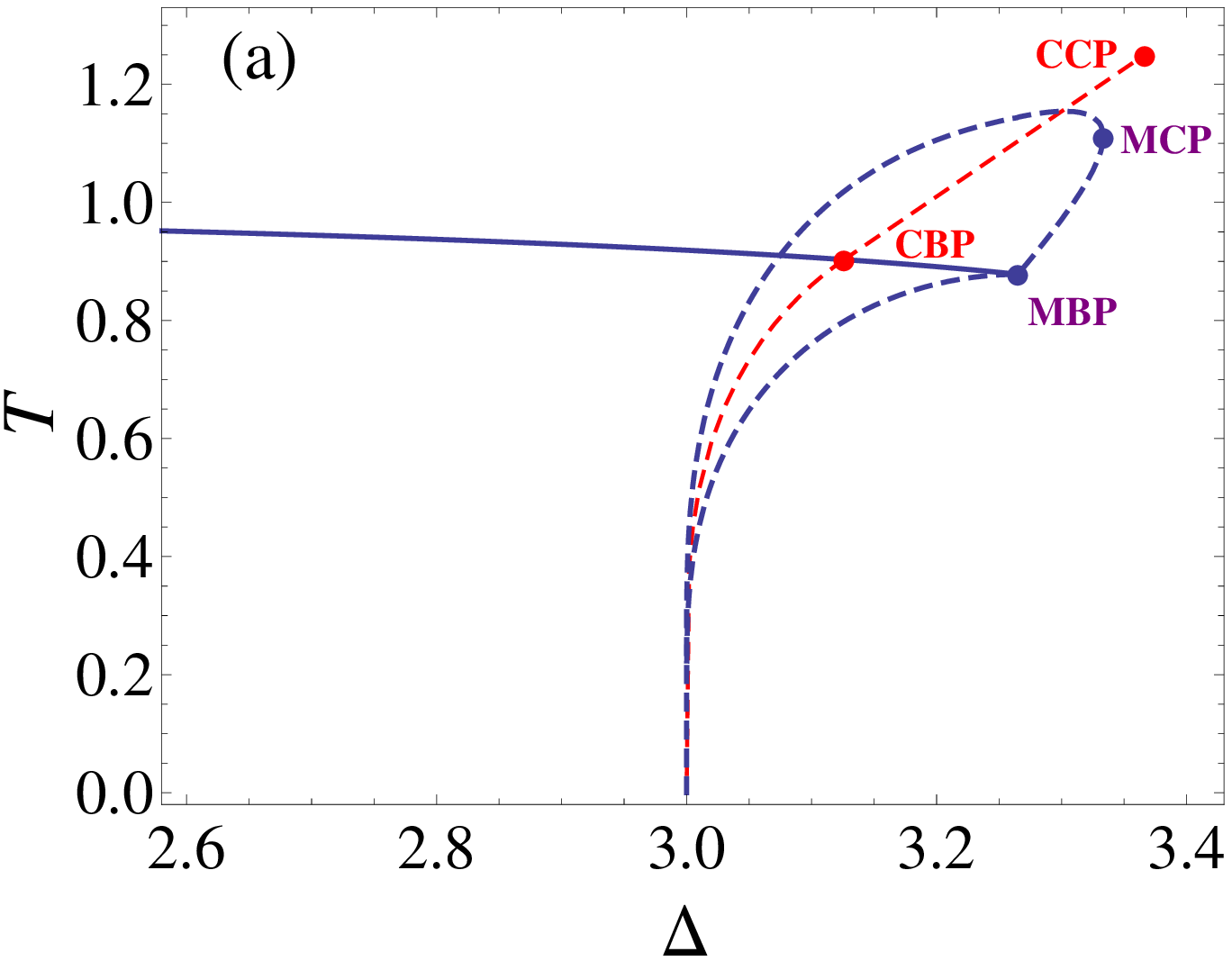}\\ \vskip 0.2cm
\includegraphics[width=6.5cm]{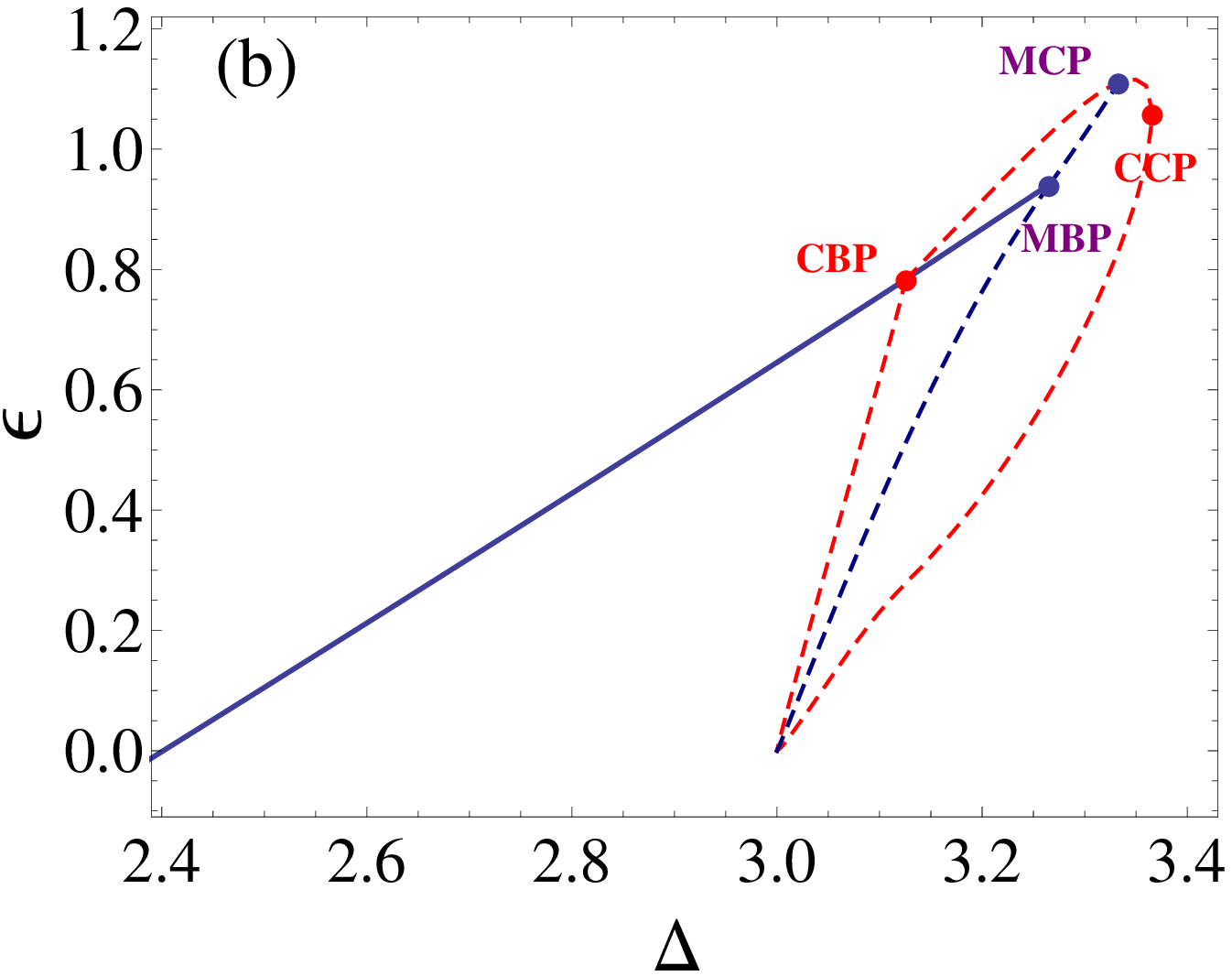}
\caption {The most interesting region of the phase diagrams of the canonical and microcanonical
ensembles at $K=5$. The solid blue line denotes the line of second order transitions, ending at the canonical
branching point (CBP) and at a microcanonical branching point (MBP) for the two ensembles, respectively.
Red (blue) dashed lines are the first order transition lines for the canonical (microcanonical) ensemble.
(a) $(\Delta,T)$ phase diagram. One canonical first order branch and the two microcanonical first order branches
meet at $T=0$ at $\Delta=(K+1)/2=3$. (b) $(\Delta,\epsilon)$ phase diagram. The first order lines meet at the point
with coordinates $\Delta=3$ and $\epsilon=0$ (the minimum allowed energy for $\Delta=3$).
The coordinates of the relevant points, CBP, MBP, MCP and CCP, are given in the text.}
\label{EnergyDeltaK5}
\end{center}
\end{figure}

For large $K$ values we have yet another kind of structure of the phase
diagram. We have chosen the value $K=5$ as representative for this
case. In Fig.~\ref{EnergyDeltaK5} we show both the $(\Delta,T)$ and
the $(\Delta,\epsilon)$ phase diagrams.
Now also the canonical ensemble does not present the tricritical
point anymore. Analogously to what happens for the microcanonical
ensemble in the previous $K$ range and also in this range, the line
of canonical second order transition ends before the quantity $B_c$
of Eq.~(\ref{bccan}) reaches the value $0$. That is because the states
corresponding to the critical line become metastable for values of
$\Delta$ larger than those where the canonical critical line
ends. It branches, at the canonical branching point (CBP), in two
lines of first order transition (in the $(\Delta,\epsilon)$ diagram
the lines departing from the CBP are those of the higher energy
associated to the energy gap corresponding to the first order
transition). As for the microcanonical case, for $\Delta$ values
smaller than that of the CBP the transition is between a
ferromagnetic and a paramagnetic state, while for $\Delta$ values
larger than that of CBP it is between two paramagnetic states (in
the literature the CBP is usually called canonical critical end
point; see also the last Subsection). The line of first order
transitions between paramagnetic states ends at the canonical
critical point (CCP).

\begin{figure}[!h]
\begin{center}
\begin{tabular}{cccc}
\includegraphics[width=4.2cm]{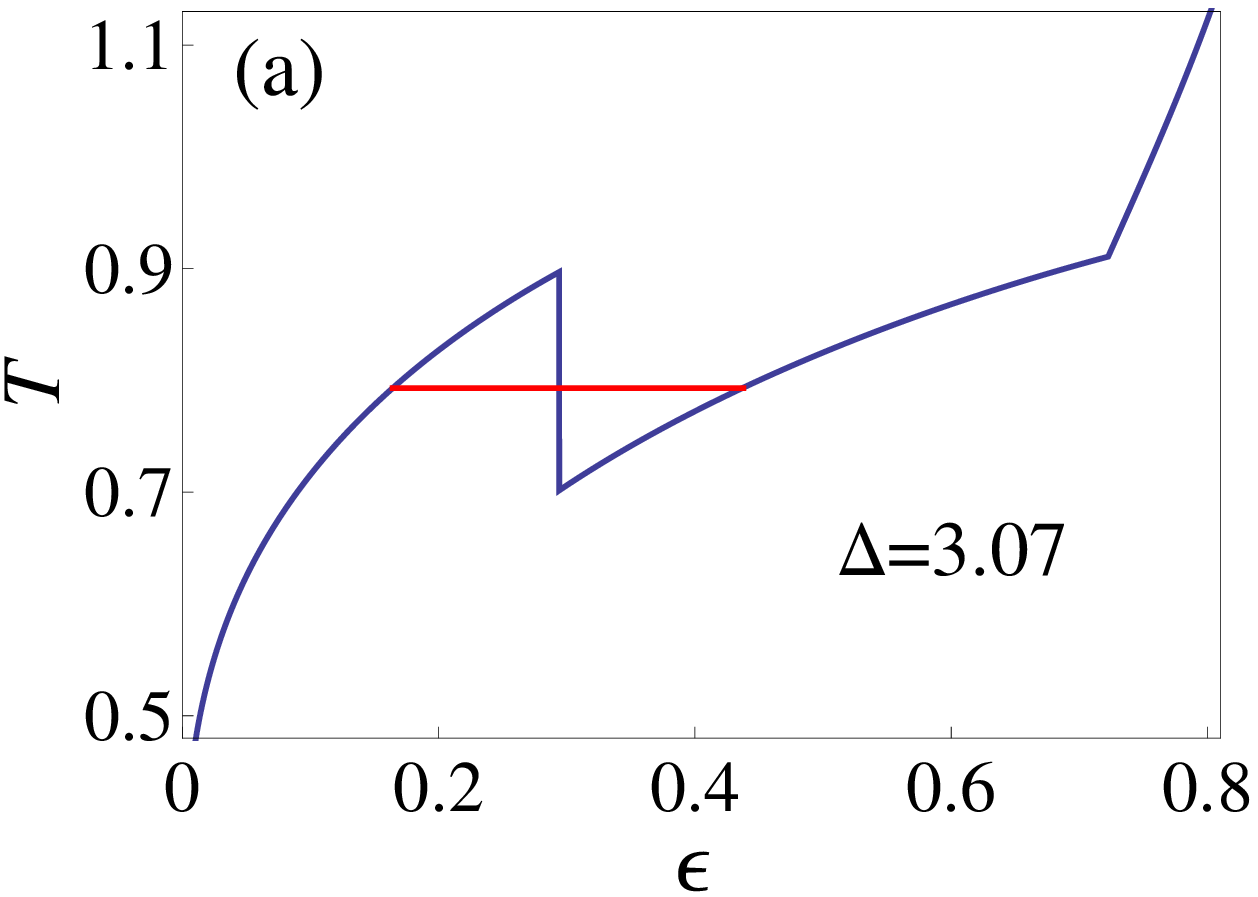} &
\includegraphics[width=4.2cm]{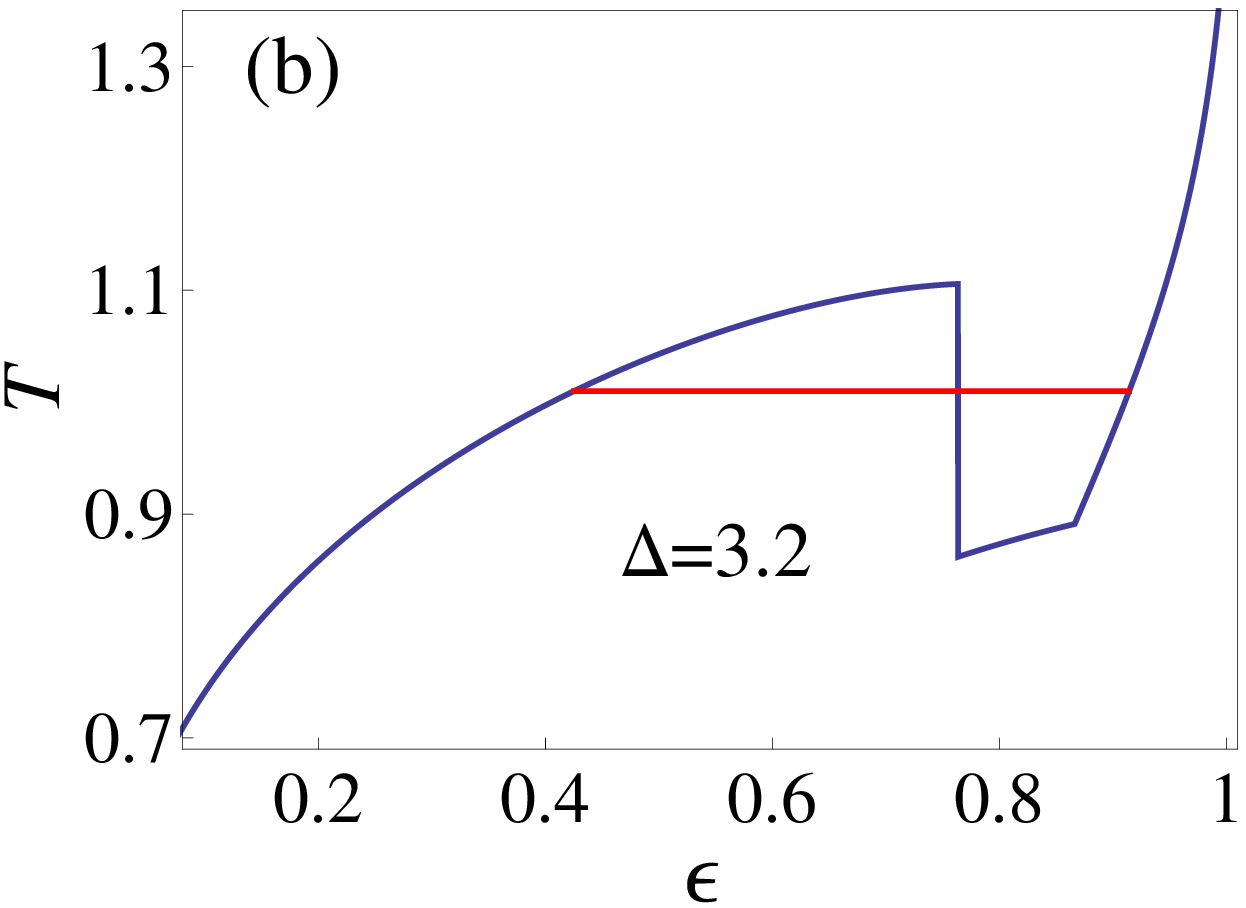}
\nonumber \\
\includegraphics[width=4.2cm]{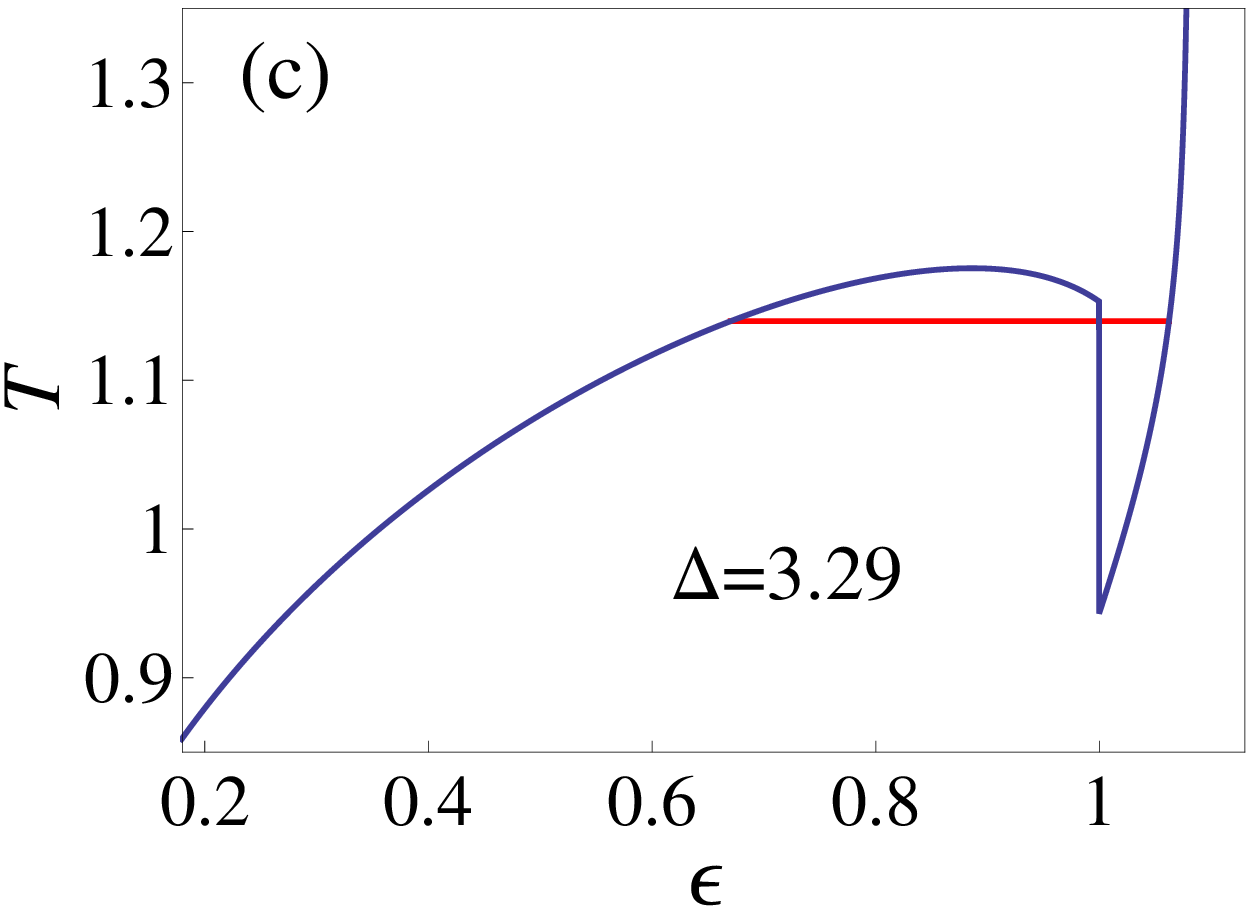} &
\includegraphics[width=4.2cm]{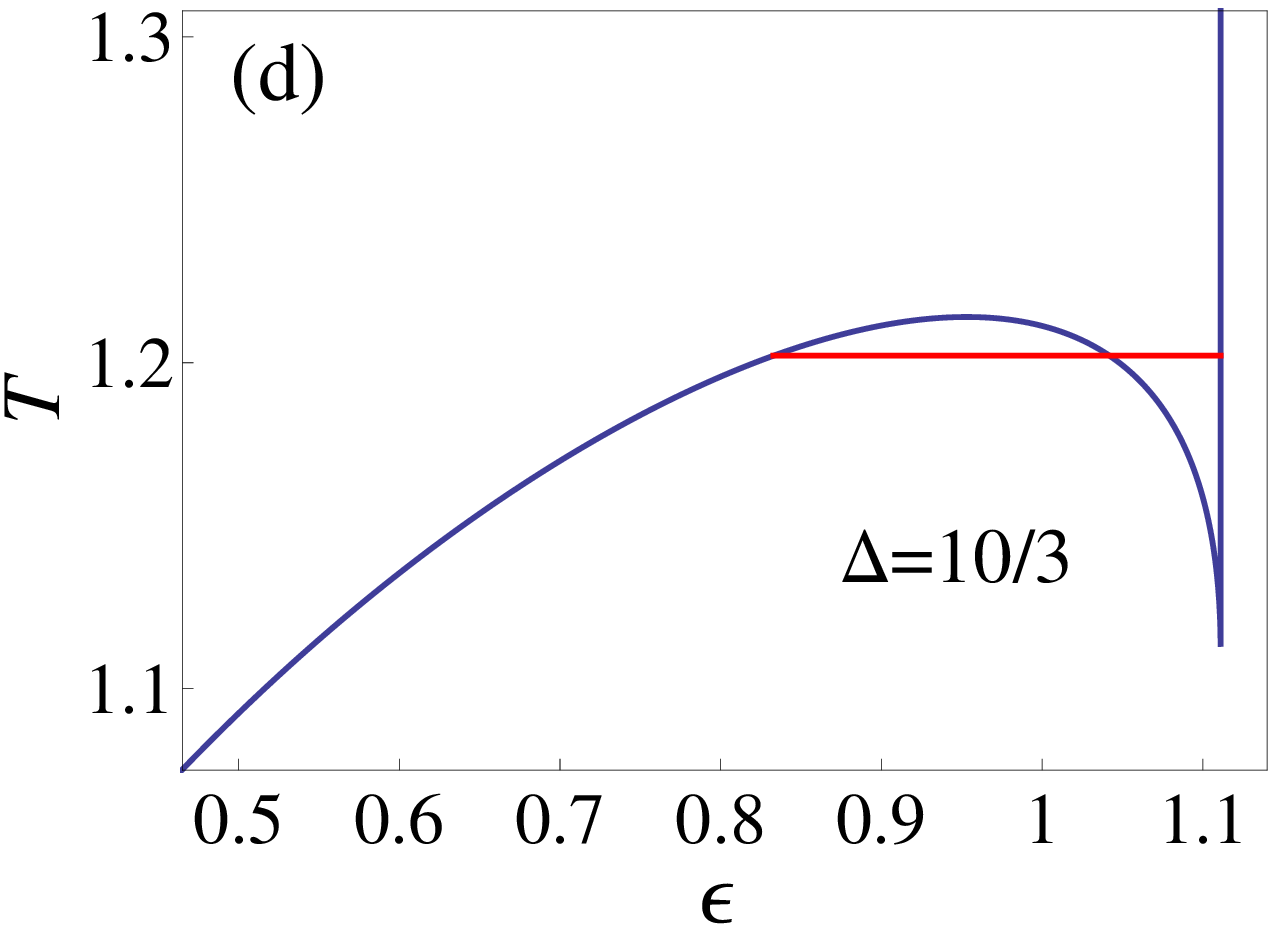}
\end{tabular}
\caption {Caloric curves for $K=5$ and different values of the
single-spin energy parameter $\Delta$. The meaning of the lines is the same as in
Fig. \ref{Energy}.}
\label{CaloricK5}
\end{center}
\end{figure}

In Fig.~\ref{CaloricK5} we show the caloric curves at four different
values of $\Delta$.
Fig.~\ref{CaloricK5}(a) refers to a $\Delta$ value smaller than the
canonical branching point, but larger than that where the first
order lines meet the $T=0$ (or $\epsilon=0$) line. By increasing the
temperature or the energy, the system goes from a paramagnetic
to a ferromagnetic state and then back to an paramagnetic state.
The first transition is first order in both ensembles, while the
second transition is second order and coincides in both ensembles.
For a $\Delta$ value between the canonical and the microcanonical
branching points (Fig.~\ref{CaloricK5}(b)) there is, in the
canonical ensemble, only a first order transition between two
paramagnetic states; in the microcanonical ensemble there is a first
order transition between the paramagnetic and the ferromagnetic state and
then a second order transition between the ferromagnetic and the other
paramagnetic state. For a $\Delta$ value between the MBP and the MCP
(Fig.~\ref{CaloricK5}(c)) there is, in both ensembles, a first order
transition between two paramagnetic states, with a region of
negative specific heat before the microcanonical transition.
Finally, at exactly the microcanonical critical point
(Fig.~\ref{CaloricK5}(d)) the canonical first order transition is
accompanied, in the microcanonical ensemble, by a second order
transition, preceded by a region of negative specific heat.

The coordinates of the relevant points of the phase diagrams for $K=5$ are:

$\bullet$ Canonical Branching Point: $T_{CBP} \backsimeq 0.9031$,
$\Delta _{CBP} \backsimeq 3.1256$, $\epsilon_{CBP}\backsimeq
0.7839$,

$\bullet$ Canonical Critical Point: $T_{CCP} = 1.25$, $\Delta _{CCP}
\backsimeq 3.3664$, $\epsilon_{CCP}\backsimeq 1.0582$,

$\bullet$ Microcanonical Branching Point: $T_{MBP} \backsimeq 0.8781$,
$\Delta_{MBP} \backsimeq 3.2651$, $\epsilon_{MBP}\backsimeq 0.9395$,

$\bullet$  Microcanonical Critical Point: $T_{MCP} = 10/9$,
$\Delta_{MCP} = 10/3$, $\epsilon_{MCP} = 10/9$.

\subsection{Negative temperatures}
\label{negtempsec}

We have already noticed that this system can achieve negative temperatures~\cite{TGBtemp}, since the energy is upper bounded. 
In this Subsection we evaluate which is the energy range in which the temperature is negative. Talking about negative temperatures, 
their physical meaning must be clearly kept in mind. In the microcanonical ensemble, they are a consequence of a negative derivative 
of the entropy with respect to the energy at the fixed energy at which the isolated system is considered, this negative derivative in
turn being a consequence of the existence of an energy upper bound. Obviously, in the canonical ensemble a heat bath cannot have a negative
temperature, forbidding a physical discussion of ensemble inequivalence. However, one could formally define a negative $\beta$, since 
the compact support of the phase space and the existence of an upper bound in energy give a finite partition function.

First of all, once we accept the formal definition of negative $\beta$, we can see that for negative temperatures the
ensembles are equivalent. This could be determined with a complete analysis through calculations similar to those of Secs.~\ref{model}
and~\ref{Microcanonical}. For the canonical case we would arrive at an expression for $\tilde{f}(\beta,x,y)$ different from 
Eq.~(\ref{free_energy}), since for negative $\beta$ the imaginary unit would appear in the second term of the exponent of the 
Hubbard-Stratonovich transformation~(\ref{HST}). However, we prefer, for brevity, to resort to a heuristic argument.

We have seen that, for large enough positive $T$, the system, for any
$\Delta$ and $K$, is in a paramagnetic state ($m=0$) and no further
phase transition occurs. We know that, when there are no transitions
in the canonical ensemble, the ensembles are equivalent, therefore we
immediately conclude in favour of ensemble equivalence for negative
temperatures, since by definition these temperatures are ``hotter''
than any positive temperature, favouring higher energy with
respect to lower energy states. In particular, $T=-\infty$ coincides
with $T=+\infty$, and by going from $T=-\infty$ to $T=0^-$ we
progressively go to ``hotter'' and ``hotter'' temperatures, while
energy will increase (positive specific heat) up to the upper
bound~\cite{foot4}. This is sufficient to determine the energy range
where negative temperatures are realized. We just need to compute
which is the energy of the system when $T=\infty$; between this value
and the upper bound of the energy, temperatures will be negative.

For $T=\infty$ we have $m=0$ and $q=2/3$, since the three spin
states are equally populated. Then the energy per particle will be
(from, e.g., Eq.~(\ref{enerDQM})) $\epsilon_{T=\infty}=
\frac{2}{3}\Delta -\frac{2}{9}K$. The upper bound of the energy can
be deduced as a consequence of the more detailed calculations in
Sec.~\ref{ergbreak} and Appendix~\ref{appenerrange} for the
bounds of the energy as a function of $m$. This
energy upper bound is $\epsilon_{\rm max}= \frac{\Delta^2}{2K}$ for
$\Delta/K \le 1$, realized for $m=0$ and $q=\Delta/K$, and
$\epsilon_{\rm max}= \Delta - \frac{K}{2}$ for $\Delta/K \ge 1$,
realized for $m=0$ and $q=1$. It is straightforward to see that it
is always $\epsilon_{\rm max} > \epsilon_{T=\infty}$, except for
$\Delta/K = 2/3$, when the two quantities are equal. Only in this
last case, negative temperatures are not realized. This has a
consequence on the determination of the microcanonical critical
point (MCP), as shown in Appendix~\ref{appmicrtric}.

\subsection{Analysis of the phase diagrams}
\label{phase_analysis}

The phase diagrams in the various $K$ ranges show different
features. They can be analyzed in the framework of the classification 
of phase transitions given in Ref.~\cite{bouchet05}. We do
this here very briefly.

The phase transitions of a system, both in the microcanonical and in
the canonical ensemble, can be deduced from the structure of the
function $s(\epsilon)$ computed in the microcanonical ensemble, more
precisely from its singularities. In turn, the singularities can be
classified according to their codimension, to be defined below. Let
us apply this concept to our system. We see from
Eq.~(\ref{entropyqm}) that the function $S/N$ depends on the two
order parameters $m$ and $q$, while Eq.~(\ref{enerqmeps}) gives a
relation between $m$, $q$, the energy $\epsilon$ and the two
parameters of the Hamiltonian $\Delta$ and $K$. Solving Eq.~(\ref{enerqmeps}) with respect
to $q$, allows us to write $S/N$ as a function of $m$, $\epsilon$, $\Delta$ and $K$, obtaining the two
functions $\tilde{s}_{\pm}$ (since Eq.~(\ref{enerqmeps}) has two
solutions). The maximization with respect to $m$ gives the
equilibrium entropy. In Sec.~\ref{Microcanonical} we
have indicated explicitly only the dependence of $\tilde{s}_{\pm}$
on $m$ and $\epsilon$, and the dependence of the equilibrium entropy
$s$ only on $\epsilon$. However, now it is useful to indicate
explicitly also the dependence of $s$ on the Hamiltanian parameters
$\Delta$ and $K$: $s=s(\epsilon;\Delta,K)$. Thus, we have a
dependence on the energy plus two parameters. Accordingly, the
thermodynamics phase space will have three dimensions, corresponding
to $(\epsilon,\Delta,K)$ or $(T,\Delta,K)$. In our presentation of
the results we have chosen to give the phase diagrams in the
$(\Delta,\epsilon)$ or $(\Delta,T)$ space at various fixed values of
$K$, since this was convenient for the central argument of our paper,
namely to show ensemble inequivalence and the different structures
of the phase diagram in the two ensembles in all ranges of $K$.
However, to follow the picture depicted in Ref.~\cite{bouchet05} we
have to consider the three dimensional phase diagram.

If $R$ is the number of parameters the entropy depends on, $2$ in
our case, i.e. $\Delta$ and $K$, a singularity is said to be of
codimension $n$ if it spans a hypersurface of dimension $R-n$ in the
thermodynamic phase space. Since $R=2$, we can have singularities of
codimension $0$, $1$ and $2$, spanning hypersurfaces of dimension
$2$, $1$ and $0$, respectively, in the three-dimensional
thermodynamic phase space. Therefore, all the  first order and
second order transitions lines in Figs.~\ref{Phasec}, \ref{Phase},
\ref{Energy-Delta}, \ref{TempDeltaK2p85}, \ref{EnergyDeltaK2p85},
\ref{TempDeltaK3p6} and \ref{EnergyDeltaK5} represent codimension
$0$ singularities: they are lines in the two-dimensional plots, but
they exist for given ranges of $K$, and therefore they are
two-dimensional hypersurfaces in the three-dimensional phase space.
Analogously, the singularities that appear as points in the various
two-dimensional plots, but that exist for given ranges of $K$, like
the tricritical, critical, branching and triple points are codimension 
$1$ singularities. On the other hand, codimension $2$ singularities are not visible in our
plots. They occur in isolated points in the three-dimensional phase
space, exactly for the values of $K$ where the structure of the
phase diagram changes; these are the boundaries of the $K$ ranges that
we have indicated in the titles of the previous Subsections.

Singularities in the function $s(\epsilon;\Delta,K)$ can originate
from different mechanisms, and we now present briefly their origin. The
reader interested in the complete general theory finds full details
in Ref.~\cite{bouchet05}. Let us begin with the microcanonical first
order transitions, associated to codimension $0$ singularities. They
arise from the process of maximization of
$\tilde{s}_{\pm}(\epsilon,m;\Delta,K)$ with respect to $m$ to get
$s(\epsilon;\Delta,K)$. When the functions $\tilde{s}_{\pm}$ have
more than one local maximum, one of which is the absolute maximum,
the singularities, i.e. the first order transitions, occur when one
local maximum becomes, changing $\epsilon$ or the parameters, the
global one. This singularity spans a two dimensional hypersurface
since it is defined by one condition, i.e. the equality of two
maxima. This transition is always associated with a negative
temperature jump (by increasing energy), as can be understood by the
following. The inverse temperature $\beta$ is given by $(\partial
s)/(\partial \epsilon)$, which is also equal to $(\partial
\tilde{s}_+)/(\partial \epsilon)$ or $(\partial
\tilde{s}_-)/(\partial \epsilon)$ (the one giving the global
maximum) taken at the equilibrium magnetization. At the point of
transition, when one local maximum becomes global, necessarily this
must have a rate of increase with $\epsilon$ faster than the maximum
becoming local from global; this means that the new maximum has a
larger $\beta$, i.e., a smaller $T$.

The canonical first order transitions are other codimension $0$ singularities, and they have 
another origin. They come from the construction of the concave envelope of a function $s$ 
that has convex portions as a function of $\epsilon$. Also in this case there is only one condition,
namely one has to locate the point in which the entropy $s$ is no more globally concave.

The final codimension $0$ singularity present in our model is
represented by the critical line associated to second order
transitions, either in both ensembles or only in the microcanonical
ensemble. It can be shown~\cite{bouchet05} that this kind of
codimension $0$ singularity is present when there is a symmetry in
the system, and the transition breaks the symmetry in the
equilibrium state. In our case the symmetry is the $m \to -m$
invariance under change of sign, and the symmetry is between symmetric paramagnetic $m=0$
states that undergo a continuous transitions to states with $m \ne
0$ that break the symmetry. As explained in Appendix~\ref{appmicrtric} in this case there is a jump in $(\dd
\beta)/(\dd \epsilon)$, i.e. the caloric curve presents a
discontinuity in the derivative. Now, the condition to satisfy is the
vanishing of the second derivative of $\tilde{s}$ with respect to
$m$ in $m=0$ for the microcanonical ensemble, and the vanishing of
the second derivative of $\tilde{f}$ with respect to $x$ in $x=0$
for the canonical ensemble. As shown in Sec.~\ref{Microcanonical}
the latter condition is actually identical to the former.

We now consider codimension $1$ singularities, beginning from those
due to the change of sign symmetry of the order parameter $m$. They
are the tricritical points and the branching points in both
ensembles. The two conditions giving the microcanonical tricritical
point are the vanishing of the second and fourth order derivative of
$\tilde{s}$ with respect to $m$ in $m=0$, while for the canonical
tricritical point they are the vanishing of the second and fourth
order variation of $\tilde{f}$ in $x=0$. It can be
shown~\cite{Campa09} that the vanishing of the fourth order variation of
$\tilde{f}$ in $x=0$ is simply related to a
condition on $\tilde{s}$ in $m=0$, confirming that the transitions
in both ensembles are related to the structure of the microcanonical
entropy.

We have already noticed that what we have called branching point is
defined in the literature as critical end point. This is due to the
fact that they are the intersection point of a critical line with a
line of first order transitions. The critical line
must end at the intersection point because the states related to the 
first order transitions are thermodynamically favoured; those associated to the
critical line become metastable. The conditions giving rise to this
situation are different in both ensembles, so the canonical and
microcanonical branching points are located in different positions.
For the microcanonical case, the point arises from the equality of
three maxima with respect to $m$, two located on $\tilde{s}_+$ and
one on $\tilde{s}_-$ (or viceversa). For the canonical case, it is
defined by the appearance of a convex portion of $s$ at exactly the
point of second order transition.

The remaining codimension $1$ singularities in our system are not
associated to the change of sign symmetry. Let us begin with the
canonical ensemble. The two conditions giving rise to the canonical
critical point (Fig.~\ref{CaloricK2p85}(d)) are the vanishing at the
same point of both the second and third order derivatives of $s$ with
respect to $\epsilon$. On the other hand, the canonical triple point
is related to the merging, in the function $s(\epsilon)$, of two
regions not globally concave in a single one; this can be seen in
Fig.~\ref{CaloricK2p85}, going from panel (c) to panel (b)
(the triple point) and then to panel (a). We finally consider the
microcanonical critical point. Normally, as explained in
Appendix~\ref{appmicrtric}, it is associated, in systems without
symmetry, to the vanishing at the same point of the second and third
order derivatives of $\tilde{s}$ with respect to $m$. But we have seen in
the same Appendix that our system has a rather peculiar
microcanonical critical point, located at the upper bound of the
energy. In this case the two conditions are given by a relation
between the two parameters, i.e., $\Delta=2K/3$, and the energy
upper bound. This critical point is present when there is a line of
first order transitions between two different paramagnetic states,
and this occurs for $K>3$.

Codimension $2$ singularities are not explicitly treated in
Ref.~\cite{bouchet05}. As mentioned above, they are located at the
$K$ values that mark the change of structure of the $(\Delta,T)$ or
$(\Delta,\epsilon)$ phase diagram, i.e. at the point we have
indicated with $K_1 \backsimeq 2.775$, $K_2=3$ and $K_3 \backsimeq
3.801$. At $K_1$ we have the appearance of the canonical triple
point and the canonical critical point. The three conditions in this
case are that the canonical critical point (two conditions) appears
at the same point of the canonical first order transition (the other
condition). It is as if the panels (b) and (d) of
Fig.~\ref{CaloricK2p85} would merge. At $K_2$ we have the disappearance of
the microcanonical tricritical point and the appearance of the
microcanonical branching and critical points. The three
conditions defining this point are that the microcanonical
tricritical point (two conditions) occurs at the upper bound of the
energy (the other condition). Finally, at $K_3$ we have the
disappearance of the canonical tricritical point and triple point
and the appearance of the canonical branching point. This is due
to the fact that the canonical tricritical point (two conditions)
reaches the canonical first order transition line (the other
condition) between the two paramagnetic states.

\section{Ergodicity breaking}
\label{ergbreak}

Now, let us turn our attention to another interesting property of this system. That
is ergodicity breaking in the microcanonical dynamics with fixed energy~\cite{Borgonovi,Schreiber}. This statement
assumes that the dynamics of the system, whose Hamiltonian has only discrete spin variables,
is defined in such a way that, as in systems with continuous dynamical variables, thermodynamic
parameters can vary only continuously.
A microcanonical dynamics for spin systems, in which at each step only one spin is updated,
has been introduced by Creutz~\cite{creutz}. Then, in the thermodynamic limit, magnetization $m$
can vary only continuously.

Ergodicity breaking at thermodynamic limit naturally occurs in nonadditive systems, like our mean-field
model, where the absence of phase separation makes it possible to have a nonconvex
region delimitating the accessible values of thermodynamic parameters. In this case
breaking of ergodicity manifests itself by the presence of disjoint ranges of magnetization $m$
for a given energy value $\epsilon$. Therefore, a dynamics
starting in one of the disjoint ranges cannot access states belonging to the other ranges,
although these would be allowed energetically.

To see how this situation arises, we can adopt the following strategy. For a given value of the
magnetization $m$, we look for the minimum and maximum values that can be achieved by the energy
$\epsilon$; this will also determine, for a given energy $\epsilon$, which are the possible
values of $m$.

\begin{figure*}[ht]
\centerline{\psfig{figure=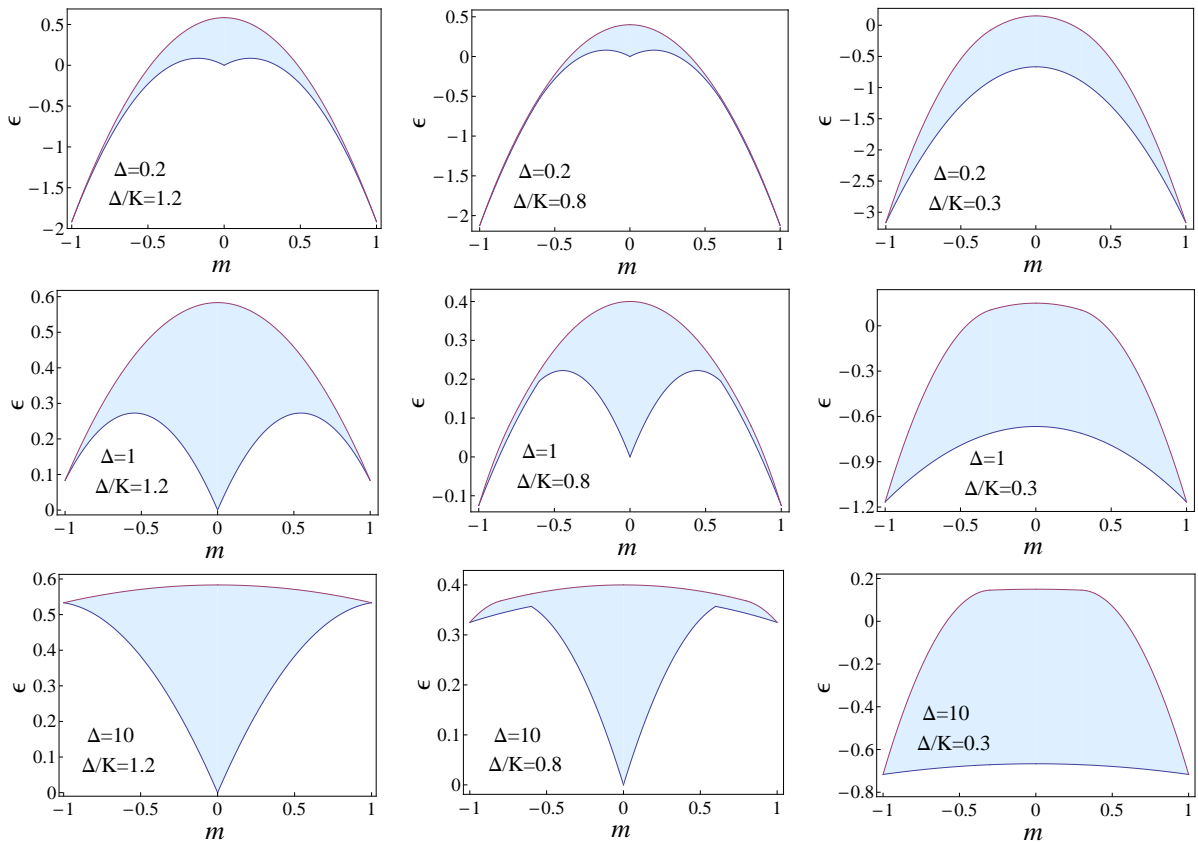,width=16 cm}}
\caption {A schematic representation of the accessible region in the
$m-\epsilon$ plane for fixed values of $\Delta$ and
$\Delta/K$.}
\label{AccRegs}
\end{figure*}

Let us rewrite Eq.~(\ref{enerqmeps}) as
\begin{equation}
\epsilon = \Delta q -\frac{K}{2} q^2  -\frac{m^2}{2}  \, .
\label{redhamil}
\end{equation}
Magnetization $m$ varies in the range $-1 \le m \le 1$,
but by symmetry we can restrict the analysis to the range $0\le m
\le 1$. As mentioned above, we are interested in the range of variability
of energy $\epsilon$ for any given fixed value of
$m$; in particular, we want to determine the minimum and maximum
values of $\epsilon$ for any given value of $m$. We first notice that
for a given $m$, the quantity $q$ varies in the range $m \le q \le
1$; therefore we have to determine, for any given $m$, the minimum
and maximum values of $\epsilon$ given by Eq.~(\ref{redhamil}) when
$q$ varies in the range $m\le q \le 1$. We denote these values as
$\epsilon_{\rm min}(m)$ and $\epsilon_{\rm max}(m)$, respectively.
Detailed calculations are shown in Appendix~\ref{appenerrange}.

The functions $\epsilon_{\rm min}(m)$ and $\epsilon_{\rm max}(m)$ depend on the
ratio $\Delta/K$. They are the following.

\vskip 0.4cm
Case $\frac{\Delta}{K}\ge 1$
\begin{eqnarray}
\epsilon_{\rm min}(m) &=& \Delta m - \frac{K}{2} m^2 -\frac{m^2}{2}
\,\,\,\,\,\,\,\,\,\,\,\,\,\,\,\,\,\,\,\,\,\,\,\,\,\,\,\,\,\,\,\,\,\,
0\le m \le 1
\nonumber \\
\epsilon_{\rm max}(m) &=& \Delta - \frac{K}{2}  -\frac{m^2}{2}
\,\,\,\,\,\,\,\,\,\,\,\,\,\,\,\,\,\,\,\,\,\,\,\,\,\,\,\,\,\,\,\,\,\,\,\,\,\,\,\,\,\,\,\,\,\,\,
0\le m \le 1 \nonumber \label{sum1}
\end{eqnarray}
\vskip 0.4cm
Case $\frac{1}{2} \le \frac{\Delta}{K}\le 1$
\begin{eqnarray}
\epsilon_{\rm min}(m) &=& \Delta m - \frac{K}{2} m^2  -\frac{m^2}{2}
\,\,\,\,\,\,\,\,\,\,\,\,\,\,\,\,\,\,\,\,\,
0\le m \le \frac{2\Delta}{K} -1
\nonumber \\
\epsilon_{\rm min}(m) &=& \Delta - \frac{K}{2}  -\frac{m^2}{2}
\,\,\,\,\,\,\,\,\,\,\,\,\,\,\,\,\,\,\,\,\,\,\,\,\,\,\,\,\,\,\,\,\,\,
\frac{2\Delta}{K} -1 \le m \le 1
\nonumber \\
\epsilon_{\rm max}(m) &=& \frac{\Delta^2}{2K}  -\frac{m^2}{2}
\,\,\,\,\,\,\,\,\,\,\,\,\,\,\,\,\,\,\,\,\,\,\,\,\,\,\,\,\,\,\,\,\,\,\,\,\,\,\,\,\,\,\,\,
0\le m \le \frac{\Delta}{K}
\nonumber \\
\epsilon_{\rm max}(m) &=& \Delta m - \frac{K}{2} m^2 -\frac{m^2}{2}
\,\,\,\,\,\,\,\,\,\,\,\,\,\,\,\,\,\,\,\,\,\,
\frac{\Delta}{K} \le m \le 1 \nonumber \label{sum1b}
\end{eqnarray}
\vskip 0.4cm
Case $0 \le \frac{\Delta}{K} \le \frac{1}{2}$
\begin{eqnarray}
\epsilon_{\rm min}(m) &=& \Delta - \frac{K}{2} -\frac{m^2}{2}
\,\,\,\,\,\,\,\,\,\,\,\,\,\,\,\,\,\,\,\,\,\,\,\,\,\,\,\,\,\,\,\,\,\,\,\,\,\,
0 \le m \le 1
\nonumber \\
\epsilon_{\rm max}(m) &=& \frac{\Delta^2}{2K}  -\frac{m^2}{2}
\,\,\,\,\,\,\,\,\,\,\,\,\,\,\,\,\,\,\,\,\,\,\,\,\,\,\,\,\,\,\,\,\,\,\,\,\,\,\,\,\,\,\,\,\,\,\,
0\le m \le \frac{\Delta}{K}
\nonumber \\
\epsilon_{\rm max}(m) &=& \Delta m - \frac{K}{2} m^2 -\frac{m^2}{2}
\,\,\,\,\,\,\,\,\,\,\,\,\,\,\,\,\,\,\,\,\,\,\,\,
\frac{\Delta}{K} \le m \le 1 \nonumber \label{sum1c}
\end{eqnarray}
The functions $\epsilon_{\rm min}(m)$ and $\epsilon_{\rm max}(m)$
are always continuous. Also their derivatives are continuous, except
the derivative of $\epsilon_{\rm min}(m)$ in the case $\frac{1}{2}
\le \frac{\Delta}{K}\le 1$, which has a discontinuity at
$m=\frac{2\Delta}{K}-1$. The results are extended by symmetry to
negative values of $m$. In Fig.~\ref{AccRegs} we report an example
for each one of the three cases.

The structure of the plots determines if an ergodicity breaking is
present and in which way it is realized. We observe in the plots
that, although the functional form of $\epsilon_{\rm min}(m)$ and
$\epsilon_{\rm max}(m)$ depends only on the ratio $\Delta/K$, some
relevant features depend on both $\Delta/K$ and $\Delta$. For
example, from the above results, we obtain that the upper bound of
the energy is always realized for $m=0$, and its value depends only
on the ratio $\Delta/K$, being equal to $\Delta^2/(2K)$ for
$\Delta/K \le 1$ and equal to $\Delta-K/2$ for $\Delta/K \ge 1$. On
the other hand, for $\Delta/K \le 1/2$ the minimum possible value of
the energy is always realized for $m=1$, and it is equal to $\Delta-
(K+1)/2 <0 $; while for $\Delta/K \ge 1/2$ this occurs only when
$\Delta - (K+1)/2 <0$ (this condition being always realized when
$\Delta/K < 1/2$), otherwise the minimum is obtained for $m=0$ and
it is equal to $0$. We remark that this result on the minimum energy
is consistent with the first order transition at $T=0$ between a
fully magnetized ferromagnetic state and a paramagnetic state at
$\Delta=(K+1)/2$ (see Sec.~\ref{model}).

These results show that, when $\Delta/K < 1/2$, ergodicity breaking is
due to the fact that, for some values of the energy $\epsilon$,
magnetization $m$ can achieve values in two disjoint ranges $m_1\le
m \le m_2$ and $-m_2 \le m \le -m_1$, with $0<m_1<m_2$. When
$\Delta/K >1/2$ one can have this situation, but also a more complex
situation in which magnetization can take values in three
disjoint ranges, $-m_2 \le m \le -m_1$, $-m_0 \le m \le m_0$ and
$m_1 \le m \le m_2$, with $0<m_0<m_1<m_2$. It is also possible, for
$\Delta/K>1$, to fall into a case in which ergodicity breaking is
absent, as in the plot in the left bottom panel of Fig.~\ref{AccRegs}; it
is not difficult to compute that this occurs, for a given ratio
$\Delta/K=r>1$, when $\Delta > r/(r-1)$.

\section{Conclusions}
\label{conclusions}

The Blume-Emery-Griffiths (BEG) model was completely solved in the
mean-field approximation in the canonical ensemble~\cite{Blume71}.
The study has been extended to the case where the Hamiltonian was
augmented with an external magnetic field term~\cite{mukamel74},
revealing the complexity of the canonical phase diagram. The
microcanonical solution of the Blume-Capel model~\cite{Blume66}, a
simplified version of the BEG model, was first obtained in
Ref.~\cite{Barre01} and ensemble inequivalence was discussed in this
context.

In this paper, we have extended the microcanonical solution to the original BEG model,
showing a wealth of different features in the phase diagrams pertaining to the canonical and
microcanonical ensembles.

As we have remarked in Sec.~\ref{phase_analysis}, the phase diagram of the model should be
represented in terms of three parameters $(\Delta,T,K)$ or $(\Delta,\epsilon,K)$, where $T$ is
temperature and $\epsilon$ the energy density. However, for convenience we have fixed
the value of the biquadratic interaction coefficient $K$ and represented the $(\Delta,T)$ and $(\Delta,\epsilon)$
phase diagrams. For different values of $K$ one observes various facets under which ensemble
inequivalence occurs.

At small values of $K$, the behavior is the one observed for the Blume-Capel model and for
other models: the two ensembles have a common critical line of continuous
transitions, terminating at tricritical points that are distinct in both ensembles. In the
region where the transition is first order in the canonical ensembles the phase transition lines
do not coincide. In conclusion, whenever the transition is continuous in both ensembles they are equivalent,
while they are not equivalent when the transition is first order in the canonical or in both.

This generally occurs when there is a symmetry of the entropy under a sign change of the
order parameter ($m \to -m$)~\cite{bouchet05}, as also displayed by an analysis of the neighborhood
of the transition points through a Landau expansion~\cite{ocohen}.

For larger values of $K$, above a given threshold $K_1$ for canonical ensemble and $K_2$ for the microcanonical
ensemble, the model has also a transition between different paramagnetic states ($m=0$). This additional transition
is characterized by a change in the quadrupole moment $q$ and is not associated to a symmetry under change of
sign of $q$. As we have seen, while a symmetry breaking second order canonical transition point is also a second order
microcanonical transition point, this is no more the case when the transition is not associated to a symmetry breaking.
Moreover, critical points in different ensembles do not necessarily coincide for non symmetry breaking transitions.
This remark was already partly contained in Refs.~\cite{bouchet,ocohen}.

The interest of the BEG model is that within the same simple model both situations occur, i.e. symmetry breaking and
non symmetry breaking transitions. The latter type is the typical situations that occurs for self-gravitating
systems~\cite{campa2016,chavanis02}, namely, for a value of the control parameter ($\Delta$ in our case) intermediate between those pertaining to
the canonical and the microcanonical critical points, the system has a first order transition in the canonical ensemble, while
in the microcanonical ensemble it has a region of negative specific heat without singularities.

Although at first glance it might seem the contrary, the fact that the canonical critical point is a generic point
of the phase diagram for the microcanonical ensemble does not spoil the general fact that when the canonical ensemble
has a continuous transition the ensembles are equivalent~\cite{Barre01,touch04,foot5}. In fact, at exactly the canonical critical
point the canonical and microcanonical caloric curves are identical (see, e.g. Fig.~\ref{CaloricK2p85}(d)). Therefore, the
thermodynamic functions and their derivatives of all orders with respect to the thermodynamic variables are identical in both
ensembles, showing in particular a diverging specific heat at the transition temperature or energy. As a consequence of this
divergence, as soon as the parameter $\Delta$ decreases from the critical value, the canonical ensemble shows a first order
transition, while the microcanonical ensemble has a region of negative specific heat.

We have based our analysis on the study of the singularities of the microcanonical entropy function. In the literature
this has been characterized as the thermodynamic level of ensemble inequivalence, and it has been
compared to the macrostate level, based on the properties of the correspondence between the equilibrium states occurring in the
two ensembles~\cite{ellis04,touch04,ellis2000}. At the energies where the microcanonical entropy coincides with the
Legendre-Fenchel transform of $\beta f(\beta)$, the ensembles are equivalent at the thermodynamic level, otherwise they
are not equivalent at this level and also at the macrostate level. At the energies where equivalence at the thermodynamic level
occurs, the ensemble can be, at the macrostate level, either fully equivalent or partially equivalent~\cite{ellis2000}. According
to this classification, in a case like ours, in which there are no ranges of energy where the microcanonical entropy is a straight
line (due to the absence of phase separation), there is  partial equivalence at the macrostate level only at the energies
where a first order canonical phase transition occurs (thus there are no ranges of partial equivalence, but only isolated energy
values); more specifically, the points of partial equivalence are those at the two extremes of each segment marking a Maxwell
construction.

The study of the BEG model in the canonical ensemble has been performed in the past on a Bethe lattice~\cite{anan91,anan96}.
In this case, the model can be exactly solved using recursive relations for the partition function.
Inequivalence for Bethe lattices was studied for the Potts model~\cite{barre07}. Using the cavity technique applied in the
latter paper, one could solve the BEG model on a Bethe lattice and analyze ensemble inequivalence.
The Blume-Capel model on long-range random networks was recently studied in Ref.~\cite{Levon}. Inequivalence was shown to
occur near first order phase transitions using extensive numerical simulations.
A class of models sometimes considered is the one of equivalent-neighbor models. When the cutoff distance
in this model is infinite they become identical to mean-field models, although for any large but finite cutoff distance their
critical behavior is different from that of mean-field models~\cite{lui1996,qian2016}. This should not be surprising, since
a finite cutoff distance makes the interaction integrable, as in short-range systems.

Monte Carlo evaluations and renormalization group analyses have been performed in the past to investigate the BEG model
with nearest-neighbor interaction (see, eg., Refs.~\cite{wang1987,berker1976,hoston1991,netz1993}). We note that these interesting
and relevant studies concern models with short-range interactions, where it suffices to study the canonical ensemble, since
ensemble equivalence occurs.

As noted in the introductory remarks, negative specific heat in the microcanonical ensemble was found in self-gravitating systems
long time ago; then, an important review by Padmanabhan~\cite{padma} was dedicated almost three decades ago to the
statistical mechanics of these systems (see also, e.g., Refs.~\cite{padma89,chava02}). In the study of self-gravitating systems,
an important role is played by spherical shell models, used to investigate the consequences of the singularity that is present in Newtonian
gravity. Model systems composed
of irrotational or rotating concentric self-gravitating shells have been employed to study phase transitions in the different
ensembles (see, e.g., Refs.~\cite{miller98,klinko01}). Ensemble inequivalence was found using mean-field theory, and confirmed
with molecular dynamics simulations. Another important aspect is the role played by boundary conditions.
Recently, it has been shown that in a onedimensional self-gravitating system in which periodic boundary conditions are imposed,
a second order phase transition at finite temperature is present, contrary to what happens with free boundary conditions, where no transition is
found~\cite{kumar2017}. Also in this case the study was conducted using mean-field theory, and the results were confirmed with
dynamical simulations.

Furthermore, we have considered the occurrence of ergodicity
breaking by studying the structure of the accessible regions in the
magnetization-energy phase space. Ergodicity breaking naturally occurs
in non additive systems. We have seen that also for this
feature the overall structure is strongly dependent on the relative
values of $\Delta$ and $K$.

Concluding, we have shown that even a simple model like the BEG model can present many different singular points in the phase diagrams
of the two ensembles and, correspondingly, ensemble inequivalence shows up in different ways.  This confirms the usefulness of
simple models as benchmarks for a detailed study of the possible physical situations that can arise in more complex and realistic
non additive systems.

\section*{Acknowledgments}

V.V.H. acknowledges financial support from project No. SCS
16YR-1C075. N.S.A. acknowledges financial support from the MC--IRSES
No. 612707 (DIONICOS) under FP7--PEOPLE--2013 and from project no.
SCS 15T-1C114. This work was made possible in part by a research
grant from the Armenian National Science and Education Fund (ANSEF, grant no. 4473)
based in New York, USA. A.C. and S.R. acknowledge financial support
from the Istituto Nazionale di Fisica Nucleare (INFN) through the
project DYNSYSMATH.

\appendix

\section{The canonical critical line, tricritical point and isolated critical point}
\label{appcantric}

Following Landau theory, canonical second order
phase transition points are characterized by the vanishing of the Hessian
of $\tilde{f}$. These points satisfy
Eqs.~(\ref{mag}) and (\ref{quad}), that we rewrite here in the
following form, using the common notation for partial derivatives
\begin{eqnarray}
\tilde{f}_x &=&0
\label{fx0} \\
\tilde{f}_y &=&0.
\label{fy0}
\end{eqnarray}
We are interested in the solutions for which $x=0$, $y=1/\beta$, characterizing the transition
between ferromagnetic and paramagnetic states. The vanishing of the Hessian is given by
\begin{equation}
\tilde{f}_{xx}\tilde{f}_{yy} - \tilde{f}_{xy}^2 = 0.
\label{hesso}
\end{equation}
Since the critical point itself is an equilibrium point, $\tilde{f}$
must have a minimum at this point. On the other hand, the vanishing of the
Hessian implies that there is a direction in the $(x,y)$ plane along
which the second order variation of $\tilde{f}$ vanishes. This
direction is exactly given by either one of the Eqs.~(\ref{fx0}) and
(\ref{fy0}). In our particular case $\tilde{f}$ is even in $x$, and
thus $\tilde{f}_{xy}=0$ at $x=0$, and Eq.~(\ref{hesso}) implies that
the critical points with $x=0$ are characterized by either
$\tilde{f}_{xx}=0$ or $\tilde{f}_{yy}=0$ \cite{foot3}. The critical
line corresponds to the case $\tilde{f}_{xx}=0$ (the case
$\tilde{f}_{yy}=0$ will be treated below). The condition
$\tilde{f}_{xx}=0$ is expressed by Eq.~(\ref{accan}). In this case
the direction of vanishing second order variation is the $x$ axis.
We use Eq.~(\ref{fy0}) to define implicitly a function $y(x)$. The
path in the $(x,y)$ plane defined by this function is tangent to the
$x$ axis in $x=0$, since $y_x=-\tilde{f}_{xy}/\tilde{f}_{yy}$ (then
$y_x=0$ at $x=0$), so that the tangent to this path in $x=0$ gives
the direction of vanishing Hessian. Moreover, for any given fixed
$x$ around $x=0$, $\tilde{f}_y=0$ gives the $y$ values for which
$\tilde{f}$ is minimum. It is then clear that we now have to study
the third order variation of $\tilde{f}$ as a function of $x$ when
$y$ is the function of $x$ implicitly given by Eq.~(\ref{fy0}).
Then, a straightforward calculation shows that the third order
variation is given by $(\dd x)^3/6$ multiplied by the expression
\begin{eqnarray}
&&\tilde{f}_{xxx}+3\tilde{f}_{xxy}y_x+3\tilde{f}_{xyy}y_x^2+\tilde{f}_{yyy}y_x^3+3\tilde{f}_{xy}y_{xx}
\nonumber \\
&+&3\tilde{f}_{yy}y_xy_{xx}+\tilde{f}_yy_{xxx},
\label{coeffthird}
\end{eqnarray}
computed at $x=0$. Since in our case $\tilde{f}$ is even in $x$, $\tilde{f}_y=0$ and $y_x=0$, then this expression vanishes.
An analogous longer but again straightforward calculation gives the expression of the fourth order variation.
Writing only the terms that do not vanish in $x=0$ we obtain the expression
\begin{equation}
\tilde{f}_{xxxx}+6\tilde{f}_{xxy}y_{xx}+3\tilde{f}_{yy}y_{xx}^2 \, .
\label{coefffourth}
\end{equation}
In this expression we substitute $y_{xx}$ as obtained by
Eq.~(\ref{fy0}). Again keeping only the terms that do not vanish we
get
\begin{equation}
\tilde{f}_{xxxx} - 3\frac{\tilde{f}_{xxy}^2}{\tilde{f}_{yy}}.
\label{coefffourhtb}
\end{equation}
The values of the terms of this expression at $x=0$, $y=1/\beta$ are easily computed, and we get the condition
for the positiveness of the fourth order variation
\begin{equation}
\beta \frac{-\beta \left( 1+2K \right) + \left( 3 + 2K \right)}{\beta - K\beta +K} >0 \, .
\label{appbccan}
\end{equation}
The numerator is exactly the expression in Eq.~(\ref{bccan}). It can
be easily checked that, as a function of $\beta$, the denominator is
positive when the numerator is positive, therefore the condition of
positiveness is given by the numerator, and we thus get
Eq.~(\ref{bccan}). When the numerator vanishes we obtain the
tricritical point.

Eqs.~(\ref{fx0}) and (\ref{fy0}) can admit other solutions with
$x=0$ but $y\ne 1/\beta$, and it is possible to have transitions
between paramagnetic states characterized by different values of the
quadrupole moment $q$. A critical point is then obtained when this
transition becomes continuous. Since these states have $x=0$, we can
treat this problem as if we had the single order parameter $y$. The
critical points are then characterized by
\begin{equation}
\tilde{f}_y=\tilde{f}_{yy}=\tilde{f}_{yyy}=0 \, ,
\label{ycritical}
\end{equation}
computed at $x=0$. There is no symmetry of $\tilde{f}$ as a function
of $y$, and then the odd derivatives with respect to $y$ do not
vanish identically. For a fixed $K$ value the three equations in
(\ref{ycritical}) determine the values of $y$, $\beta$ and $\Delta$,
and therefore there is a single critical point, at variance with the
line of critical points obtained before. This difference is a
consequence of the symmetry of $\tilde{f}$ with respect to
$x$~\cite{bouchet05}. The solution of the three equations
(\ref{ycritical}) is easily obtained
\begin{equation}
y=\frac{1}{2} \,\,\,\,\,\,\,\,\,\,\,\,\,\,\, \beta=\frac{4}{K} \,\,\,\,\,\,\,\,\,\,\,\,\,\,\,
\Delta=\frac{K}{4}\left( 2 + \ln 2 \right).
\label{solcritpoint}
\end{equation}
This solution holds for any $K>0$. However, since the former analysis is local, it does not determine if the
point is a local or global minimum of $\tilde{f}$; only in the latter case the critical point actually exists.
We have found that this happens only for $K$ larger than a threshold $K_1$, determined numerical to be about $2.775$.

In principle one could have a critical point in which
Eq.~(\ref{hesso}) is satisfied with all three second derivatives
$f_{xx}$, $f_{yy}$ and $f_{xy}$ different from $0$. This would imply
the presence of transitions between different ferromagnetic states
($f_{xy}$ can be different from $0$ only for $x$ different from
$0$). The associated isolated critical point would occur when this
transition becomes continuous. However, this system does not present
transitions between different ferromagnetic states, and therefore
such a critical point does not occur.

\section{The microcanonical tricritical point and isolated critical point}
\label{appmicrtric}

The values of $\Delta$ and $T$ at the microcanonical tricritical
point can be obtained by Eqs.~(\ref{entropycoeffa}),
(\ref{entropycoeffb}) and (\ref{microtempb}). The relation $B_m=0$,
used together with $A_m=0$ and the definition of $T$ in Eq.
(\ref{microtempb}), gives:
\begin{equation}
-\frac{3\beta}{\beta-1} + 6(\beta \Delta -K) -2(\beta \Delta -K)^2 +3K = 0 \, .
\label{bm01}
\end{equation}
On the one hand, this equation can be solved for $\Delta$ as a function of $\beta$,
by writing it in the form
\begin{eqnarray}
&&2\beta^2(\beta-1)\Delta^2 -2\beta(\beta-1)(3+2K)\Delta \nonumber \\
&+&3\beta + (\beta-1)(2K^2+3K)=0 \, ,
\label{bm02}
\end{eqnarray}
whose solution in terms of $T=1/\beta$ is
\begin{eqnarray}
\Delta_{MTP}&=&\frac{(2K+3)T_{MTP}}{2}
\label{DeltaMTP} \\
&-&\frac{\sqrt{3}T_{MTP}}{2}\sqrt{\frac{1-3T_{MTP}+2K(1-T_{MTP})}{1-T_{MTP}}}
\nonumber
\end{eqnarray}
(the relevant solution being the one giving the smaller value of
$\Delta_{MTP}$). On the other hand, by using again
Eq.~(\ref{microtempb}) in the first term of Eq.~(\ref{bm01}) allows
to write the latter equation as
\begin{equation}
-3\frac{e^{\beta \Delta-K}+2}{e^{\beta \Delta -K}} + 6(\beta \Delta -K) -2(\beta \Delta -K)^2 +3K = 0 \, .
\label{bm03}
\end{equation}
Defining $w=\beta \Delta -K$ we then have an equation for $w$
\begin{equation}
2w^2 -6w +6e^{-w} +3 -3K = 0 \, .
\label{bm04}
\end{equation}
Solving numerically this equation for $w$ one obtains $\beta_{MTP}
\Delta_{MTP}$. Dividing both members of Eq.~(\ref{DeltaMTP}) by
$T_{MTP}$ and plugging the obtained value in the left hand side one
can obtain $T_{MTP}$, after which one gets $\Delta_{MTP}$ by
$T_{MTP}(\beta_{MTP} \Delta_{MTP})=T_{MTP}(w+K)$.

As far as the dependence on order parameters is concerned, the functions $\tilde{s}_{\pm}(\epsilon,m)$ depend only on $m$,
and they are even in this variable. The points of microcanonical continuous transition should then occur
when the necessary condition $(\partial^2 \tilde{s}_{\pm})/(\partial m^2) =0$ is verified.
However, this is not the case as we now explain, and, as we will see, it is possible to have an isolated critical point, at the end of a line
of first order transitions, by another mechanism; however, this will be a rather peculiar critical point.
To have a lighter notation, we neglect, in the following, the subscript $\pm$, denoting with $\tilde{s}(\epsilon,m)$ the function to be maximized with respect
to $m$ to have $s(\epsilon)$; the argument does not depend on which of the functions $\tilde{s}_{\pm}$ realizes the absolute
maximum.

Using, as in the previous Appendix, the common notation for partial derivatives, the equilibrium magnetization, as we know,
is given by $\tilde{s}_m=0$ with $\tilde{s}_{mm}<0$. The former relation defines the function $m(\epsilon)$ at equilibrium, and
from it we get
\begin{equation}
\frac{\dd m}{\dd \epsilon} = -\frac{\tilde{s}_{m\epsilon}}{\tilde{s}_{mm}} \, .
\label{dmdeps}
\end{equation}
From this relation one obtains
\begin{equation}
\frac{\dd \beta}{\dd \epsilon} = \frac{\dd^2 s}{\dd \epsilon^2} =
\tilde{s}_{\epsilon \epsilon} - \frac{\tilde{s}_{m\epsilon}^2}{\tilde{s}_{mm}} \, .
\label{d2sdeps2}
\end{equation}
This relation shows that the points of continuous transition, where
$\beta(\epsilon)$ is continuous but $(\dd \beta)/(\dd \epsilon)$ has
a singularity, are characterized by $\tilde{s}_{mm}=0$, with the
further conditions $\tilde{s}_{mmm}=0$ and $\tilde{s}_{mmmm}<0$. The
critical line in the $(\Delta, T)$ plane for the continuous
transition between a ferromagnetic state and a paramagnetic state,
evaluated in Sec.~\ref{Microcanonical}, is included in this case,
since we have obtained it from a power expansion of
$\tilde{s}(\epsilon,m)$ at $m=0$, posing $A_m=\tilde{s}_{mm}/2=0$
and $B_m=\tilde{s}_{mmmm}/24<0$; the function
$\tilde{s}(\epsilon,m)$ is even in $m$, so that $\tilde{s}_{mmm}$
identically vanishes in $m=0$. Since this implies that also
$\tilde{s}_{m\epsilon}$ identically vanishes at $m=0$, the
singularity of $(\dd \beta)/(\dd \epsilon)$ at the transition point
is represented by a jump (except at the tricritical point, where
this quantity diverges). On the other hand, if a continuous
transition would occur at $m \ne 0$, where $\tilde{s}_{m\epsilon}\ne
0$ and $\tilde{s}_{mmm}$ does not vanish identically, one should
explicitly require that $\tilde{s}_{mmm}=0$ at the point of
continuous transition; this would then be the critical point at the
end of a line a first order transitions, and it would have $(\dd
\beta)/(\dd \epsilon) = +\infty$ ($\tilde{s}_{mm}$ tends to $0$ from
below). However, as in the canonical ensemble, this system does not
present transitions between different ferromagnetic states, and thus
such a critical point does not occur.

Nevertheless, Eq.~(\ref{d2sdeps2}) shows that a singularity in $(\dd
\beta)/(\dd \epsilon)$ can occur when there is a singularity in
$\tilde{s}_{\epsilon \epsilon}$. The presence of this singularity is
a consequence of the existence of an upper bound in the energy for
our system. We start by computing the first and second derivatives
of $\tilde{s}$ with respect to $\epsilon$; in doing so, we assume
from the start that $m=0$, since we have to find the critical point of
the transition between two paramagnetic states. Therefore we have:
\begin{eqnarray}
\frac{\partial \tilde{s}_{\pm}}{\partial \epsilon} &=& \beta = \frac{\partial \tilde{s}_{\pm}}{\partial q}\frac{\partial q_{\pm}}{\partial \epsilon}
\nonumber \\
&=& \mp \frac{1}{K}\ln \left[ \frac{2(1-q_{\pm})}{q_{\pm}}\right] \frac{1}{\sqrt{\left(\frac{\Delta}{K}\right)^2 -\frac{2\epsilon}{K}}} \, ,
\label{tdsdeps}
\end{eqnarray}
and
\begin{eqnarray}
\frac{\partial^2 \tilde{s}_{\pm}}{\partial \epsilon^2} &=&
\frac{\partial^2 \tilde{s}_{\pm}}{\partial q^2}\left(\frac{\partial q_{\pm}}{\partial \epsilon}\right)^2
+\frac{\partial \tilde{s}_{\pm}}{\partial q}\frac{\partial^2 q_{\pm}}{\partial \epsilon^2} \nonumber \\
&=& -\frac{1}{K^2 q_{\pm}(1-q_{\pm})\left[\left(\frac{\Delta}{K}\right)^2 -\frac{2\epsilon}{K}\right]}
\nonumber \\
&\mp& \frac{1}{K^2} \ln \left[ \frac{2(1-q_{\pm})}{q_{\pm}}\right]
\frac{1}{\left[\left(\frac{\Delta}{K}\right)^2 -\frac{2\epsilon}{K}\right]^{\frac{3}{2}}}
\label{tdsdeps2}
\end{eqnarray}
where for convenience we rewrite the expression of $q_{\pm}$ for $m=0$:
\begin{equation}
q_{\pm} = \frac{\Delta}{K} \pm \sqrt{\left(\frac{\Delta}{K}\right)^2-\frac{2\epsilon }{K}} \, .
\label{qpmsolapp}
\end{equation}
We are interested in the value of Eqs.~(\ref{tdsdeps}) and
(\ref{tdsdeps2}) at the upper bound $\epsilon_{\rm max}$ of the
energy. Let us begin with the case $\Delta/K \ge 1$. Obviously in
this case $q_+$ is not acceptable, and we have to consider only
$q_-$. When $\epsilon \to \epsilon_{\rm max} =\Delta - K/2$, then
$q_- \to 1^-$. Thus, from Eq.~(\ref{tdsdeps}) we see that $\beta \to
-\infty$, and the upper bound of the energy coincides with $T=0^-$.
Since the first derivative diverges, also the second does, but then
this is not associated to a critical point; this point of the phase
diagram does not mark the end of a line of first order transitions.

We now consider $\Delta/K <1$. We separate this analysis in three
parts, corresponding to $\Delta/K < 2/3$, $\Delta/K > 2/3$ and
$\Delta/K = 2/3$. The reason is the following. For $\Delta/K <1$ the
upper bound $\epsilon_{\rm max}=\Delta^2/(2K)$ is realized for
$q=\Delta/K$, and the value $q=2/3$ is the one separating positive
and negative values of the logarithm appearing in
Eq.~(\ref{tdsdeps}). We know that the entropy of the system will be
given by the largest between $\tilde{s}_+$ and $\tilde{s}_-$. For
$\Delta/K < 2/3$ (respectively $>2/3$) and an energy sufficiently
close to $\epsilon_{\rm max}$ also $q$ will be $<2/3$ (respectively
$>2/3$). Since $(\partial \tilde{s})/(\partial q) = \ln
\left[2(1-q)/q\right]$ is positive (respectively negative) for $q<
2/3$ (respectively $>2/3$) we have to choose $q_+$, i.e.
$\tilde{s}_+$ (respectively $q_-$, i.e. $\tilde{s}_-$) for $\Delta/K
<2/3$ (respectively $\Delta/K >2/3$). Then we see from
Eq.~(\ref{tdsdeps}) that, for both $\Delta/K <2/3$ and $\Delta/K >
2/3$, $\beta \to -\infty$ when $\epsilon \to \epsilon_{\rm max}$. As
before, the first derivative diverges, and the divergence of the
second derivative is not associated to a critical point.

On the other hand, for $\Delta/K=2/3$ the situation is different.
For $q=2/3$, $(\partial \tilde{s})/(\partial q) = \ln
\left[2(1-q)/q\right]=0$, so to determine the choice between $q_+$
and $q_-$ we have to compute higher derivatives of $\tilde{s}$ with
respect to $q$. Without showing the calculations, we give the result
that we have to choose $q_-$, i.e. $\tilde{s}_-$, so that, for
$\epsilon \to \epsilon_{\rm max}$, $q \to \Delta/K = 2/3$ from
below. Then, from Eq.~(\ref{tdsdeps}) we get that $\beta$ tends to a
positive value; the important thing is that this value is finite.
This can be obtained performing the limit, in which the argument of
the logarithm tends to $1$ and the argument of the square root in
the denominator tends to $0$. One gets that the limit of $\beta$ is
equal to $9/(2K)$. Performing the same limit evaluation in
Eq.~(\ref{tdsdeps2}) one finds that the second derivatives diverges
to $+\infty$ (thus the derivative of $T$ with respect to $\epsilon$
diverges to $-\infty$). This is exactly the divergence giving rise
to the critical point, that occurs at $\Delta=2K/3$, $T=2K/9$ and
$\epsilon = 2K/9$. This point is at the end of a line of first order
transitions. In Fig.~\ref{CaloricK5}(d) we show the caloric curve
for that value of $\Delta$ for $K=5$. Since the critical point has
the peculiarity of occurring at exactly the upper bound of the
energy, in that graph the temperature has a vertical jump to
$T=+\infty$ at the same energy. This critical point can be present
only when there is a line of microcanonical first order transitions
between different paramagnetic states; as described in
Sec.~\ref{phase_analysis} this occurs for $K>K_2=3$. For $K<3$
the point of the microcanonical $(\Delta,T)$ phase diagram
with coordinates $\Delta=2K/3$ and $T=2/3$, corresponding in the
$(\Delta,\epsilon)$ diagram to $\epsilon=2K/9$, belongs to the
critical line of second order transitions. In that case, a limiting
procedure in Eq.~(\ref{d2sdeps2}), tedious but straightforward,
shows that the caloric curve has a finite derivative for $\epsilon$
tending to the upper bound.

\section{The energy range as a function of the magnetization}
\label{appenerrange}

Considering the right hand side of Eq.~(\ref{redhamil}) as a
function of $q$, it is a downward parabola with vertex in
$q=\frac{\Delta}{K}$, then with positive derivative for
$q<\frac{\Delta}{K}$ and negative derivative for
$q>\frac{\Delta}{K}$. We are thus led to consider two main cases,
i.e. $\frac{\Delta}{K} \ge 1$ and $\frac{\Delta}{K} \le 1$,
respectively.

In the first case the range $m\le q \le 1$ is entirely in the region of the parabola with positive derivative, then for any
given $m$ the minimum and maximum values of $\epsilon$ when $q$ varies in
$m\le q \le 1$ are obtained for $q=m$ and $q=1$, respectively; namely
\begin{eqnarray}
\epsilon_{\rm min}(m) &=& \Delta m - \frac{K}{2} m^2 -\frac{m^2}{2}
\nonumber \\
\epsilon_{\rm max}(m) &=& \Delta - \frac{K}{2}  -\frac{m^2}{2}
\nonumber \\
\frac{\Delta}{K} \ge 1 &&\,\,\,\,\,\,\,\,\,\,\, 0\le m \le 1 \, .
\nonumber
\label{case1}
\end{eqnarray}

When $\frac{\Delta}{K} \le 1$ we have to distinguish between two
situations, i.e. $0\le m \le \frac{\Delta}{K}$ and $\frac{\Delta}{K}
\le m \le 1$. The latter is simpler, since the range $m \le q \le 1$
is entirely in the region of the parabola with negative derivative,
then the minimum and maximum values of $\epsilon$ are obtained for
$q=1$ and $q=m$, respectively; namely
\begin{eqnarray}
\epsilon_{\rm min}(m) &=& \Delta - \frac{K}{2}  -\frac{m^2}{2}
\nonumber \\
\epsilon_{\rm max}(m) &=& \Delta m - \frac{K}{2} m^2 -\frac{m^2}{2}
\nonumber \\
\frac{\Delta}{K} \le 1 &&\,\,\,\,\,\,\,\,\,\,\, \frac{\Delta}{K} \le
m \le 1 \, .
\nonumber
\label{case2b}
\end{eqnarray}
In the second situation, $0\le m \le \frac{\Delta}{K}$, the range $m
\le q \le 1$ is partly in the region with positive derivative of the
parabola and partly in the region with negative derivative.
Therefore the maximum value of $\epsilon$ occurs for
$q=\frac{\Delta}{K}$, and thus it is given by
\begin{eqnarray}
\epsilon_{\rm max}(m) &=& \frac{\Delta^2}{2K}  -\frac{m^2}{2}
\nonumber \\
\frac{\Delta}{K} \le 1 &&\,\,\,\,\,\,\,\,\,\,\, 0\le m \le
\frac{\Delta}{K} \, .
\nonumber
\label{case2amax}
\end{eqnarray}
To determine the minimum we have to see if $\epsilon$ is smaller for
$q=m$ or for $q=1$. Substituting respectively $q=m$ and $q=1$ in
Eq.~(\ref{redhamil}) we have that $\epsilon(q=m)\le \epsilon(q=1)$
when
\begin{equation}
\frac{K}{2} m^2 +\Delta(1-m) - \frac{K}{2} \ge 0 \, .
\nonumber
\label{discr}
\end{equation}
The roots of the left hand side of this inequality are
$m=\frac{2\Delta}{K} - 1$ and $m=1$; the former is smaller than
$\frac{\Delta}{K}$ since $\frac{\Delta}{K} \le 1$. Therefore
$\epsilon(q=m)\le \epsilon(q=1)$ for $m \le  \frac{2\Delta}{K} - 1$
and for $m\ge 1$. The latter is not of interest, while for the
former we have to determine if $\frac{2\Delta}{K} - 1 \ge 0$. This
is not verified if $\frac{\Delta}{K} < \frac{1}{2}$, and in this
subcase we have that $\epsilon(q=m) > \epsilon(q=1)$ in particular
for any $0 \le m\le \frac{\Delta}{K}$. Then
\begin{eqnarray}
\epsilon_{\rm min}(m) &=& \Delta - \frac{K}{2}  -\frac{m^2}{2}
\nonumber \\
\frac{\Delta}{K} \le \frac{1}{2} &&\,\,\,\,\,\,\,\,\,\,\, 0\le m \le
\frac{\Delta}{K} \, .
\nonumber
\label{case2amin1}
\end{eqnarray}
On the contrary, when $\frac{1}{2}\le \frac{\Delta}{K}\le 1$, then
$\frac{2\Delta}{K} - 1 \ge 0$, and $\epsilon(q=m)\le \epsilon(q=1)$
for $0\le m \le \frac{2\Delta}{K} -1$, while $\epsilon(q=m)\ge
\epsilon(q=1)$ for $\frac{2\Delta}{K} - 1 \le m \le
\frac{\Delta}{K}$. So
\begin{eqnarray}
\epsilon_{\rm min}(m) &=& \Delta m - \frac{K}{2} m^2  -\frac{m^2}{2}
\nonumber \\
\frac{\Delta}{K} \ge \frac{1}{2} &&\,\,\,\,\,\,\,\,\,\,\, 0\le m \le
\frac{2\Delta}{K} -1 \, ,
\nonumber
\label{case2amin2a}
\end{eqnarray}
and
\begin{eqnarray}
\epsilon_{\rm min}(m) &=& \Delta - \frac{K}{2}  -\frac{m^2}{2}
\nonumber \\
\frac{\Delta}{K} \ge \frac{1}{2} &&\,\,\,\,\,\,\,\,\,\,\,
\frac{2\Delta}{K} -1 \le m \le \frac{\Delta}{K} \, .
\nonumber
\label{case2amin2b}
\end{eqnarray}
The main text summarizes the results.


\begin{thebibliography}{99}
\bibitem{foot7}
While in the canonical case $x$ and $y$ are auxiliary variables used in the Hubbard-Stratonovich transformation, and they are equal
to the magnetization $m$ and the quadrupole moment $q$, respectively, only at the equilibrium values, in the microcanonical case
we use directly $m$ and $q$. However, as explained in the text, $(m,q)$ are the equilibrium values only when
they optimize the $S$ given in Eq. (\ref{entropyqm}).
\bibitem{foot1}
Actually, in models with an upper bound on the energy (due to the absence of a kinetic energy term),
energy values close to the upper bound could be associated to negative temperatures. This is what occurs for
the BEG model. However, this happens for energy values higher than those occurring on the critical line.
The physical meaning of negative temperatures is briefly reminded in Sec.~\ref{negtempsec}. See also
footnote~\cite{TGBtemp}.
\bibitem{foot2}
The allowed range of $\epsilon$ depends on $\Delta$,
and for $\Delta<(K+1)/2$ the minimum $\epsilon$ is negative (see Sec.~\ref{ergbreak}).
Fig. \ref{Energy-Delta}, showing only the most interesting region of the phase diagram, contains only
positive $\epsilon$ values.
\bibitem{foot4}
We remind that, at negative temperatures, the free energy of the system reaches the maximum value compatible 
with the imposed constraints; not the minimum value, at variance with the standard
positive temperature. That is why for $T=0^-$ the energy, coincident in that case with the free energy, achieves 
the upper bound.
\bibitem{foot5}
We remind the origin of this fact. If the microcanonical entropy function has a singularity, its Legendre-Fenchel transform
$\phi(\beta) \equiv \beta f(\beta)$, where $f(\beta)$ is the canonical free energy, has a discontinuous derivative. Therefore,
if $\phi(\beta)$ has a continuous derivative, i.e. if the canonical ensemble has no transition or a continuous transition,
the microcanonical entropy has no singularity.
\bibitem{foot3}
In this case it is necessary to perform a limiting procedure in one of the
Eqs. (\ref{fx0}) and (\ref{fy0}) to see that they both define the same direction at the critical point.
\end{thebibliography}
\end{document}